\documentclass{article}
\usepackage[letterpaper, margin=1in]{geometry} 
\usepackage{amssymb}
\usepackage{amsmath}
\allowdisplaybreaks
\usepackage{amsthm}
\usepackage{xspace}
\usepackage{physics}
\usepackage[dvipsnames]{xcolor}
\usepackage{tikz}
\usetikzlibrary{positioning, calc, decorations.pathmorphing}
\usepackage{mathrsfs}
\usepackage{mathtools}
\usepackage{dirtytalk}
\usepackage{stmaryrd}
\usepackage{enumitem}
\usepackage{relsize}
\usepackage{graphicx}
\usepackage[normalem]{ulem}
\usepackage{bbm}
\usepackage[colorlinks=true, allcolors=blue]{hyperref}
\usepackage[capitalise,nameinlink]{cleveref}
\usepackage{tcolorbox}
\tcbuselibrary{skins, breakable}

\usepackage[font=small]{caption}

\newtheorem{theorem}{Theorem}[section]

\newtheorem{lemma}[theorem]{Lemma}
\newtheorem{definition}[theorem]{Definition}

\crefname{assumption}{Assumption}{Assumptions}
\newtheorem{proposition}[theorem]{Proposition}

\crefname{conjecture}{Conjecture}{Conjectures}
\newtheorem{observation}[theorem]{Observation}

\newcommand{\cOp}[1]{\ensuremath{\mathrm{#1}}\xspace}
\newcommand{\cpO}{\cOp{cpO}}
\newcommand{\pC}{\cOp{pC}}
\newcommand{\cpwt}{\cOp{cpWT}}

\newcommand{\cpwtZ}{\cOp{cpWT_{\circ}}}
\newcommand{\cpLabInv}{\cOp{cpO^{inv}_{lab}}}
\newcommand{\cpLabFor}{\cOp{cpO^{for}_{lab}}}
\newcommand{\cpPerm}{\cOp{cpO_{edge}}}
\newcommand{\xorOP}{\cOp{P}}
\newcommand{\wtO}{\ensuremath{O^{(\mathrm{wt})}}\xspace}
\newcommand{\wtOZ}{\ensuremath{O^{\mathrm{wt}}_{\circ}}}
\newcommand{\wtF}{\ensuremath{F^{(\mathrm{wt})}}\xspace}
\newcommand{\middleCol}{\ensuremath{\mathrm{B}_{\mathrm{mid}}}\xspace}
\newcommand{\middleEdge}{\mathrm{B}_{\mathrm{edge}}\xspace}
\newcommand{\middleNodes}{\mathrm{V}_{\mathrm{mid}}\xspace}
\newcommand{\edgeIndic}{\mathrm{b_{\mathrm{edge}}}\xspace}
\newcommand{\icpwt}{\ensuremath{\widehat{\cpwt}}\xspace}

\newcommand{\edgeVertex}[2]{\mathbf{e}_{#1}(#2)\xspace}
\newcommand{\probAssignEdge}[2]{\ensuremath{\kappa}_{#1}(#2)}
\newcommand{\colorVertMap}{\phi}
\newcommand{\goodMap}{G}
\newcommand{\canVec}[1]{\ket{g_{#1}}}
\newcommand{\distWeld}{l}
\newcommand{\errorTerm}{\epsilon_{\distWeld}}
\newcommand{\permSize}{N_{\distWeld}}

\newcommand{\pCLabelInv}{\cOp{pC_{lab}^{inv}}}
\newcommand{\pCPerm}{\cOp{pC_{edge}}}
\newcommand{\pCLabel}{\cOp{pC_{lab}^{for}}}
\newcommand{\pCfromPlus}{\cOp{pCF}}
\newcommand{\stepFunct}{\mathrm{sF}}

\newcommand{\labelPerm}{\ensuremath{\sigma_{\textup{label}}}\xspace}
\newcommand{\colorPerm}{\ensuremath{\sigma_{\textup{edge}}}\xspace}
\newcommand{\colorPresSet}{\ensuremath{S_{\textup{edge}}}\xspace}
\newcommand{\head}{\ensuremath{\mathrm{head}}}
\newcommand{\tail}{\ensuremath{\mathrm{tail}}}
\newcommand{\treeEdges}{E_{\mathrm{tr},c,\alpha}}
\newcommand{\weldedEdges}{E_{\mathrm{wld}, \alpha}}

\newcommand{\multDistr}{\ensuremath{\mathcal{D}_{\mathrm{prod}}}\xspace}
\newcommand{\singDistr}{\ensuremath{\mathcal{D}_{\mathrm{sing}}}\xspace}
\newcommand{\evSingle}{\ensuremath{\mathcal{S}}\xspace}

\newcommand{\register}[1]{\ensuremath{\mathrm{#1}}\xspace}
\newcommand{\X}{\register{X}}
\newcommand{\Y}{\register{Y}}
\newcommand{\Z}{\register{Z}}
\newcommand{\A}{\register{A}}
\newcommand{\I}{\register{I}}
\newcommand{\C}{\register{C}}
\newcommand{\Lab}{\register{L}}
\newcommand{\regP}{\register{P}}
\newcommand{\db}{\ensuremath{\mathbf{I}}\xspace}
\newcommand{\col}{\ensuremath{\textup{col}}\xspace}

\newcommand{\rG}{\ensuremath{\hat{G}}}
\newcommand{\rV}{\ensuremath{\hat{V}}}
\newcommand{\rE}{\ensuremath{\hat{E}}}

\newcommand{\Sym}{\ensuremath{\mathrm{Sym}}}

\newcommand{\prv}[1]{\ensuremath{\Pi_{#1, \I}}}
\newcommand{\prCross}{\ensuremath{\Pi_{\mathrm{Cross}}}}

\newcommand{\probWTO}{\ensuremath{p_{\mathrm{wt}}}}
\newcommand{\probCross}{\ensuremath{p_{\mathrm{cross}}}}

\newcommand{\Gram}{K}

\newcommand{\Vclass}[1]{\mathrm{V}_{\mathrm{class}}(#1)}
\newcommand{\Tdiff}{T^{\mathrm{diff}}}
\newcommand{\Err}{\mathrm{Err}}
\newcommand{\Psrc}{\Pi^{\mathrm{src}}}
\newcommand{\Pwork}{\Pi^{\mathrm{work}}}
\newcommand{\tcut}{t^\prime}
\newcommand{\Vweld}{\mathrm{V}_{\mathrm{weld}}}
\newcommand{\Vbd}{\mathrm{V}_{\mathrm{bd}}}
\newcommand{\forestcol}{\mathrm{Color}_H}
\newcommand{\forestlab}{\lambda_H}
\newcommand{\Ctree}{C^{\mathrm{tree}}}
\newcommand{\Sfix}{S^{\mathrm{fix}}}
\newcommand{\Wfix}{W^{\mathrm{fix}}}
\newcommand{\muref}{\mu^{\mathrm{ref}}}
\newcommand{\Zref}{Z^{\mathrm{ref}}}
\newcommand{\Fbd}{F^{\mathrm{bd}}}
\newcommand{\Scoef}{S}
\newcommand{\Pdist}{\ensuremath{\Pi^{\mathrm{inj}}}}
\newcommand{\Pgood}{\Pi^{\mathrm{good}}}
\newcommand{\parentop}{\mathrm{par}}
\newcommand{\Udesc}{U^{\mathrm{desc}}}
\newcommand{\Pcoin}{\Pi^{\mathrm{coin}}}
\newcommand{\Kdist}{\ker(\Pdist|_{\Scoef})}
\newcommand{\Aut}{\mathrm{Aut}}
\newcommand{\etastep}{\eta^{\mathrm{step}}}

\DeclareMathOperator{\EX}{\mathbb{E}}

\DeclareMathOperator{\dom}{\mathrm{Dom}}
\DeclareMathOperator{\im}{\mathrm{Im}}
\DeclareMathOperator{\ran}{\mathrm{Ran}}
\DeclareMathOperator{\Id}{\mathbbm{1}}
\DeclarePairedDelimiter\ceil{\lceil}{\rceil}
\DeclarePairedDelimiter\floor{\lfloor}{\rfloor}
\newcommand{\groot}{\ensuremath{\mathrm{r}}} % I am g(raph) root :)
\newcommand{\depth}{\ensuremath{\mathrm{depth}}}
\newcommand{\strucIdx}{\ensuremath{\mathrm{w}}\xspace}
\newcommand{\alg}{\ensuremath{\mathcal{A}}\xspace}
\newcommand{\calE}{\ensuremath{\mathcal{E}}}
\newcommand{\calH}{\ensuremath{\mathcal{H}}}

\newcommand{\calA}{\ensuremath{\mathcal{A}}}
\newcommand{\calI}{\ensuremath{\mathcal{I}}}
\newcommand{\calV}{\ensuremath{\mathcal{V}}}
\newcommand{\calK}{\ensuremath{\mathcal{K}}}
\newcommand{\calS}{\ensuremath{\mathcal{S}}}
\newcommand{\calD}{\ensuremath{\mathcal{D}}}
\newcommand{\calL}{\ensuremath{\mathcal{L}}}
\newcommand{\calG}{\ensuremath{\mathcal{G}}}
\newcommand{\pairPath}{\mathbf{x_{\textup{path}}}}
\providecommand{\Span}{\mathsf{Span}}
\newcommand{\colorAlpha}{{\color{colorA}\alpha}}

\newcommand{\Ext}{\ensuremath{\mathrm{Ext}}}

\colorlet{colorA}{RoyalBlue}
\colorlet{colorB}{ForestGreen}
\colorlet{colorC}{BrickRed}
\newcommand{\colorSet}{\ensuremath{\mathcal{C}}}
\newcommand{\upcolor}{\ensuremath{\chi}}
\newcommand{\colF}{\ensuremath{\mathrm{Color}}\xspace}
\newcommand{\ncF}{\ensuremath{\mathrm{Nb}}}
\newcommand{\colClass}[1]{\ensuremath{\mathcal{C}_{#1}}}
\newcommand{\colClassSize}[1]{\ensuremath{\mathrm{N}_{#1}}}

\usepackage{mdframed}
  {%
  \begin{mdframed}%
  [leftmargin=.5em,rightmargin=0em,#1,bottomline=false,topline=false,rightline=false,linewidth=2pt,linecolor=black!20,innerleftmargin=.5em,innertopmargin=0pt,innerrightmargin=0pt]\begingroup\vspace*{0pt}}%
  {\endgroup
  \end{mdframed}
  }

  {%
  \begin{mdframed}%
  [leftmargin=.5em,rightmargin=0em,#1,bottomline=false,topline=false,rightline=false,linewidth=2pt,linecolor=red!20,innerleftmargin=.5em,innertopmargin=0pt,innerrightmargin=0pt]\begingroup\vspace*{-2pt}}%
  {\endgroup
  \end{mdframed}
  }

\definecolor{algbg}{HTML}{F7F7F7}
\definecolor{algframe}{HTML}{CCCCCC}
\definecolor{alglabel}{HTML}{555555}
\definecolor{inputlabel}{HTML}{888888}

\newcounter{algoctr}[section]

\crefname{algoctr}{Algorithm}{Algorithms}
\Crefname{algoctr}{Algorithm}{Algorithms}

\newcommand{\alglabel}[1]{%
  \setcounter{algoctr}{\value{theorem}}%
  \addtocounter{algoctr}{-1}%
  \refstepcounter{algoctr}%
  \label{#1}%
}

\newtcolorbox{algobox}[2][]{%          #1 = tcolorbox keys, #2 = title text
  enhanced,
  breakable,
  colback    = algbg,
  colframe   = algframe,
  boxrule    = 0.3pt,
  arc        = 1.5pt,
  left       = 10pt,
  right      = 10pt,
  top        = 2pt,
  bottom     = 10pt,
  fontupper  = \small,
  title      = {\refstepcounter{theorem}%
                \small\textsc{Algorithm \thetheorem}\;\;#2},
  coltitle   = alglabel,
  colbacktitle = algbg,
  toptitle   = 6pt,
  bottomtitle = 6pt,
  #1
}

\let\originalleft\left
\let\originalright\right
\renewcommand{\left}{\mathopen{}\mathclose\bgroup\originalleft}
\renewcommand{\right}{\aftergroup\egroup\originalright}

\usepackage{authblk}

\makeatletter
\renewcommand\AB@authnote[1]{\rlap{\textsuperscript{\normalfont#1}}}

\makeatother

\title{Algorithmic Wave-Particle Duality:\\ Pathfinding in Welded Trees is Hard}

\author[1]{Joseph Carolan}
\author[1]{Andrew M.~Childs}
\author[1]{Matthew Coudron}
\author[1]{\authorcr Amin Shiraz Gilani}
\author[2]{David Miloschewsky}
\author[2]{Supartha Podder}
\affil[1]{Department of Computer Science, Institute for Advanced Computer Studies, and \protect\\
Joint Center for Quantum Information and Computer Science, University of Maryland}
\affil[2]{Department of Computer Science, Stony Brook University}

\date{}

\begin{document}

\maketitle

\begin{abstract}
    In the welded trees problem, an algorithm navigates a randomly labeled graph formed from two binary trees joined at their leaves through a “weld” of connecting edges. A quantum walk can navigate from root to root exponentially faster than any classical algorithm. However, recording its path destroys the constructive interference responsible for the speedup, leaving open whether some quantum algorithm can efficiently find a path between the roots.

    We prove that, for trees of height $n$, the quantum query complexity of finding a path between the roots with constant probability is $\tilde{\Theta}(2^{n/3})$. 
    This establishes an algorithmic wave-particle duality, adapting the double-slit experiment to the high-complexity regime: 
    interference between many paths lets a quantum algorithm reach the exit efficiently, but finding a single path 
    (the algorithmic analog of the route of a particle) 
    requires exponentially many queries.
    Furthermore, the bound is tight, matching the cost of a quantum algorithm that finds collisions between paths from each root to the weld. 
    Our proof uses the compressed permutation oracle to track the information about the graph that an algorithm retains and forgets. We show that the compressed database remains close to a space of states that do not record a path between the roots by bounding the error introduced by each compressed oracle query.
\end{abstract}

\tableofcontents

\pagebreak
\section{Introduction}

Quantum computers can solve certain problems dramatically faster than classical ones, but the extent of this advantage remains poorly understood. Notable families of algorithms with superpolynomial speedup include those based on finding periodic structures using Fourier transforms~\cite{Sho97} and simulating the dynamics of quantum systems~\cite{Llo96}. Identifying other mechanisms for quantum speedup, and understanding their limitations, remain central challenges.

\begin{figure}
    \center
    % The graph of glued_tree.tex, in the layout of fresh_oracle.tex.
\begin{tikzpicture}[
    scale=0.8,
    x=1cm, y=1cm,
    basicVertex/.style={circle, draw=black!75, fill=black!20,
        minimum size=14pt, inner sep=0pt, line width=0.9pt},
    basicEdge/.style={draw=black!80, line width=0.9pt}
]

% Two binary trees facing one another across the weld.
\foreach \v/\x/\y in {
    0/0/0, 1/1.8/1.8, 2/1.8/-1.8,
    3/3.6/2.7, 4/3.6/0.9, 5/3.6/-0.9, 6/3.6/-2.7,
    7/5.4/3.15, 8/5.4/2.25, 9/5.4/1.35, 10/5.4/0.45,
    11/5.4/-0.45, 12/5.4/-1.35, 13/5.4/-2.25, 14/5.4/-3.15,
    22/8.2/3.15, 23/8.2/2.25, 24/8.2/1.35, 25/8.2/0.45,
    26/8.2/-0.45, 27/8.2/-1.35, 28/8.2/-2.25, 29/8.2/-3.15,
    18/10/2.7, 19/10/0.9, 20/10/-0.9, 21/10/-2.7,
    16/11.8/1.8, 17/11.8/-1.8, 15/13.6/0
} { \node[basicVertex] (v\v) at (\x,\y) {}; }

% Tree edges.
\foreach \u/\v in {
    0/1, 0/2, 1/3, 1/4, 2/5, 2/6,
    3/7, 3/8, 4/9, 4/10, 5/11, 5/12, 6/13, 6/14,
    15/16, 15/17, 16/18, 16/19, 17/20, 17/21,
    18/22, 18/23, 19/24, 19/25, 20/26, 20/27, 21/28, 21/29%
} { \draw[basicEdge] (v\u) -- (v\v); }

% The same weld edges as in glued_tree.tex.
% White underlays keep crossing edges visually distinct.
\foreach \u/\v in {
    7/27, 7/28, 8/29, 8/24, 9/22, 9/28, 10/26, 10/29,
    11/25, 11/23, 12/23, 12/27, 13/22, 13/24, 14/25, 14/26%
} {
    \draw[basicEdge, preaction={draw=white, line width=2.9pt}]
        (v\u) -- (v\v);
}
% Distinct colors for the entrance, exit, and leaves of the two trees.
\node[basicVertex, draw=colorA, fill=colorA!25] at (v0) {};
\node[basicVertex, draw=colorB, fill=colorB!25] at (v15) {};
\foreach \v in {7,8,9,10,11,12,13,14,22,23,24,25,26,27,28,29} {
    \node[basicVertex, draw=colorC, fill=colorC!25] at (v\v) {};
}
\node[anchor=east, font=\small\bfseries] at (v0.west) {Entrance};
\node[anchor=west, font=\small\bfseries] at (v15.east) {Exit};
\node[anchor=south, yshift=14pt, font=\small\bfseries] at (6.8,3.15) {Weld};
\end{tikzpicture}
    \caption{An instance of a welded trees graph. The \textcolor{colorA}{blue} node is the entrance, the \textcolor{colorB}{green} node is the exit and the \textcolor{colorC}{red} nodes are the leaves. The edges between the leaves are the weld edges.}
    \label{fig:basic_welded_tree}
\end{figure}
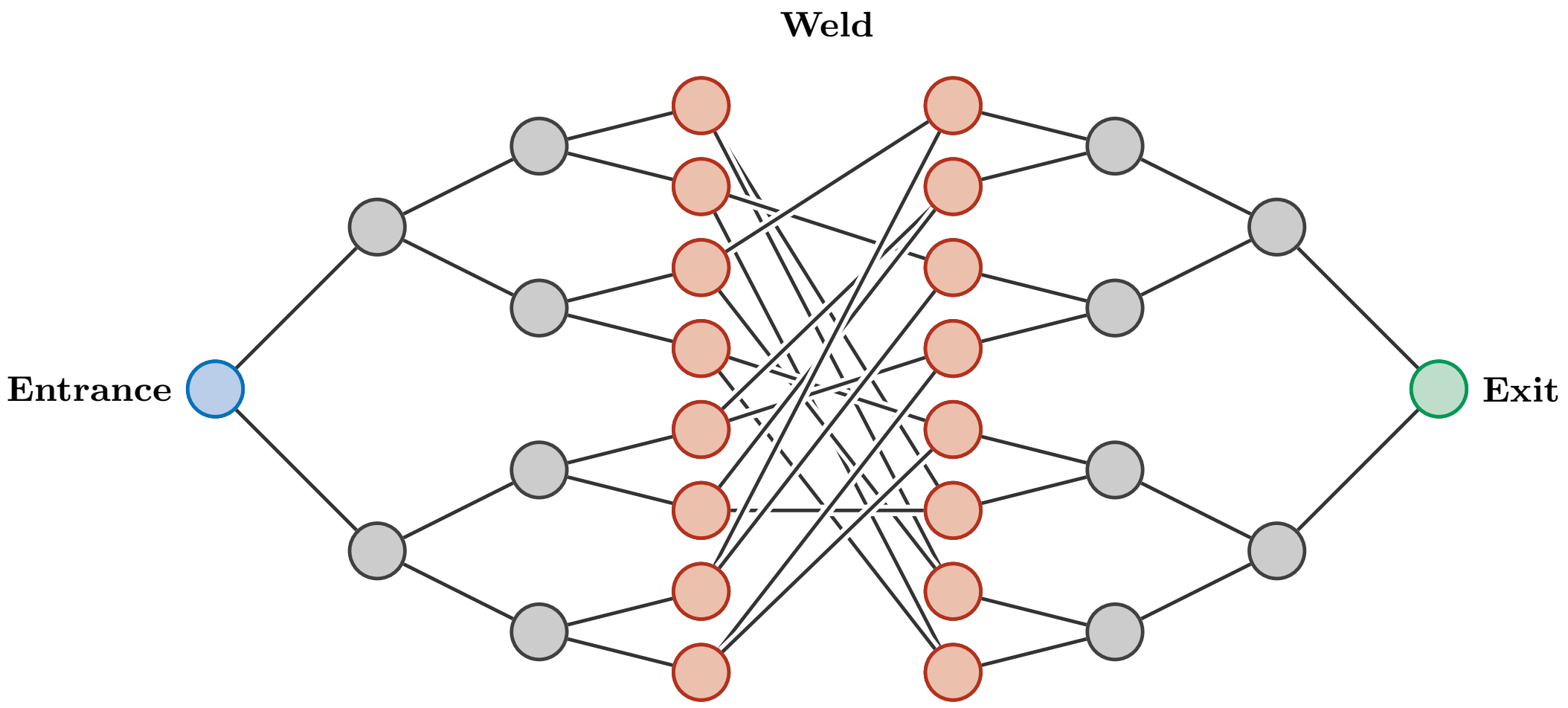

The \emph{welded trees problem} provides an example of a different kind of speedup that is due to interference when coherently exploring a graph. The graph, illustrated in \cref{fig:basic_welded_tree}, consists of two height-$n$ balanced binary trees whose leaves are joined by a random cycle alternating between the two trees (the \say{weld}). Vertices have random labels and an algorithm accesses the graph through an adjacency-list oracle. Given the label of one root, the \emph{entrance}, a quantum walk finds the other root, the \emph{exit}, in time polynomial in $n$, whereas every classical algorithm requires exponentially many queries~\cite{CCD+03}. The quantum walk exploits the layered structure of the graph. Its evolution stays in the span of uniform superpositions over successive layers, effectively reducing the problem to traversal of a path of length $O(n)$. On the other hand, after crossing the weld, a classical traversal has little information about which edges lead towards the exit. Reaching a previously visited node along an unexplored edge could help identify a way towards the exit, but even this task requires exponentially many queries.

Beyond providing a new mechanism for quantum speedup, the welded trees problem has yielded insight into a variety of other questions pertaining to the power of quantum computation. It was used to show that some quantum speedups cannot be achieved with only logarithmic quantum depth~\cite{CM19}, and was a building block for results on the power of adiabatic computation and optimization~\cite{GV20,LWWZ25}, the quantum query complexity of graph property testing~\cite{BCGKPW24}, and exponential separations for machine learning tasks~\cite{GL24}. Understanding what quantum algorithms can and cannot do on these graphs therefore informs a broader set of questions about quantum advantage.

The problem also suggests a striking distinction between reaching a destination and finding a route to it. A classical traversal can record its route at little additional cost, so in particular, exit finding and pathfinding have the same classical query complexity, which is $\tilde{\Theta}(2^{n/2})$~\cite{FZ03,BD14,AA18}. On the other hand, the quantum walk traverses many paths in superposition without outputting any one of them. Retaining distinguishable records of these paths suppresses the interference responsible for the speedup, similarly to how recording which slit a particle passes through destroys interference in the double-slit experiment. This explains why simply recording the quantum walk eliminates the speedup, but leaves open whether a different quantum algorithm could find a path efficiently. This has been a longstanding open problem, alluded to in the original paper~\cite{CCD+03} and highlighted in a list of open problems in quantum query complexity~\cite{Aar21}.

Prior work ruled out efficient pathfinding algorithms that are both \emph{rooted} and \emph{genuine}~\cite{CCG23}. Informally, a genuine algorithm is restricted to only query the oracle at (superpositions of) known vertex labels, and a rooted algorithm always retains a path from the entrance to every vertex it records. These conditions capture natural approaches, including versions of the quantum snake walk~\cite{Ros11}, whose state describes a polynomially long path in the graph. However, general algorithms need not satisfy either condition. We combine and strengthen our independent preliminary works~\cite{MP26,CCCG26} in order to obtain a quantum query lower bound for pathfinding on the welded trees graph that is tight up to logarithmic factors.

\begin{theorem}[Informal version of \cref{thm:pathfinding_hard}]\label{thm:main_result}
    For any quantum algorithm $\alg$ making $q\geq1$ queries to a welded trees graph oracle,
    \begin{align*}
        \Pr[\alg \text{ outputs an entrance-exit path}] = {O\left( \frac{q}{2^{n/3}} + \frac{n\cdot q^3}{2^n} \right)},
    \end{align*}
    with the oracle sampled from a distribution over welded trees graphs with a single cycle.
\end{theorem}

Hence in order to achieve constant success probability, an algorithm must make $\Tilde{\Omega}(2^{n/3})$ queries. This lower bound is nearly tight since an algorithm can find a path in $O(n\cdot 2^{n/3})$ queries, as follows. First, run the quantum walk algorithm to find the exit. Then observe that a nonbacktracking walk of length $n$ from the entrance finds a leaf on the left side of the weld, and a nonbacktracking walk of length $n$ from the exit finds a leaf on the right side of the weld. Furthermore, we can determine with two queries whether a given left leaf is connected to a given right leaf. 
Using the collision algorithm of~\cite{BHT98}, with each root-to-leaf evaluation costing $O(n)$ queries, we can then find a pair of connected leaves in $O(n\cdot 2^{n/3})$ queries. Therefore, in both classical and quantum settings, pathfinding reduces to collision finding.
Furthermore, the pathfinding algorithm produces a shortest entrance-exit path by joining two root-to-leaf paths across a single weld edge. While in principle it might be easier to find a non-shortest path, our lower bound shows that finding any entrance-exit path is just as difficult as finding a shortest one up to factors polynomial in $n$.

Our proof uses the compressed permutation oracle of~\cite{Car25,CM26}, which provides a method for showing quantum lower bounds when querying an oracle that implements a random permutation. Conceptually, we use several compressed permutations as building blocks by implementing the welded trees oracle using separate permutations for the labels of vertices and edges between two adjacent columns.
The compressed oracle method provides a way of reversibly recording an algorithm's queries in a certain database state. Intuitively, this database provides a similar representation of the algorithm's knowledge of the graph as in the genuine algorithms of \cite{CCG23}, but in a way that applies to general algorithms.

\subsection{Technical overview}

Briefly, the structure of the proof is as follows. The key idea is to track the information an algorithm retains about the structure of the graph during computation. In \cref{sec:wt_oracle}, we construct two compressed versions of the welded trees oracle and relate successful pathfinding to recording a path across the graph. In \cref{sec:good_space}, we define the \emph{good space}, which is a collection of states that do not record such a path. The key step is showing in \cref{sec:comp_goodness,sec:ordinary_to_exact} that a single query can move the state only slightly outside the good space. As the computation starts with an empty record, it starts in the good space and hence we find that $\Tilde{\Omega}(2^{n/3})$ queries are necessary to find a path with constant probability. 

\paragraph{Recording graph information.} In order to record what an algorithm knows about the graph, we use the compressed oracle framework of~\cite{Zha19}. Introduced as a technique for proving quantum indifferentiability (i.e., showing that some cryptographic construction is essentially as good as an ideal object), it has found a variety of other uses~\cite{CFHL21,HM23,BKW26,MH25,ACMT25,BHNZ25,GWWZ26,Car25}. For an introduction, see~\cite{Ham25}. Intuitively, the technique represents the information an algorithm retains about a random oracle using a database of input-output records. Note that the database is itself quantum, with records stored in superposition. These records can also be erased when the algorithm uncomputes queries, reflecting its ability to forget information.

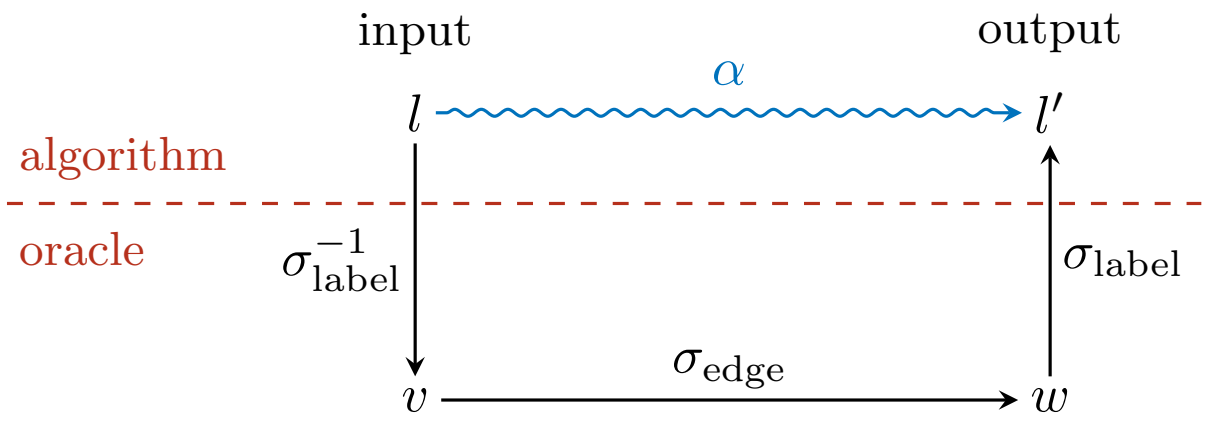
\begin{figure}[h]
    \center
    \begin{tikzpicture}[>=stealth, line width=0.6pt,
    every node/.style={inner sep=2pt}]
    \node (l) at (0,1.9) {$l$};
    \node (lp) at (4.2,1.9) {$l^\prime$};
    \node (v) at (0,0) {$v$};
    \node (w) at (4.2,0) {$w$};
    \node[above=3pt of l, font=\small] {input};
    \node[above=3pt of lp, font=\small] {output};
    \draw[colorC, dashed, line width=0.6pt] (-2.7,1.3) -- (5.2,1.3);
    \node[anchor=west, text=colorC, font=\small] at (-2.7,1.6) {algorithm};
    \node[anchor=west, text=colorC, font=\small] at (-2.7,1.0) {oracle};
    \draw[->, colorA, decorate,
        decoration={snake, amplitude=0.65pt, segment length=5pt,
            pre length=2pt, post length=3pt}]
        (l.east) -- node[above=3pt] {$\alpha$} (lp.west);
    \draw[->] (l) -- node[left, midway] {$\labelPerm^{-1}$} (v);
    \draw[->] (v) -- node[above] {$\colorPerm$} (w);
    \draw[->] (w) -- node[right, midway] {$\labelPerm$} (lp);
\end{tikzpicture}
    \caption{Answer to the query $(l,\colorAlpha)$. The oracle maps the label $l$ to vertex $v$, follows its edge using $\colorPerm$ to vertex $w$ and outputs the label $l^\prime$ of $w$. The algorithm only receives $l^\prime$.}\label{fig:algo_oracle_split}
\end{figure}

We apply this framework to welded trees using the recent construction of compressed permutation oracles~\cite{Car25,CM26}. Specifically, we consider access to the graph via a neighbor oracle \wtO that takes a vertex label $l$ and color $\colorAlpha$ as input and returns the label $l^\prime$ of the corresponding neighbor, if it exists. By revealing an edge coloring, this oracle gives the algorithm more information than the original adjacency-list oracle (so in particular, a lower bound in the model with colors implies a lower bound for the standard adjacency-list model).\footnote{Our result also resolves an open problem of \cite{CCG23} by showing that exit finding is hard for a classical algorithm when provided with a 3-coloring.} On the other hand, the coloring gives us a natural way to randomize the graph using permutations. 
We instantiate this oracle by fixing a welded trees graph along with a proper edge coloring and applying two independent families of permutations. The first, $\labelPerm$, assigns random labels to vertices. The second, $\colorPerm$, permutes edges within each column and color in a certain region around the weld. To answer the query, the oracle identifies the corresponding vertex via $v = \labelPerm^{-1}(l)$, follows its edge of color $\colorAlpha$ using $\colorPerm$ to node $w$ and then returns the label of the neighboring vertex, outputting $l^\prime = \labelPerm(w)$. This computation is illustrated in \cref{fig:algo_oracle_split}.

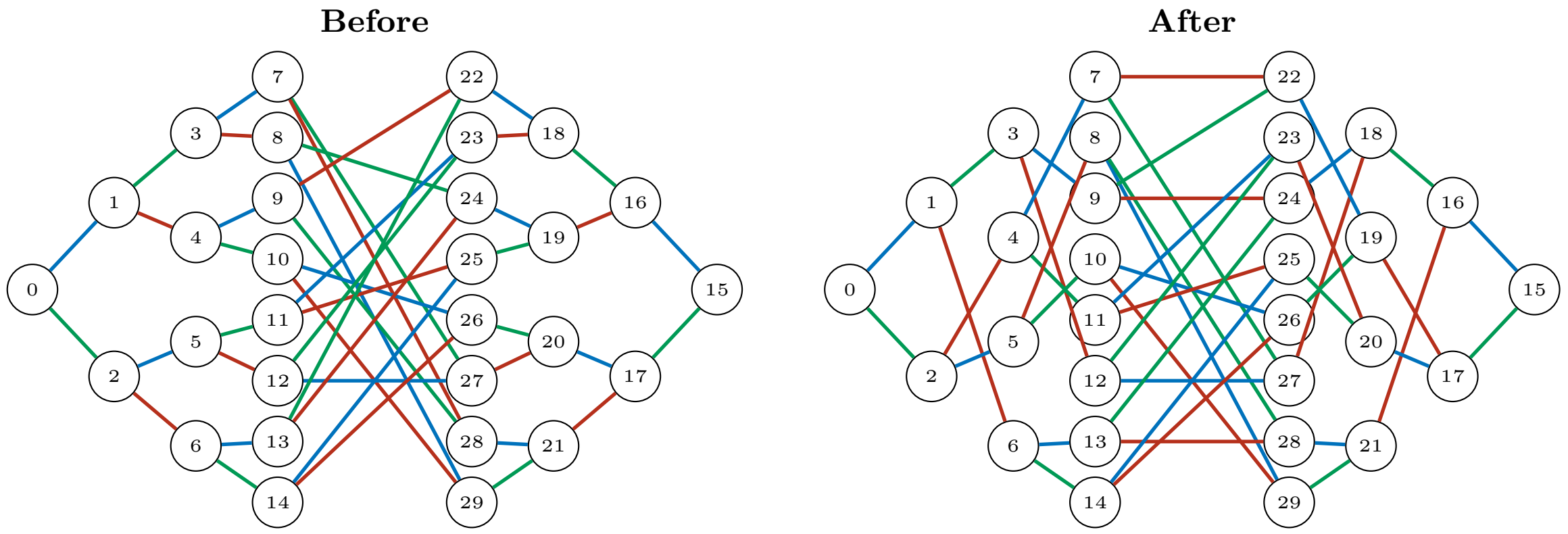
\begin{figure}[h]
    \center
    \begin{tikzpicture}[
    every node/.style={circle, draw, fill=white, inner sep=0pt, minimum size=14pt, font=\tiny},
    treelabel/.style={rectangle, draw=none, fill=none, font=\small\bfseries},
    >=latex
]
 
\def\edgeThickness{1pt} % Adjust the width of all colored edges in this figure.
\tikzset{
    edgeA/.style={line width=\edgeThickness, colorA},
    edgeB/.style={line width=\edgeThickness, colorB},
    edgeC/.style={line width=\edgeThickness, colorC},
}
 
\def\xA{0}
\def\xB{0.8}
\def\xC{1.6}
\def\xD{2.4}
\def\xE{4.3}
\def\xF{5.1}
\def\xG{5.9}
\def\xH{6.7}
 
\def\yshrink{0.85}
\def\hgap{8.0}
 
% %%%%%%%%%%%%%%%%%%%%%%%%%%%%%%%%%%%%%%%%%%%%%%%%%%%%%%%%%%%%
%   PANEL 1:  Before pi  (original edges)
% %%%%%%%%%%%%%%%%%%%%%%%%%%%%%%%%%%%%%%%%%%%%%%%%%%%%%%%%%%%%
 
\begin{scope}
 
\node[treelabel] at ({(\xD+\xE)/2}, \yshrink*2.45+0.55) {Before};
 
% -- Nodes --
\node (L0) at (\xA, 0) {0};
\node (L1a) at (\xB,  \yshrink*1.0) {1};
\node (L1b) at (\xB, -\yshrink*1.0) {2};
\node (L2a) at (\xC,  \yshrink*1.8) {3};
\node (L2b) at (\xC,  \yshrink*0.6) {4};
\node (L2c) at (\xC, -\yshrink*0.6) {5};
\node (L2d) at (\xC, -\yshrink*1.8) {6};
\node (L3a) at (\xD,  \yshrink*2.45) {7};
\node (L3b) at (\xD,  \yshrink*1.75) {8};
\node (L3c) at (\xD,  \yshrink*1.05) {9};
\node (L3d) at (\xD,  \yshrink*0.35) {10};
\node (L3e) at (\xD, -\yshrink*0.35) {11};
\node (L3f) at (\xD, -\yshrink*1.05) {12};
\node (L3g) at (\xD, -\yshrink*1.75) {13};
\node (L3h) at (\xD, -\yshrink*2.45) {14};
\node (R3a) at (\xE,  \yshrink*2.45) {22};
\node (R3b) at (\xE,  \yshrink*1.75) {23};
\node (R3c) at (\xE,  \yshrink*1.05) {24};
\node (R3d) at (\xE,  \yshrink*0.35) {25};
\node (R3e) at (\xE, -\yshrink*0.35) {26};
\node (R3f) at (\xE, -\yshrink*1.05) {27};
\node (R3g) at (\xE, -\yshrink*1.75) {28};
\node (R3h) at (\xE, -\yshrink*2.45) {29};
\node (R2a) at (\xF,  \yshrink*1.8) {18};
\node (R2b) at (\xF,  \yshrink*0.6) {19};
\node (R2c) at (\xF, -\yshrink*0.6) {20};
\node (R2d) at (\xF, -\yshrink*1.8) {21};
\node (R1a) at (\xG,  \yshrink*1.0) {16};
\node (R1b) at (\xG, -\yshrink*1.0) {17};
\node (R0)  at (\xH,  0)            {15};
 
% -- Original left tree edges --
\draw[edgeA] (L0) -- (L1a);  \draw[edgeB] (L0) -- (L1b);
\draw[edgeB] (L1a) -- (L2a); \draw[edgeC] (L1a) -- (L2b);
\draw[edgeA] (L1b) -- (L2c); \draw[edgeC] (L1b) -- (L2d);
\draw[edgeA] (L2a) -- (L3a); \draw[edgeC] (L2a) -- (L3b);
\draw[edgeA] (L2b) -- (L3c); \draw[edgeB] (L2b) -- (L3d);
\draw[edgeB] (L2c) -- (L3e); \draw[edgeC] (L2c) -- (L3f);
\draw[edgeA] (L2d) -- (L3g); \draw[edgeB] (L2d) -- (L3h);
 
% -- Original right tree edges --
\draw[edgeA] (R0) -- (R1a);  \draw[edgeB] (R0) -- (R1b);
\draw[edgeB] (R1a) -- (R2a); \draw[edgeC] (R1a) -- (R2b);
\draw[edgeA] (R1b) -- (R2c); \draw[edgeC] (R1b) -- (R2d);
\draw[edgeA] (R2a) -- (R3a); \draw[edgeC] (R2a) -- (R3b);
\draw[edgeA] (R2b) -- (R3c); \draw[edgeB] (R2b) -- (R3d);
\draw[edgeB] (R2c) -- (R3e); \draw[edgeC] (R2c) -- (R3f);
\draw[edgeA] (R2d) -- (R3g); \draw[edgeB] (R2d) -- (R3h);
 
% -- Original weld edges --
\draw[edgeB] (L3a) -- (R3f); \draw[edgeC] (L3a) -- (R3g);
\draw[edgeA] (L3b) -- (R3h); \draw[edgeB] (L3b) -- (R3c);
\draw[edgeC] (L3c) -- (R3a); \draw[edgeB] (L3c) -- (R3g);
\draw[edgeA] (L3d) -- (R3e); \draw[edgeC] (L3d) -- (R3h);
\draw[edgeC] (L3e) -- (R3d); \draw[edgeA] (L3e) -- (R3b);
\draw[edgeB] (L3f) -- (R3b); \draw[edgeA] (L3f) -- (R3f);
\draw[edgeB] (L3g) -- (R3a); \draw[edgeC] (L3g) -- (R3c);
\draw[edgeA] (L3h) -- (R3d); \draw[edgeC] (L3h) -- (R3e);
 
\end{scope}

% %%%%%%%%%%%%%%%%%%%%%%%%%%%%%%%%%%%%%%%%%%%%%%%%%%%%%%%%%%%%
%   PANEL 2:  After pi  (edges rewired by pi . ncF . pi^{-1})
%
%   pi:  5<->7,  8->10->14->8,  9<->13,  11->15->12->11,
%        20<->22,  23->29->25->23,  24<->28,  26->30->27->26
% %%%%%%%%%%%%%%%%%%%%%%%%%%%%%%%%%%%%%%%%%%%%%%%%%%%%%%%%%%%%
 
\begin{scope}[xshift=\hgap cm]
 
\node[treelabel] at ({(\xD+\xE)/2}, \yshrink*2.45+0.55) {After};
 
% -- Nodes (same positions and labels) --
\node (L0p) at (\xA, 0) {0};
\node (L1ap) at (\xB,  \yshrink*1.0) {1};
\node (L1bp) at (\xB, -\yshrink*1.0) {2};
\node (L2ap) at (\xC,  \yshrink*1.8) {3};
\node (L2bp) at (\xC,  \yshrink*0.6) {4};
\node (L2cp) at (\xC, -\yshrink*0.6) {5};
\node (L2dp) at (\xC, -\yshrink*1.8) {6};
\node (L3ap) at (\xD,  \yshrink*2.45) {7};
\node (L3bp) at (\xD,  \yshrink*1.75) {8};
\node (L3cp) at (\xD,  \yshrink*1.05) {9};
\node (L3dp) at (\xD,  \yshrink*0.35) {10};
\node (L3ep) at (\xD, -\yshrink*0.35) {11};
\node (L3fp) at (\xD, -\yshrink*1.05) {12};
\node (L3gp) at (\xD, -\yshrink*1.75) {13};
\node (L3hp) at (\xD, -\yshrink*2.45) {14};
\node (R3ap) at (\xE,  \yshrink*2.45) {22};
\node (R3bp) at (\xE,  \yshrink*1.75) {23};
\node (R3cp) at (\xE,  \yshrink*1.05) {24};
\node (R3dp) at (\xE,  \yshrink*0.35) {25};
\node (R3ep) at (\xE, -\yshrink*0.35) {26};
\node (R3fp) at (\xE, -\yshrink*1.05) {27};
\node (R3gp) at (\xE, -\yshrink*1.75) {28};
\node (R3hp) at (\xE, -\yshrink*2.45) {29};
\node (R2ap) at (\xF,  \yshrink*1.8) {18};
\node (R2bp) at (\xF,  \yshrink*0.6) {19};
\node (R2cp) at (\xF, -\yshrink*0.6) {20};
\node (R2dp) at (\xF, -\yshrink*1.8) {21};
\node (R1ap) at (\xG,  \yshrink*1.0) {16};
\node (R1bp) at (\xG, -\yshrink*1.0) {17};
\node (R0p)  at (\xH,  0)            {15};
 
% -- Rewired left tree edges --
% {1,2}A  {1,3}B  (unchanged)
\draw[edgeA] (L0p) -- (L1ap);  \draw[edgeB] (L0p) -- (L1bp);
% {2,4}B (same)   {2,7}C (was {2,5})
\draw[edgeB] (L1ap) -- (L2ap); \draw[edgeC] (L1ap) -- (L2dp);
% {3,6}A (same)   {3,5}C (was {3,7})
\draw[edgeA] (L1bp) -- (L2cp); \draw[edgeC] (L1bp) -- (L2bp);
% {4,10}A (was {4,8})   {4,13}C (was {4,9})
\draw[edgeA] (L2ap) -- (L3cp); \draw[edgeC] (L2ap) -- (L3fp);
% {5,8}A (was {5,10})   {5,12}B (was {5,11})
\draw[edgeA] (L2bp) -- (L3ap); \draw[edgeB] (L2bp) -- (L3ep);
% {6,11}B (was {6,12})  {6,9}C (was {6,13})
\draw[edgeB] (L2cp) -- (L3dp); \draw[edgeC] (L2cp) -- (L3bp);
% {7,14}A (same)  {7,15}B (same)
\draw[edgeA] (L2dp) -- (L3gp); \draw[edgeB] (L2dp) -- (L3hp);
 
% -- Rewired right tree edges --
% {16,17}A  {16,18}B  (unchanged)
\draw[edgeA] (R0p) -- (R1ap);  \draw[edgeB] (R0p) -- (R1bp);
% {17,19}B (same)   {17,22}C (was {17,20})
\draw[edgeB] (R1ap) -- (R2ap); \draw[edgeC] (R1ap) -- (R2dp);
% {18,21}A (same)   {18,20}C (was {18,22})
\draw[edgeA] (R1bp) -- (R2cp); \draw[edgeC] (R1bp) -- (R2bp);
% {19,29}A (was {19,23})   {19,28}C (was {19,24})
\draw[edgeA] (R2ap) -- (R3cp); \draw[edgeC] (R2ap) -- (R3fp);
% {20,25}A (same)   {20,27}B (was {20,26})
\draw[edgeA] (R2bp) -- (R3ap); \draw[edgeB] (R2bp) -- (R3ep);
% {21,26}B (was {21,27})   {21,24}C (was {21,28})
\draw[edgeB] (R2cp) -- (R3dp); \draw[edgeC] (R2cp) -- (R3bp);
% {22,23}A (was {22,29})   {22,30}B (same)
\draw[edgeA] (R2dp) -- (R3gp); \draw[edgeB] (R2dp) -- (R3hp);
 
% -- Rewired weld edges --
% {8,23}C   {8,28}B
\draw[edgeC] (L3ap) -- (R3ap); \draw[edgeB] (L3ap) -- (R3fp);
% {9,30}A   {9,29}B
\draw[edgeA] (L3bp) -- (R3hp); \draw[edgeB] (L3bp) -- (R3gp);
% {10,23}B   {10,25}C
\draw[edgeB] (L3cp) -- (R3ap); \draw[edgeC] (L3cp) -- (R3cp);
% {11,27}A   {11,30}C   (unchanged)
\draw[edgeA] (L3dp) -- (R3ep); \draw[edgeC] (L3dp) -- (R3hp);
% {12,26}C   {12,24}A
\draw[edgeC] (L3ep) -- (R3dp); \draw[edgeA] (L3ep) -- (R3bp);
% {13,24}B   {13,28}A   (unchanged)
\draw[edgeB] (L3fp) -- (R3bp); \draw[edgeA] (L3fp) -- (R3fp);
% {14,25}B   {14,29}C
\draw[edgeB] (L3gp) -- (R3cp); \draw[edgeC] (L3gp) -- (R3gp);
% {15,26}A   {15,27}C   (unchanged)
\draw[edgeA] (L3hp) -- (R3dp); \draw[edgeC] (L3hp) -- (R3ep);
 
\end{scope}
 
\end{tikzpicture}
    \caption{An example of an application of \colorPerm on the welded trees graph.}\label{fig:color_preserving}
\end{figure}

The algorithm never directly receives the vertices $v$ or $w$. The edge permutation $\colorPerm$ (illustrated in \cref{fig:color_preserving}) consists of independent permutations, each acting on edges of the same color $\colorAlpha$ between two adjacent columns $c$ and $c+1$, which we call an \emph{edge class}. Within each such class, we keep the endpoints in one column fixed and permute their partners in the other column. We apply these edge permutations only in a region around the weld extending to approximately $n/3$ edges away from each root, leaving the edges closer to the roots fixed. As this assignment is a bijection, every vertex retains the same number of edges of each color to each adjacent column. Thus the two binary trees retain their structure and the edge coloring remains proper.

We define this formally in \cref{sec:construction_welded_trees}. Replacing $\labelPerm$ and $\colorPerm$ with compressed permutation oracles gives us two databases, a label database $L$ that maps vertices to labels and an edge-color database $C$ for the edge permutations. In \cref{sec:compressed_welded_tree}, we construct the \emph{ordinary} compressed welded trees oracle \cpwt, which we use to understand what information is tracked. For some fixed input $x$, let $\pC$ denote an operator that swaps partial injective database states $\ket{I}$ in which $x$ is not recorded with the uniform superposition over extensions of $I$ where $x$ maps to some unused output $y$. We can write each step of the computation in $\wtO$ described above using a corresponding $\pC$ operator. Specifically, $\pCLabelInv$ associates some vertex $v = L^{-1}(l)$ to the label $l$, $\pCPerm$ handles recording an edge from $v$ to $w$ in the database $C$ and $\pCLabel$ assigns the label $L(w)$ to the vertex $w$. Lastly, we define $\xorOP$ to be the operation that XORs $L(w)$ into the output register, assuming all of the corresponding database values are in $L$ and $C$, otherwise acting as the identity. Combining these gives
\begin{align}
    \cpwt \coloneq \pCLabelInv \cdot \pCPerm \cdot \pCLabel \cdot \xorOP \cdot \pCLabel \cdot \pCPerm \cdot \pCLabelInv.\label{eq:definition_cpwt} 
\end{align}
We also construct the \emph{exact} compressed welded trees oracle \icpwt, which can be viewed as an intermediate step between \wtO and \cpwt: it uses the same database states as \cpwt, but exactly preserves the success probability of the original oracle \wtO. 

In \cref{sec:bridge_oracles}, we compare these oracles. First, we characterize the relationship between \cpwt and \icpwt, which allows us to transfer our bounds between them. Second, we show that the success probability of an algorithm using \wtO is bounded by the probability that a path across the region with permuted edges is recorded in the database $C$ when using \icpwt, with an additive error growing linearly with the number of queries.

\paragraph{Quantum walk speedup.}

Before turning to the proof of \cref{thm:main_result}, we explain how the databases $L,C$ develop during the quantum walk that finds the exit vertex exponentially faster than any classical algorithm~\cite{CCD+03}. We analyze this process in more detail in \cref{sec:quantum_walk_analysis}.

The quantum walk algorithm evolves with a Hamiltonian given by the adjacency matrix of the graph. As described above, starting at the entrance, the walk remains in the span of uniform superpositions over columns, so its evolution reduces to a continuous-time quantum walk on a weighted path of length $O(n)$.

For welded tree instances obtained by sampling $\labelPerm$ and $\colorPerm$ uniformly (as described in \cref{sec:construction_welded_trees}), the evolution of the database during a run of the walk is especially simple. At any evolution time, $C$ is empty and $L$ has at most one record. Moreover, up to norm error $O(2^{-n/2})$, this is the current vertex-label record (except that no label is needed at the entrance since its label is fixed). This follows because the amplitude at a given vertex depends only on its column, independently of the edge permutations. The purified state therefore has no dependence on those permutations and depends on the label permutation only through the current label. More precisely, the compression isometry produces an empty edge database and at most one label record. This record can be identified with the current vertex-label pair up to a small normalization correction, as described in detail in \cref{sec:quantum_walk_analysis}. Since the database record does not distinguish different routes by which the quantum walk reached its current state, contributions from different routes can interfere.

A simulation of the continuous-time dynamics using a finite number of queries introduces bounded simulation error, and intermediate queries can temporarily create additional records. However, this does not change the fundamental conclusion that the database does not store information about the algorithm's history and is compatible with producing the exit label without recording a path.

\paragraph{Good space.} Returning to the hardness of pathfinding, the goal is to show that recording a path in the databases $(L,C)$ is difficult. In \cref{sec:good_space}, we define the good space, which is a space of states that have not recorded such a path. As mentioned above, the pair of databases $(L,C)$ can be viewed as a representation of the known graph, where $L$ represents the retained labels and $C$ the edges. However, an algorithm only interacts with $(L,C)$ through \cpwt. Each query takes in a label-color pair and outputs another label that represents its neighbor, meaning the query does not directly reveal vertex positions in the graph. We describe the labels and edges the algorithm retains by an \emph{abstract forest} $H$, whose connected components are trees $H_i$. Note that $H$ does not explicitly specify vertex positions. Instead, the actual vertices are written down using the map $\colorVertMap$ from abstract vertices in $H$ to vertices on the welded trees graph. We call the vertices of the welded trees graph structural vertices.

While $H$ accurately represents the labels and the connections the algorithm retains, knowing where the exploration began can help the algorithm infer the vertices' positions. For example, suppose that the algorithm randomly walks from the root. Until it crosses the weld, we know which column each node is in. However, after crossing the weld, the labels and edge colors along the abstract path may no longer tell the algorithm whether its subsequent steps move towards or away from the other root (unless they hit previously explored nodes or a root). To represent this information, for each connected component $H_i$ of $H$, we define the partial assignment $\Wfix_i$ on $\colorVertMap$. Fix an assignment $\colorVertMap$ and a vertex $v$ in $H_i$. Starting from $v$, follow every path in $H_i$ until just after crossing the first weld edge under $\colorVertMap$ or reaching a leaf. The vertices encountered and their positions form the fixed data $\Wfix_i$. We call these vertices \emph{fixed} and the remaining vertices in $H_i$ \emph{free}. 

Under some fixed $H_i$ and $\Wfix_i$, we assign the free vertices to structural vertices on the welded trees graph as follows. Starting from a leaf in the fixed data $\Wfix_i$, we assign successive vertices along the edges of $H_i$. At each step, the edge color and preceding vertex's position determine the next column and the allowed positions in it, from which we choose uniformly. Let $\mu$ denote the distribution over assignments $\colorVertMap$ according to this rule restricted to assignments that remain in the permuted region.

Finally, we define the boundary $B$ to have two parts, one at each end of the permuted region, each of which is approximately $n/3$ steps away from each root. Notice that any entrance-to-exit path must cross through this permuted region, so we solely focus on recording a path between the two boundary columns.

\begin{figure}[h]
    \center
    % Fixed data and two successive free assignments on the original smaller graph.
\begin{tikzpicture}[
    scale=0.8,
    x=1cm, y=1cm,
    basicVertex/.style={circle, draw=black!30, fill=black!6,
        minimum size=14pt, inner sep=0pt, line width=0.9pt},
    basicEdge/.style={draw=black!25, line width=0.9pt},
    fixedVertex/.style={circle, draw=colorA, fill=colorA!25,
        minimum size=14pt, inner sep=0pt, line width=0.9pt},
    freeVertex/.style={circle, draw=colorA, fill=white, dashed,
        minimum size=14pt, inner sep=0pt, line width=0.9pt,
        font=\small},
    fixedEdge/.style={draw=colorA, line width=2pt},
    freeEdge/.style={draw=colorA, line width=1.2pt, dashed,
        preaction={draw=white, solid, line width=3.2pt}},
    boundary/.style={fill=colorC!10, draw=colorC!60, line width=0.9pt,
        rounded corners=7pt}
]

% Each boundary contains the root column and its adjacent column.
\draw[boundary] (-0.45,-3.45) rectangle (2.25,3.45);
\draw[boundary] (11.35,-3.45) rectangle (14.05,3.45);

% Two binary trees facing one another across the weld.
\foreach \v/\x/\y in {
    0/0/0, 1/1.8/1.8, 2/1.8/-1.8,
    3/3.6/2.7, 4/3.6/0.9, 5/3.6/-0.9, 6/3.6/-2.7,
    7/5.4/3.15, 8/5.4/2.25, 9/5.4/1.35, 10/5.4/0.45,
    11/5.4/-0.45, 12/5.4/-1.35, 13/5.4/-2.25, 14/5.4/-3.15,
    22/8.2/3.15, 23/8.2/2.25, 24/8.2/1.35, 25/8.2/0.45,
    26/8.2/-0.45, 27/8.2/-1.35, 28/8.2/-2.25, 29/8.2/-3.15,
    18/10/2.7, 19/10/0.9, 20/10/-0.9, 21/10/-2.7,
    16/11.8/1.8, 17/11.8/-1.8, 15/13.6/0
} { \coordinate (v\v) at (\x,\y); }

% Tree edges.
\foreach \u/\v in {
    0/1, 0/2, 1/3, 1/4, 2/5, 2/6,
    3/7, 3/8, 4/9, 4/10, 5/11, 5/12, 6/13, 6/14,
    15/16, 15/17, 16/18, 16/19, 17/20, 17/21,
    18/22, 18/23, 19/24, 19/25, 20/26, 20/27, 21/28, 21/29%
} { \draw[basicEdge] (v\u) -- (v\v); }

% Weld edges.
\foreach \u/\v in {
    7/27, 7/28, 8/29, 8/24, 9/22, 9/28, 10/26, 10/29,
    11/25, 11/23, 12/23, 12/27, 13/22, 13/24, 14/25, 14/26%
} { \draw[basicEdge] (v\u) -- (v\v); }

\foreach \v in {0,...,29} { \node[basicVertex] at (v\v) {}; }

% Fixed positions, including the first endpoint across the weld.
\foreach \u/\v in {0/1, 1/4, 4/10, 10/26} { \draw[fixedEdge] (v\u) -- (v\v); }

% The same two-edge continuation has color B followed by color A
% in glued_tree.tex. Vertex a can be at 19,20,21. At 19 and 21 the
% next edge returns towards the weld, placing b at 22,24,28.
% At 20 it instead reaches 17 in the boundary.
\foreach \a in {19,21} {
    \draw[freeEdge] (v26) -- (v\a);
    \foreach \b in {22,24,28} {
        \draw[freeEdge] (v\a) -- (v\b);
    }
}
\draw[freeEdge, draw=colorC] (v26) -- (v20) -- (v17);
\foreach \a in {19,21} { \node[freeVertex] at (v\a) {$a$}; }
\foreach \b in {22,24,28} { \node[freeVertex] at (v\b) {$b$}; }
\node[freeVertex, draw=colorC] at (v20) {$a$};
\node[freeVertex, draw=colorC] at (v17) {$b$};

\foreach \v in {0, 1, 4, 10, 26} { \node[fixedVertex] at (v\v) {}; }

\node[anchor=east, xshift=-10pt, font=\small\bfseries] at (v0.west) {Entrance};
\node[anchor=west, xshift=10pt, font=\small\bfseries] at (v15.east) {Exit};
\node[anchor=south, yshift=14pt, font=\small\bfseries] at (6.8,3.15) {Weld};
\node[anchor=south, font=\small\bfseries] at (0.9,3.5) {$B$};
\node[anchor=south, font=\small\bfseries] at (12.7,3.5) {$B$};
\end{tikzpicture}
    \caption{An abstract tree $H_i$ consisting of the \textcolor{colorA}{blue} vertices and edges, and the nodes labeled $a,b$. The \textcolor{colorA}{solid} vertices denote $\Wfix_i$, while dashed vertices are free, assigned according to $\mu$. \textcolor{colorA}{Blue} assignments to $a,b$ are good, while the \textcolor{colorC}{red} assignment is bad. The boundary $B$ is represented by the \textcolor{colorC!60}{shaded} region.
    In general, the boundary on each side consists of two columns of vertices, which happen to include the roots in this small example.}
    \label{fig:good_space}
\end{figure}
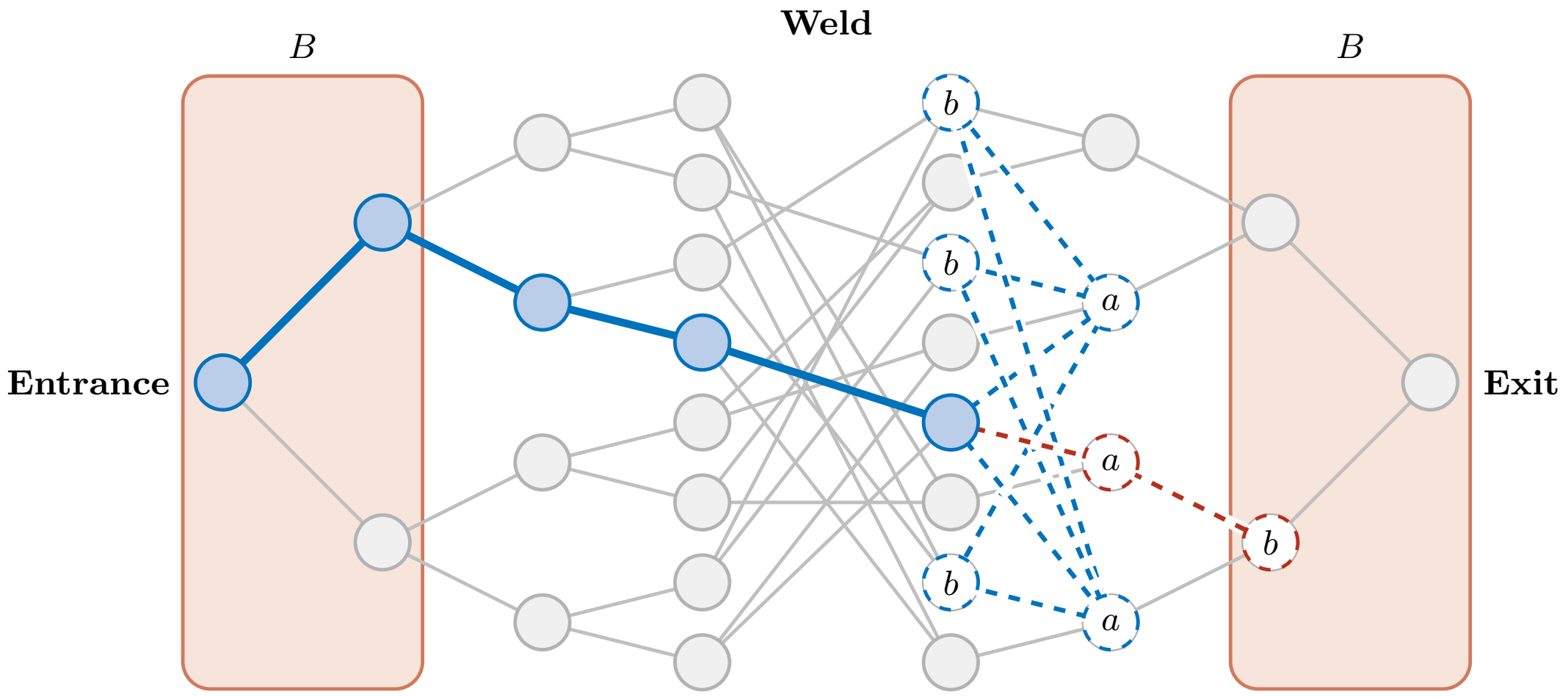

An assignment to the variable $\colorVertMap$ is good if it is in the support of $\mu$,
is injective and does not hit the boundary $B$ outside the fixed part $\Wfix$. Since each $\Wfix_i$ of each $H_i$ can meet $B$ only on one side of the weld, a path between the two boundary columns would have to hit $B$ at a free vertex. This means that if $\colorVertMap$ is good, it cannot contain a path across the region 
with permuted edges, meaning the corresponding database $(L,C)$ it represents cannot contain an entrance-exit path. For each choice of $H$ and fixed data, we form a superposition over good assignments with amplitudes weighted by $\sqrt{\mu}$. The good space is the span of these states.

\paragraph{Remaining in the good space.} In \cref{sec:comp_goodness}, we show that each application of \cpwt keeps us in the good space up to a small error by analyzing each operator in its decomposition in \cref{eq:definition_cpwt}. The main difficulty lies in the edge compression operator $\pCPerm$, where we have to consider adding, deleting and replacing edge records. Adding an edge may violate goodness by hitting the boundary $B$ outside of the fixed nodes, creating a cycle or joining two trees in the forest $H$ outside of their fixed parts. We use the distribution $\mu$ over assignments to bound these events. Deletion requires a separate argument because databases that differ only in the erased record become identical, which means their amplitudes add. As we saw above, erasing records allows different paths of the quantum walk to interfere. Nonetheless, we show that the resulting state remains in the good space, up to a small error. Finally, controlling both addition and deletion lets us control replacing an edge record.

First, fix some abstract forest $H$ and data $\Wfix$, and consider the assignments under $\mu$. The first free vertex after the fixed weld edge is assigned randomly among $\Omega(N)$ possible positions, where $N=2^n$. Each subsequent step averages over possible assignments, which cannot increase the maximum probability of assignment to a particular vertex. Thus any free vertex has probability $O(1/N)$ of being assigned to a particular structural vertex. Furthermore, even though the positions of any two distinct vertices are not necessarily independent, we show that the probability they are 
assigned to the same value is $O(n/N)$ when at least one of them is free. After $q$ queries, we only have $O(q)$ recorded vertices, so
\begin{align}
    \Pr_{\mu}[\text{new vertex hits $B$ or another recorded vertex}] = O\left( \frac{\abs{B}}{N} + \frac{nq}{N} \right).\label{eq:pr_bad_events}
\end{align}
For an edge permutation over a column containing $s$ nodes where $s\ll N$, using the $O(q/s)$ collision estimate at every query would only give an $\Omega(s^{1/3})$ query lower bound which could not be improved using this estimate alone due to the collision lower bound~\cite{AS04}. However, a free queried vertex is unlikely to lie in this column in the first place, which is why instead we average over its position, allowing us to obtain the $\tilde{\Theta}(N^{1/3})$ bound instead. 
Fix some queried color and let $s_b$ be the size of the permutation for an edge class $b$ between adjacent columns. By the argument above, if the queried vertex is free, it lies in class $b$ with probability $O(s_b/N)$. When this occurs, it will hit one of the $O(q)$ recorded vertices with probability at most $O(q/s_b)$. 
Averaging over all $O(n)$ edge classes,
\begin{align}
    \sum_b O\left( \frac{s_b}{N} \right) \cdot O\left( \frac{q}{s_b} \right) = O\left( \frac{nq}{N} \right),\label{eq:averaging_over_nodes}
\end{align}
which gives us the $O(nq/N)$ term in \cref{eq:pr_bad_events}. Intuitively, a smaller permutation increases the chance of collision, but this is counteracted by the decreasing probability a random walker moving according to $\mu$ gets this far up the tree. Otherwise, if the queried vertex is fixed, the same bound applies as the events that violate goodness are either coinciding with another assignment or creating a cycle.

An arbitrary good state is a linear combination of the states defined with a fixed $H$ and $\Wfix$, which may overlap as different fixed data may produce the same database. We group the complex coefficients in this combination according to which part of each $H_i$ is fixed and reorganize them without changing the state. The goal is to combine the error bounds of each group without paying a factor for the number of groups. For a component separated from the queried vertex by a single weld edge $\rho$, we average its coefficients under $\mu$ while fixing $\rho$. We subtract this average from those coefficients and add it to the coefficients of the component with the queried vertex. This is possible because both choices of fixed data include $\rho$. Repeating this over all tree components leaves the average $0$ over all components besides the one containing the queried vertex. Fixing $\rho$ and the assignments on the side containing the queried vertex makes one group's coefficient a constant while the other averages to $0$ under $\mu$. Therefore, their product, which is the sum of cross terms in the squared norm, also averages to $0$. We also show that the sum of the averaged squared coefficients is at most a constant times the state's squared norm, allowing us to combine the preceding bounds with only a constant loss.

For deletion, we use the same grouping of coefficients and check whether the deleted edge's non-queried endpoint remains recorded in $H$. If it does, the group with the queried component fixed remains good. For the remaining groups,
at most $O(q)$ deleted edges lead to the same database and each deletion carries a factor of $O(1/\sqrt{s_b})$, so Cauchy-Schwarz gives an $O(q/s_b)$ factor in the squared error, which we average as in \cref{eq:averaging_over_nodes}. Otherwise, if deleting the edge also erases its endpoint, the amplitudes for all its possible positions add together. The amplitudes from different possible assignments would add up to a good state, except that some are missing because those assignments would have already been bad before the deletion. However, this is a reverse addition, which the bounds above show can only add a small error.

Since replacing an edge is just deleting it followed by adding a new one, the same arguments follow for this case, completing the analysis of \pCPerm. The remaining operators in \cpwt are simpler. The operators \pCLabel and \xorOP preserve goodness exactly, while an analogous averaging argument bounds the error from \pCLabelInv. Adding all of these bounds, an application of \cpwt moves the norm outside of the good space by at most $O(\sqrt{\abs{B}/N + nq/N})$.

\paragraph{Completing the proof.} In \cref{sec:ordinary_to_exact}, we transfer the good-space bound from \cpwt to \icpwt using the comparison in \cref{sec:bridge_oracles}. Replacing exact label queries with ordinary ones is simple as it incurs $O(1/N)$ error, but the edge operator \pCPerm requires more care. Applying the standard compression techniques from~\cite{CM26} accrues an error of $O(1/\sqrt{s})$ for permutations of size $s$. As mentioned earlier, the smallest permutation may have size $s=\Theta(N^{1/3})$, which would give us a weaker bound. Instead, we bound how much the operator moves the state outside of the good space. If $s$ is too small, we add its points to the boundary $B$, shrinking the part of the graph where we may walk. This causes two errors, one from adding points that are not fixed to the new boundary $B$ and another because the new good states need not belong to the original good space. We show that these errors are both at most $O(\sqrt{q/N})$. As a free vertex has an $O(1/N)$ probability of being at a particular point, the probability any of the $O(q)$ vertices hits one of the $O(s)$ boundary points is $O(qs/N)$. This gives us an $O(\sqrt{qs/N})$ bound on the difference between the good spaces, which we multiply by the $O(1/\sqrt{s})$ difference between the oracle operators to bound each error by $O(\sqrt{q/N})$. For larger permutations, the transfer from~\cite{CM26} suffices, giving us a transfer within the permitted bounds.

We finally obtain \cref{thm:main_result}. An empty database is trivially good and the algorithm's unitaries do not affect the database. Therefore summing the errors over $q$ queries, we get the bound. In \cref{sec:single_cycle}, we also compare independent weld permutations with a weld that forms a single cycle. By showing an algorithm cannot distinguish them efficiently, their acceptance probability cannot differ too much. Placing the boundary at distance $n/3$ away from both roots gives us the final success bound.

\paragraph{Distributions over graphs.} Our sampling distribution differs from the original distribution over welded trees in~\cite{CCD+03}. The original chooses a uniformly random alternating cycle between the leaves, while we fix a proper edge coloring and permute edges between adjacent columns, separately for each color. Different uncolored graphs can be obtained from different numbers of compatible colored graphs, so forgetting the colors does not make the distribution uniform. For example, if two leaves on opposite sides have parent edges colored $\alpha$ and $\beta$, any weld edge joining them must have color $\gamma$. On the other hand, if both parent edges have color $\alpha$, their connecting edge can instead have color $\beta$ or $\gamma$. Furthermore, even when conditioning on a single cycle, as we do in \cref{sec:single_cycle}, there is a sampling bias. Consequently, \cref{thm:main_result} gives a lower bound for this sampling distribution and for pathfinding on welded trees in the worst case---which suffices to illustrate an exponential separation between exit finding and pathfinding---but not on average for the original distribution.

Nonetheless, our techniques can be altered to recover an average-case bound for the original distribution by abandoning the proper edge coloring and instead assigning distinct labels to the edges at each vertex, without requiring labels at opposite ends of an edge to be equal. Therefore, we permute all edges between adjacent columns together. Every weld cycle has the same number of label assignments, since each leaf has two ways to assign its welded edges to its two available labels. Conditioning on a single cycle gives us the original distribution. The same proof technique applies, with an additional error due to the fact that the edge labels occur in slightly uneven numbers within each column that is absorbed in the final bound. We omit the details of this variant of the result.

\subsection{Open problems}
\label{sec:open-intro}

This work raises several natural questions for further investigation.

\paragraph{Instantiating the welded trees oracle.}
Both the classical hardness of exit finding~\cite{CCD+03}

and our lower bound for pathfinding are stated in the black-box model where they use oracles which we do not know how to implement efficiently.
A natural question is whether the oracle can be \emph{instantiated}: is there
a family of graphs with a succinct and efficiently computable description
(such as a 

keyed construction, possibly combined with
obfuscation) for which the exit can still be found by a quantum walk but
finding an entrance--exit path remains hard for efficient quantum algorithms?
The difficulty is that an explicit circuit for the neighbor function can leak
structure that the random-oracle model hides. For example, if labels are
computed 

in a way that reveals the column of a vertex,
then the exit is trivial to find classically. Hiding this information while
keeping the neighbor function efficiently computable seems to require
a nontrivial cryptographic assumption. Of course, non-black-box attacks exploiting an
explicit instantiation are not covered by our lower bound, in contrast to
settings where a random oracle can be heuristically replaced by a hash function
(e.g.,~\cite{YZ24}).
To the best of our knowledge this question has not been addressed in the
literature; existing work on implementing the welded trees oracle concerns
circuit synthesis for explicit, non-hidden welds (e.g.,~\cite{GLRSV13,GCJ17}), 
which does not aim to conceal the graph structure.

\paragraph{The remaining gap in the query complexity.} \Cref{thm:main_result} shows that $\Omega\bigl((2^n/n)^{1/3}\bigr)$
queries are required for constant success probability, while the straightforward collision-finding algorithm
described above uses $O(n\,2^{n/3})$ queries.

Can we close this gap?

\paragraph{Space complexity.}
The collision-based algorithm uses $\tilde\Theta(2^{n/3})$ bits of coherently addressable classical memory. 

Similar considerations hold for the basic task of collision finding: the query-optimal algorithm \cite{BHT98} uses an amount of coherently addressable classical space comparable to its query complexity, and it is unknown whether similar query complexity can be achieved with significantly less space, or what is the nature of the query--space tradeoff (though a tradeoff has been established for the problem of finding many collision pairs \cite{HM23}).
It would be interesting to prove query--space tradeoffs for pathfinding, to understand how they compare to such tradeoffs for collision finding,

and to determine the time complexity of pathfinding without coherently addressable memory.
(Such tradeoffs are explored in \cite{LS25}, but this question remains largely open.)

\paragraph{Other graphs.}

Previous work has identified examples of other graphs with exponential separations between classical and quantum traversal, including graphs that generalize the structure of welded trees~\cite{BLH25} and graphs built by combining many welded trees~\cite{Li23,LT24,LZ25}.

Our framework applies compressed permutation oracles to randomly labeled and randomly structured graphs in a general way, so it might be able to show hardness of pathfinding for other graphs.

Another open problem is to identify examples of quantum speedup for graph traversal, with or without hardness of pathfinding, that are not obtained by modifying or composing the welded trees graph.

\paragraph{Pathfinding in cryptographic graphs.}
Many hash functions and key-exchange schemes are built on expander graphs, and their
security reduces to the hardness of finding a path between two vertices. For example, the
Charles--Lauter--Goren hash function~\cite{CLG09} maps a message to the endpoint of a nonbacktracking
walk in a supersingular isogeny graph starting at a fixed curve; preimage finding is equivalent to
pathfinding in this graph, and collision finding is equivalent to finding a cycle. Pathfinding in this
graph is polynomial-time equivalent to computing the endomorphism ring of a supersingular
curve (subject to the generalized Riemann hypothesis, with a randomized reduction running in expected polynomial time)~\cite{EHLMP18,Wes22}.
The welded trees problem can be seen as providing a
black-box example of a graph for which
such a pathfinding problem essentially
requires collision finding.

We stress that there is no formal connection between welded trees and isogeny graphs. Isogeny graphs are explicit rather than randomly labeled,
their vertices have canonical names ($j$-invariants), and their algebraic structure
admits non-generic reductions (such as to the endomorphism ring problem) that have no analogue in
the black-box setting. Is it possible to make a precise connection that establishes cryptographically relevant hardness results?

\section{Preliminaries}

Let us describe the notation used in this paper. For positive integers $a,b$ such that $a\leq b$, we use $[a]$ to denote the set $\{1,2,\dots,a\}$ and $[a,b]$ for $\{a,a+1,\dots,b\}$. For operators $A,B$ we write $[A,B]=AB-BA$. For subspaces $V \subseteq U$, we write $U \ominus V$ to denote $U \cap V^\perp$. The symbol $\rho$ will be used to denote density operators and for a pure state $\ket{\psi}$, $\rho(\ket{\psi}) = \ketbra{\psi}{\psi}$. All Hilbert spaces considered here are finite-dimensional. We use $\norm{\,\cdot \,}$ for the Euclidean norm of a vector or the induced operator norm. All projectors are orthogonal projectors. Throughout the section, we use $s$ to denote an arbitrary integer such that $s\geq 3$. We use $\Sym(s)$ to denote the set of permutations of $[s]$, and $\phi\sim \Sym(s)$ denotes drawing a uniformly random permutation. Similarly, $\Sym(D)$ for a set $D$ is the set of permutations over $D$.

A graph $G = (V,E)$ consists of vertices $V$ and edges $E$ between the vertices. When $G$ is a rooted tree, we let $\groot(G)$ be the root of $G$ and for any $v\in V$, let $\depth(v)$ be the distance between $v$ and $\groot(G)$.

A quantum query algorithm $\alg^{(O)}$ that makes $q$ queries to the oracle $O$ is specified by a sequence of oracle-independent unitaries $A_0,\dots,A_q$ on its workspace, producing the final state $A_q O\cdots A_1 O A_0\ket{0}$. Access to $O$ is provided using an XOR oracle, defined as
\begin{align*}
    \ket{x,y} \xmapsto{O_f} \ket{x, y\oplus f(x)}.
\end{align*}
Note that we are assuming that every algorithm makes at least $1$ query.

\paragraph{Useful statements.} The following facts are useful in later analysis.

\begin{lemma}\label{lem:operator_norm_max}
    Let $\calH = \calH_1\oplus \dots \oplus \calH_k$ be a Hilbert space and $O: \calH\rightarrow \calH$ an operator such that $O = O_1 \oplus \dots \oplus O_k$ where for all $i\in [k]$, $O_i: \calH_i\rightarrow \calH_i$. Then,
    \begin{align*}
        \norm{O} = \max_{i\in [k]} \norm{O_i}.
    \end{align*}
\end{lemma}

In the following statement, for a subspace $\tau$, $\Pi_\tau$ denotes the orthogonal projector onto $\tau$.

\begin{lemma}[{\cite[Theorem~1.1]{CCL16}}]\label{lem:lin_map_subspace}
    Let $G$ be a linear map with $\norm{G}\leq 1$ and $\calS, \calS^\prime$ be subspaces of its domain. Suppose there exists a constant $c>0$ such that for all $z\in G\calS$, there exists some $f\in \calS$ such that $Gf= z$ and $\norm{f}\leq c\norm{z}$. Suppose the same holds for $\calS^\prime$. Then,
    \begin{align*}
        \norm{\Pi_{G\calS} - \Pi_{G \calS^\prime}} \leq c \norm{\Pi_{\calS} - \Pi_{\calS^\prime}}.
    \end{align*}
\end{lemma}

\begin{lemma}[Consequence of Theorem IX.5.11 in~\cite{Bha96}]\label{lem:orthogonal_differences}
    Let $\{T_b\}_b$ be a set of self-adjoint operators such that $T_b T_c = 0$ for $b\neq c$. Let $\Pi,\Pi_b$ be orthogonal projectors such that $[T_b, \Pi_b] = 0$. Then,
    \begin{align*}
        \norm{\left[ \Pi, \sum_b T_b \right]} \leq 2 \left( \sum_b \norm{T_b}^2 \norm{\Pi - \Pi_b}^2 \right)^{1/2}.
    \end{align*}
\end{lemma}

\begin{lemma}[Consequence of Lemma~1 of~\cite{Li95}]\label{ex:polar-normalization}
    For some spaces $\calH,\calL$, let $V:\calH \to \calL$ be a contraction (i.e., $\norm{V} \le 1$) with $K = V^\dagger V \succ 0$, $U,T$ self-adjoint operators on $\calH,\calL$, $\delta = TV - VU$ and $\calI = V K^{-1/2}$. Suppose that some orthogonal projectors $\Pi, \Pi^\prime$ commute with $K$, $[\Gram,U]\Pi=\Pi^\prime[\Gram,U]\Pi$ and $\Gram\succeq\kappa\Id$ on the ranges of $\Pi,\Pi^\prime$ for some constant $\kappa$. Then $\calI$ is an isometry and
    \begin{align*}
        \norm{(T\calI-\calI U)\Pi} \leq \frac{\norm{\delta \Pi}}{\kappa^{1/2}} + \frac{\norm{(\delta^\dagger V - V^\dagger \delta) \Pi}}{2 \kappa^{3/2}}.
    \end{align*}
\end{lemma}

\paragraph{Injective databases.} As our result extensively uses the compressed permutation oracle of~\cite{Car25}, we follow their notation described below. We use $\db_{s}$ to denote the set of partial injective functions from $[s]$ to $[s]$ (i.e., functions $I : [s] \to [s] \cup \{\perp\}$ whose outputs in $[s]$ are pairwise distinct). For $I \in \db_s$, we write $\dom(I)$ and $\im(I)$ for its domain and image (both excluding $\perp$), and define its size $|I| = |\dom(I)|$. We write $\db_{\leq t} = \{I \in \db_s : |I| \leq t\}$. For $x\in[s]$ and $y \in [s] \cup \{\perp\}$, we define $I[x \to y]$ as,
\begin{align*}
    I[x \to y](x^\prime) \coloneq \begin{cases}
        I(x^\prime) &\text{If $x\neq x^\prime$ and either $I(x^\prime)\neq y$ or $y= \perp$}\\
        \perp &\text{If $x\neq x^\prime, y\neq \perp$ and $I(x^\prime)=y$}\\
        y &\text{If $x=x^\prime$.}
    \end{cases}
\end{align*}

The Hilbert space $\calH(\db_s)$ is spanned by orthonormal basis states $\{\ket{I} : I \in \db_s\}$, and we store the injective database in the register $\I$. We write $\Pi_{\leq t,\I}$ for the projector
\begin{align*}
    \Pi_{\leq t, \I} \coloneq \sum_{I: \abs{I} \leq t} \ketbra{I}.
\end{align*}
The notation $\norm{\,\cdot \,}_{\leq t}$ means $\norm{\,\cdot \, \Pi_{\leq t, \I}}$.
When comparing two injective databases $I,J$, we say that $J\succeq I$ when for all $x\in \dom(I)$, we have $I(x)=J(x)$. We denote the set of databases that extend $I$ as $\Ext(I) \coloneq \{J: J \succeq I\}$. The projector $\Pi_{\succeq I,\I}$ on all databases that \emph{include} $I$ is defined as
\begin{align*}
    \Pi_{\succeq I,\I} \coloneq \sum_{J: J\in \Ext(I)} \ketbra{J}.
\end{align*}

\subsection{Permutation oracles}\label{sec:permutation_oracles_definition}

Let us present permutation oracles and the compressed permutation oracle framework of~\cite{Car25}. We emphasize that throughout the paper, symbols such as $\I, \C$ and $\Lab$ symbolize registers, while $I,C$ and $L$ represent instances of databases (usually in these registers). Hence $\ket{I}_{\I}$ represents the database $I$ in register~$\I$.

\paragraph{Standard permutation oracle.} First, let us describe the oracle $O_\phi$ for $\phi\sim \Sym(s)$ on $b\in \{0,1\}$, $x\in[s]$, and $y\in\{0,1\}^d$,
\begin{align*}
    O_{\phi} \ket{b, x}_{\X} \ket{y}_\Y \coloneq \ket{b,x}_\X \ket{y \oplus \phi^{1-2b}(x)}_\Y,
\end{align*}
where the registers $\X,\Y$ describe the query input and output registers. In this case, the bit $b$ controls the direction of querying $x$. Let $\alg^{O_\phi}$ be a quantum algorithm that makes $q$ queries to the oracle $O_\phi$. The final mixed state of the algorithm and its purification are described as
\begin{align*}
    \ket{\psi_q^{O_\phi}} &= A_q O_\phi \dots A_1 O_\phi A_0 \ket{0},\\
    \rho_{\alg}^{(O_{\phi})} &= \EX_{\phi \sim \Sym(s)}\left[\ketbra{\psi_q^{O_\phi}}{\psi_q^{O_\phi}}\right],\\
    \ket{\psi_q^{(O)}}_{\A\regP} &=\frac{1}{\sqrt{s!}}\sum_{\phi\in \Sym(s)} \ket{\psi_q^{O_\phi}}_{\A} \ket{\phi}_{\regP}. 
\end{align*}

\paragraph{Ordinary compressed permutation oracle.} Let us begin by defining several key terms. Let $\ket{\psi_1}, \ket{\psi_2}$ be arbitrary pure states such that $\braket{\psi_1}{\psi_2} =0$. Then $\pCfromPlus(\ket{\psi_1}, \ket{\psi_2})$ represents the compression primitive defined as
\begin{align*}
    \pCfromPlus(\ket{\psi_1}, \ket{\psi_2}) \coloneq \Id - \ketbra{\psi_1} - \ketbra{\psi_2} + \ketbra{\psi_1}{\psi_2} + \ketbra{\psi_2}{\psi_1}.
\end{align*}
This operator swaps $\ket{\psi_1}$ and $\ket{\psi_2}$ and fixes their common orthogonal complement and is a self-adjoint unitary.

The database register $\I$ holds states in $\calH(\db_s)$, initialized to $\ket{\perp}_\I$, the trivial database mapping all inputs to $\perp$. In the forward operators below, we suppress the direction bit $b$. For $I \in \db_s$ and $x \notin \dom(I)$, let
\begin{align*}
    \ket{+_{x,I}} \coloneqq \frac{1}{\sqrt{s - |I|}} \sum_{y \in [s] \setminus \im(I)} \ket{I[x \to y]}
\end{align*}
be the uniform superposition over all non-colliding extensions of $I$ at $x$. Define
\begin{align*}
    \calH_{x,I} \coloneqq \Span\bigl(\{\ket{I}\}\cup\{\ket{I[x\to y]}:y\in[s]\setminus\im(I)\}\bigr).
\end{align*}
These subspaces form an orthogonal decomposition of $\calH(\db_s)$ for fixed $x$. On $\calH_{x,I}$, the \emph{permutation compression} operator $\pC_{x,I}$ swaps the trivially undefined state $\ket{I}$ with this uniform superposition:
\begin{align}
    \pC_{x,I} &\coloneqq \Id - \ketbra{+_{x,I}}{+_{x,I}} - \ketbra{I}{I} + \ketbra{+_{x,I}}{I} + \ketbra{I}{+_{x,I}}\label{eq:pc_x_I}\\
    &=\pCfromPlus(\ket{+_{x,I}}, \ket{I})\big|_{\calH_{x,I}}.
\end{align}
The key difference from the standard (function) compression operator is that the superposition ranges only over $[s] \setminus \im(I)$, enforcing injectivity at each step. The full operator $\pC$ is defined as
\begin{align}
    \pC_x &\coloneq \bigoplus_{I: x \notin \dom(I)} \pC_{x,I}\\
    \pC &\coloneq \sum_{x\in[s]}\ketbra{x}_{\X}\otimes\pC_{x,\I}\label{eq:pc_def}
\end{align}
The \emph{extended purified oracle} $\xorOP$ XORs the current database value into the output register when the queried input is defined, and acts as the identity otherwise:
\begin{align*}
    \xorOP\ket{x}_\X\ket{y}_\Y\ket{I}_\I \coloneqq \begin{cases}
        \ket{x}_\X\ket{y \oplus I(x)}_\Y\ket{I}_\I & \text{if $x \in \dom(I)$,} \\
        \ket{x}_\X\ket{y}_\Y\ket{I}_\I & \text{otherwise.}
    \end{cases}
\end{align*}
Finally, the \emph{flip} operator $\cOp{F}$ inverts the injective database, $\cOp{F}\ket{I}_\I = \ket{I^{-1}}_\I$, where $I^{-1}(y) = x$ whenever $I(x) = y$. This allows inverse queries to be answered by first flipping the database, applying the forward-direction oracle, then flipping back.

The compressed permutation oracle \cpO is defined as
\begin{align}
    \cpO \ket{b} \coloneq \ket{b} \otimes \begin{cases}
        \pC \cdot \xorOP \cdot \pC^{\dagger} &\text{if $b=0$,}\\
        \cOp{F} \cdot \pC \cdot \xorOP \cdot \pC^{\dagger} \cOp{F}^{\dagger} &\text{otherwise.}
    \end{cases}\label{eq:cpo_definition}
\end{align}
The compressed oracle experiment initializes $\I$ to $\ket{\perp}_\I$ and replaces each call to $O_\phi$ with $\cpO$. The unitaries $A_i$ act only on $\A$, with the identity on $\I$ implicit. The joint state of the algorithm and database registers after $q$ queries is
\begin{align*}
    \ket*{\psi_q^{(\cpO)}}_{\A\I} \coloneqq A_q \cpO \cdots A_1 \cpO A_0 \ket{0}_\A\ket{\perp}_\I.
\end{align*}
Tracing out $\I$ gives the algorithm's reduced state
\begin{align*}
    \rho_\A^{(\cpO)} = \Tr_\I\!\left[\ketbra{\psi_q^{(\cpO)}}{\psi_q^{(\cpO)}}_{\A\I}\right].
\end{align*}

The following observation is useful to understand how querying the same value may behave.

\begin{observation}\label{obs:compressed_permutation_oracle_already_queried}
    Let $I\in\db_s$, $x\notin\dom(I)$, and $y\in[s]\setminus\im(I)$, and let $m=s - \abs{I}$. Then,
    \begin{align*}
        \pC_x \ket{I[x\to y]}_{\I} = \ket{I[x\to y]} - \frac{1}{\sqrt{m}} \ket{+_{x,I}} + \frac{1}{\sqrt{m}}\ket{I}.
    \end{align*}
\end{observation}

The same calculation gives the following bound when the answer is retained.

\begin{proposition}\label{prop:recompression_unlikely}
    Let $x,y\in[s]$ and $I$ be an injective database over $[s]$ such that $I(x)=y$, and $I_\bot = I\setminus \{(x,y)\}$. Then,
    \begin{align*}
        \norm{(\pC_x - \Id)\ket{I}}^2 = \frac{2}{s - \abs{I}+1}.
    \end{align*}
    More generally, fix $x$ and let $\zeta$ be a vector supported on databases $I$ with $x\in\dom(I)$ and $\abs{I}\leq r$, with an orthogonal record $\ket{I(x)}$ in a subregister of the algorithm register. Then,
    \begin{align*}
        \norm{(\pC_x - \Id)\zeta} \leq \sqrt{\frac{2}{s - r+1}}\norm{\zeta}.
    \end{align*}
\end{proposition}
\begin{proof}
    By \cref{obs:compressed_permutation_oracle_already_queried},
    \begin{align}
        \norm{(\pC_x - \Id)\ket{I}}^2 = \norm{\frac{\ket{I_\bot} - \ket{+_{x,I_\bot}}}{\sqrt{s-\abs{I_\bot}}}}^2 = \frac{2}{s - \abs{I_\bot}}=\frac{2}{s-|I|+1},\label{eq:recompression_unlikely_pt1}
    \end{align}
    where in the second equality we used the fact that $\ket{I_\bot}$ and $\ket{+_{x,I_\bot}}$ are orthogonal. For the second part of the statement, we may write $\zeta$ as
    \begin{align*}
        \zeta = \sum_{y, I_\bot} (\zeta_{y,I_\bot}\otimes\ket{y})_{\A} \ket{I_\bot[x\to y]}_{\I}.
    \end{align*}
    Therefore,
    \begin{align*}
        \norm{(\pC_x - \Id)\zeta}^2 &= \sum_{y,I_\bot} \frac{2}{s-\abs{I_\bot}} \norm{\zeta_{y,I_\bot}}^2 &\mbox{(By \cref{eq:recompression_unlikely_pt1})}\\
        &\leq \frac{2}{s - r + 1} \sum_{y,I_\bot} \norm{\zeta_{y,I_\bot}}^2 = \frac{2}{s - r + 1}\norm{\zeta}^2.
    \end{align*}
    Taking the square root completes the statement.
\end{proof}

\paragraph{Exact compressed permutation oracle.} Aside from the compressed permutation oracle from~\cite{Car25}, we use an oracle that bridges the gap between \cpO and $O_{\phi}$ based on the subspaces analyzed by Rosmanis~\cite{rosmanis2022tight} and revisited in~\cite{CM26}. This will be an exact compressed purification of the standard oracle, we call it the \emph{exact compressed permutation oracle} and denote it by $\widehat{\cpO}$. Throughout this section, we say a bound on $O_c$ holds with \emph{coherent controls} if it also holds for $\sum_c \ketbra{c} \otimes O_c$.

For $I \in \db_s$, let $\ket{T_I}$ be the uniform superposition over permutations that extend $I$,
\begin{align*}
    \ket{T_I} \coloneq \frac{1}{\sqrt{(s - \abs{I})!}} \sum_{\phi \succeq I} \ket{\phi}.
\end{align*}
For $r\in [0,s]$, define $\calH_{s,\leq r} \coloneq \Span\{\ket{T_I}: I\in \db_s, \abs{I}=r \}$. Notice that these spaces are nested as for $\abs{I}=r < s$ and $x\notin \dom(I)$,
\begin{align*}
    \ket{T_I} = \frac{1}{\sqrt{s - r}}\sum_{y\notin \im(I)} \ket{T_{I[x\to y]}}.
\end{align*} 
Let $\calH_{s,\leq -1} \coloneq 0$, $\calH_{s,r} \coloneq \calH_{s,\leq r} \ominus \calH_{s,\leq r - 1}$ and $\Pi_{s,r}, \Pi_{s,\leq r}$ be the projectors on $\calH_{s,r},\calH_{s,\leq r}$ respectively. We call $\calH_{s,r}$ level $r$. On the register $\regP$ over fully described permutations $\phi$ we define
\begin{align*}
    \Pi_{\succeq I, \regP} \coloneq \sum_{\phi \succeq I} \ketbra{\phi}_{\regP}.
\end{align*}
Notice the following.
\begin{observation}[\cite{CM26,rosmanis2022tight}]\label{obs:proj_level_bound}
    A standard XOR query changes the level by at most $1$, \cpO changes the database size by at most $1$, and
    \begin{align*}
        \Pi_{\succeq I, \regP} \calH_{s,\leq t} \subseteq \calH_{s, \leq t + \abs{I}}.
    \end{align*}
\end{observation}
On the span of databases $I$ such that $\abs{I}=r$, define the operators $\Gamma_{s,r}$ and $G_{s,r}$ as
\begin{align}
    \Gamma_{s,r}\ket I=\Pi_{s,r}\ket{T_I},\quad G_{s,r}=\Gamma_{s,r}\Gamma_{s,r}^{\dagger}.\label{ex:frame-definition}
\end{align}
By Lemma 4.9 in~\cite{CM26}, $G_{s,r}\succeq\Pi_{s,r}$ and when $2r < s$,
\begin{align}
    \Pi_{s,r}\preceq G_{s,r} \preceq\exp\left(\frac{r}{s-2r+1}\right)\Pi_{s,r}.\label{ex:frame-bounds}
\end{align}
The intertwiner $\calI_s$ that takes us from $\regP$ to $\I$ is defined as
\begin{align}
    \calI_s\coloneqq\bigoplus_{r<s}\Gamma_{s,r}^{\dagger}G_{s,r}^{-1/2}.\label{ex:isometry-definition}
\end{align}
By Theorem~4.2 and Lemma~4.10 in~\cite{CM26},
\begin{align}
    \calI_s^\dagger\calI_s&=\Id, &\calI_s\ket{T_\perp}&=\ket\perp,\notag\\
    \prv{\leq r}\calI_s&=\calI_s\Pi_{s,\leq r}. \label{eq:degree-records}
\end{align}
In particular, $\calI_s$ preserves norms and converts level bounds into record bounds. Let $K_s = \ran(\calI_s)$ and $\Pi_{K_s}=\calI_s\calI_s^\dagger$. By \cref{eq:degree-records}, $\Pi_{K_s}$ commutes with database size projectors $\Pi_{\leq t, \I}$.

For a fixed $x\in [s]$, let $O_x$ be defined by $O_x\ket{a,\phi}_{\Y\regP} = \ket{a \oplus \phi(x),\phi}_{\Y\regP}$ and let $\cpO_x$ be the corresponding ordinary compressed permutation query. Define the exact query operator $\widehat{\cpO}_x$ as
\begin{align}
    \widehat{\cpO}_x \coloneq \calI_sO_x\calI_s^{\dagger} +(\Id-\Pi_{K_s})\cpO_x(\Id-\Pi_{K_s}). \label{ex:exact-query-definition}
\end{align}
Using the proof techniques of~\cite{CM26}, we obtain the following statement. Note that in the ranges we use, we obtain an improvement over Theorem~4.2 of~\cite{CM26} as we can get rid of a $O(t/s)$ term, where $t$ is the number of records in the database.
\begin{lemma}[Exact compressed oracle query]\label{ex:exact-query}
    The operator $\widehat{\cpO}_x$ is a self-adjoint contraction that satisfies $\widehat{\cpO}_x \calI_s = \calI_s O_x$ and changes the database size by at most $1$. Furthermore, if $4(t+4) \leq s$, then
    \begin{align}
        \norm{(\widehat{\cpO}_x - \cpO_x) \Pi_{\leq t, \I}} = O\left( \frac{1}{\sqrt{s}} \right).\label{eq:exact_ordinary_oracle_calls}
    \end{align}
    The same holds for inverse queries and under coherent control.
\end{lemma}
\begin{proof}
    Let us first show the statements about $\widehat{\cpO}_x$. In \cref{ex:exact-query-definition}, the two summands act on orthogonal subspaces and are self-adjoint contractions. Furthermore, as $\calI_s^\dagger \calI_s = \Id$ and $(\Id-\Pi_{K_s})\calI_s=0$, we have that $\widehat{\cpO}_x\calI_s=\calI_sO_x$. By \cref{obs:proj_level_bound,eq:degree-records}, each summand changes the size of a database by at most one.

    The rest of the proof is concerned with \cref{eq:exact_ordinary_oracle_calls}. Proposition~4.8 in~\cite{CM26} is an analogous statement to ours, except for swap queries. Specifically, it uses the maps
    \begin{align*}
        V_s&\coloneq\bigoplus_{r<s}\Gamma_{s,r}^\dagger G_{s,r}^{-1},&
        \Gram&\coloneq V_s^\dagger V_s=\bigoplus_{r<s}G_{s,r}^{-1}.
    \end{align*}
    By \cref{ex:frame-bounds,ex:isometry-definition} and $4(t+4)\leq s$,
    \begin{align*}
        0\prec\Gram\preceq\Id,\qquad \Gram\big|_{\calH_{s,\leq t+3}}\succeq e^{-1/2}\Id,\qquad \calI_s=V_s\Gram^{-1/2}.
    \end{align*}
    For $y\in [s]$, let $S_y = \pCfromPlus(\ket{\bot}, \ket{y})$ swap $\ket{\perp}$ with $\ket{y}$ and fix the other basis states. Define the swap compressed permutation oracle as
    \begin{align*}
        O_x^{\mathrm{swap}}\ket{a,\phi}_{\Y\regP} &\coloneq(S_{\phi(x)}\ket a)_\Y\otimes\ket{\phi}_\regP,\\
        \xorOP_x^{\mathrm{swap}}\ket{a,I}_{\Y\I} &\coloneq
        \begin{cases}
            (S_{I(x)}\ket a)_\Y\otimes\ket I_\I,&x\in\dom(I),\\
            \ket{a,I}_{\Y\I},&x\notin\dom(I),
        \end{cases}\\
        \cpO_x^{\mathrm{swap}}&\coloneq\pC_x\xorOP_x^{\mathrm{swap}}\pC_x.
    \end{align*}
    Proposition~4.8 in~\cite{CM26} states that for $j\in [0,t+3]$,
    \begin{align}
        \norm{(\cpO_x^{\mathrm{swap}}V_s-V_sO_x^{\mathrm{swap}})\Pi_{s,\leq j}} \leq\frac{60}{\sqrt{s-j}}=O(s^{-1/2}),\label{ex:unnormalized-comparison}
    \end{align}
    where $O_x^{\mathrm{swap}}=\sum_{y\in[s]}S_y\otimes\Pi_{\succeq\{(x,y)\},\regP}$. By \cref{obs:proj_level_bound}, $O_x^{\mathrm{swap}}$ changes the level by at most $1$. Similarly, $\cpO_x^{\mathrm{swap}}$ also only changes the size of a database by at most $1$. Consequently, as $V_s$ maps level $r$ to databases of size $r$, for $j\in [0,t+2]$,
    \begin{align*}
        \norm{(\cpO_x^{\mathrm{swap}}V_s-V_sO_x^{\mathrm{swap}})^\dagger\Pi_{\leq j,\I}} &=O(s^{-1/2}).
    \end{align*}
    Consequently, as $\norm{V_s} \leq 1$,
    \begin{align*}
        \norm{[\Gram,O_x^{\mathrm{swap}}]\Pi_{s,\leq j}} &= \norm{\left((\cpO_x^{\mathrm{swap}}V_s-V_sO_x^{\mathrm{swap}})^\dagger V_s - V_s^\dagger(\cpO_x^{\mathrm{swap}}V_s-V_sO_x^{\mathrm{swap}}) \right) \Pi_{s,\leq j}}\\ 
        &= O(s^{-1/2}).
    \end{align*}
    As $\Gram$ preserves levels, $[\Gram,O_x^{\mathrm{swap}}]$ maps $\calH_{s,\leq j}$ into $\calH_{s,\leq j+1}$. Apply \cref{ex:polar-normalization} with $V=V_s$, $U=O_x^{\mathrm{swap}}$, $T=\cpO_x^{\mathrm{swap}}$, $\calI = \calI_s$, $\Pi = \Pi_{s,\leq j}$, $\Pi^\prime = \Pi_{s,\leq j + 1}$ and $\kappa = e^{-1/2}$ to get
    \begin{align}
        \norm{(\cpO_x^{\mathrm{swap}}\calI_s-\calI_sO_x^{\mathrm{swap}}) \Pi_{s,\leq j} } = O(s^{-1/2}),\label{eq:normalized_swap_comparison}
    \end{align}
    when $j\in [0,t+2]$. Notice that any XOR query can be implemented by a swap query into an auxiliary register $\ket{\bot}$, XORing its value into the output register $\Y$ and applying a second swap query to restore the register to $\ket{\bot}$. The same works for compressed queries. Hence by the triangle inequality and \cref{eq:normalized_swap_comparison},
    \begin{align}
        \norm{(\cpO_x\calI_s-\calI_sO_x)\Pi_{s,\leq j}} = O(s^{-1/2}),\label{eq:xor_normalized_bound}
    \end{align}
    where $j\in [0,t+1]$. The decrease in $j$ is due to the first swap query. By \cref{ex:exact-query-definition} and $\Pi_{K_s}=\calI_s\calI_s^\dagger$,
    \begin{align*}
        \widehat{\cpO}_x-\cpO_x &=-(\cpO_x\calI_s-\calI_sO_x)\calI_s^\dagger -\calI_s(\cpO_x\calI_s-\calI_sO_x)^\dagger(\Id-\Pi_{K_s}).
    \end{align*}
    By \cref{obs:proj_level_bound,eq:degree-records}, $\cpO_x\calI_s-\calI_sO_x$ maps level $r$ to databases of sizes in $[r-1,r+1]$, while $\Pi_{K_s}$ preserves the database size. Hence by \cref{eq:xor_normalized_bound},
    \begin{align*}
        \norm{(\widehat{\cpO}_x-\cpO_x)\Pi_{\leq t,\I}} &\leq\norm{(\cpO_x\calI_s-\calI_sO_x)\Pi_{s,\leq t}} +\norm{(\cpO_x\calI_s-\calI_sO_x)\Pi_{s,\leq t+1}}\\
        &=O(s^{-1/2}).
    \end{align*}
    Hence \cref{eq:exact_ordinary_oracle_calls} holds. Inversion follows analogously. The generalization under coherent controls follows from \cref{lem:operator_norm_max}.
\end{proof}

\section{Welded trees oracles}\label{sec:wt_oracle}

In this section, we use the compressed permutation oracle technique to construct the welded trees oracles and analyze their properties.

\subsection{Standard welded trees oracle}\label{sec:construction_welded_trees}

The following hold for the remainder of the paper. Assume that $n$ is even and $N\coloneq 2^n$. Let $\distWeld\in [2,n-4]$ be a fixed integer for the remainder of the paper and the middle columns be $\middleCol = [n-\distWeld,n+\distWeld]$. 
Furthermore, we use $\permSize \coloneq 2^{n-\distWeld}$ for the size of the permuted column and $\errorTerm \coloneq 2^{-\distWeld}$ as an error term.

An instance of a welded trees graph $G_n = (V_n, E_n)$ consists of two binary trees $T_1,T_2$ of height $n$ (with $n$ edges from each root to a leaf and $n+1$ vertex levels) whose leaves are joined by a bipartite cycle. We denote the set of edges on the cycle $E_W$ and call them the \say{weld edges}. For $c\in [0,2n+1]$, we define the column nodes $\col(c) \subseteq V_n$ as
\begin{align*}
    \col(c) \coloneq \begin{cases}
        \{v\in V(T_1): \depth(v) = c\} &\text{if $c\leq n$,}\\
        \{v\in V(T_2): \depth(v) = 2n+1-c\} &\text{if $c\geq n+1$.}
    \end{cases}
\end{align*}
For $v \in V_n$, we define $\col(v)$ to be the unique $c \in [0,2n+1]$ such that $v \in \col(c)$.

Notice that as $T_1, T_2$ are binary trees, $\abs{\col(c)} = 2^{\min(c, 2n+1-c)}$. Each vertex $v\in V_n$ is assigned a structural index $\strucIdx: V_n \to [0,2^{n+2}-3]$ as follows:
\begin{align*}
    \strucIdx(v) \coloneq \begin{cases}
        0 &\text{if $v=\groot(T_1)$,}\\
        2^{n+1} - 1 &\text{if $v=\groot(T_2)$,}\\
        2\cdot \strucIdx(v^\prime) + 1 &\text{$\col(v)\in [1,n]$ and $v$ is the \emph{left} child of $v^\prime$,}\\
        2\cdot \strucIdx(v^\prime) + 2 &\text{$\col(v)\in [1,n]$ and $v$ is the \emph{right} child of $v^\prime$,}\\
        2\cdot \strucIdx(v^\prime) - 2^{n+1} + 2 &\text{$\col(v)\in [n+1,2n]$ and $v$ is the \emph{left} child of $v^\prime$,}\\
        2\cdot \strucIdx(v^\prime) - 2^{n+1} + 3  &\text{$\col(v)\in [n+1,2n]$ and $v$ is the \emph{right} child of $v^\prime$.}
    \end{cases}
\end{align*}

For the remainder of the paper, we will use the structural vertex $v\in V_n$ and the structural index $\strucIdx(v)$ interchangeably. We will use $p: V_n \to V_n\cup\{\bot\}$ to denote the parent of a vertex,
\begin{align*}
    p(w) \coloneq \begin{cases}
        \floor{(w - 1)/2} &\text{if $\col(w) \in [1,n]$,}\\
        \floor{(w-2^{n+1})/2} + 2^{n+1} - 1 &\text{if $\col(w) \in [n+1, 2n]$,}\\
        \bot &\text{otherwise.}
    \end{cases}
\end{align*}

Next, let us assign colors from the set $\colorSet = \{{\color{colorA}\alpha},{\color{colorB}\beta},{\color{colorC}\gamma}\}$ to the edges.

\begin{observation}[Edge coloring $G_n$]\label{lem:color_welded_graph}
    For any $G_n$, there exists a proper edge coloring $\colF: E_n \to \colorSet$.
\end{observation}
\begin{proof}
    Notice that $G_n$ is a bipartite graph as the odd columns may form one partition and the even columns the other. As the maximum degree of $G_n$ is 3, Kőnig's theorem implies that the graph is 3-edge-colorable~\cite{Konig16}.
\end{proof}

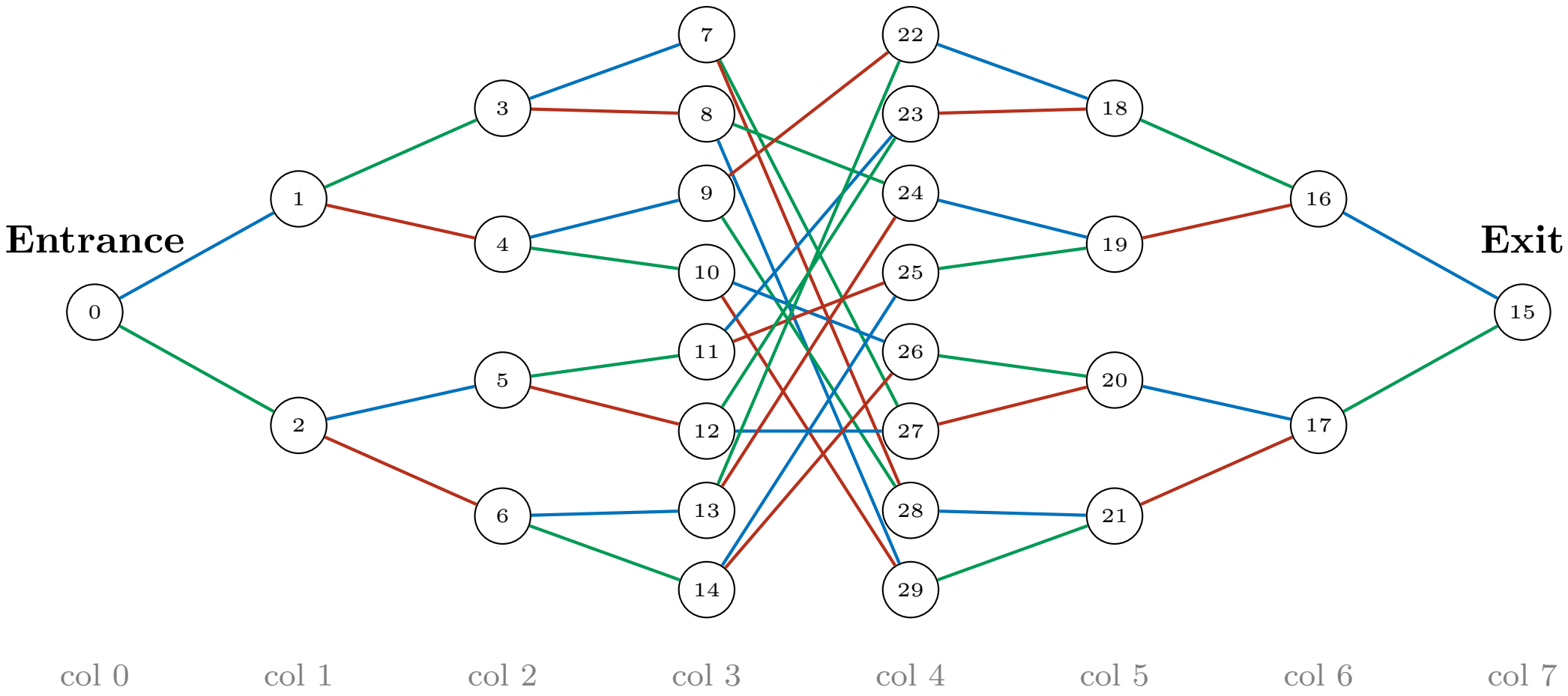
\begin{figure}
    \center
    \begin{tikzpicture}[
    every node/.style={circle, draw, fill=white, inner sep=0pt, minimum size=14pt, font=\tiny},
    treelabel/.style={rectangle, draw=none, fill=none, font=\normalsize\bfseries},
    collabel/.style={rectangle, draw=none, fill=none, font=\footnotesize, text=gray},
    >=latex
]

% ============================================================
%   Color definitions  (change these to recolor all edges)
% ============================================================

% Edge styles for each color
\tikzset{
    edgeA/.style={thick, colorA},
    edgeB/.style={thick, colorB},
    edgeC/.style={thick, colorC},
}

% ============================================================
%   Layout parameters
% ============================================================
\def\xsep{1.8}   % horizontal distance between columns
 
% ============================================================
%   LEFT TREE  (columns 0..3,  depth 3)
%   BFS labeling from root: 1, 2, 3, ..., 15
%
%   Tree-edge coloring (parent color → child colors):
%     L0:  →L1a = A,  →L1b = B
%     L1a: parent A,  →L2a = B,  →L2b = C
%     L1b: parent B,  →L2c = A,  →L2d = C
%     L2a: parent B,  →L3a = A,  →L3b = C
%     L2b: parent C,  →L3c = A,  →L3d = B
%     L2c: parent A,  →L3e = B,  →L3f = C
%     L2d: parent C,  →L3g = A,  →L3h = B
% ============================================================
 
% --- Col 0: root (entrance) ---
\node (L0) at (0, 0) {0};
 
% --- Col 1 ---
\node (L1a) at (\xsep,  1.0) {1};
\node (L1b) at (\xsep, -1.0) {2};
 
% --- Col 2 ---
\node (L2a) at (2*\xsep,  1.8) {3};
\node (L2b) at (2*\xsep,  0.6) {4};
\node (L2c) at (2*\xsep, -0.6) {5};
\node (L2d) at (2*\xsep, -1.8) {6};
 
% --- Col 3 (leaves) ---
\node (L3a) at (3*\xsep,  2.45) {7};
\node (L3b) at (3*\xsep,  1.75) {8};
\node (L3c) at (3*\xsep,  1.05) {9};
\node (L3d) at (3*\xsep,  0.35) {10};
\node (L3e) at (3*\xsep, -0.35) {11};
\node (L3f) at (3*\xsep, -1.05) {12};
\node (L3g) at (3*\xsep, -1.75) {13};
\node (L3h) at (3*\xsep, -2.45) {14};
 
% --- Left tree edges ---
\draw[edgeA] (L0) -- (L1a);
\draw[edgeB] (L0) -- (L1b);
 
\draw[edgeB] (L1a) -- (L2a);
\draw[edgeC] (L1a) -- (L2b);
\draw[edgeA] (L1b) -- (L2c);
\draw[edgeC] (L1b) -- (L2d);
 
\draw[edgeA] (L2a) -- (L3a);
\draw[edgeC] (L2a) -- (L3b);
\draw[edgeA] (L2b) -- (L3c);
\draw[edgeB] (L2b) -- (L3d);
\draw[edgeB] (L2c) -- (L3e);
\draw[edgeC] (L2c) -- (L3f);
\draw[edgeA] (L2d) -- (L3g);
\draw[edgeB] (L2d) -- (L3h);

% ============================================================
%   RIGHT TREE  (columns 4..7,  mirror of left tree)
%   BFS labeling from root: 16, 17, 18, ..., 30
%
%   Tree-edge coloring (same scheme as left tree):
%     R0:  →R1a = A,  →R1b = B
%     R1a: parent A,  →R2a = B,  →R2b = C
%     R1b: parent B,  →R2c = A,  →R2d = C
%     R2a: parent B,  →R3a = A,  →R3b = C
%     R2b: parent C,  →R3c = A,  →R3d = B
%     R2c: parent A,  →R3e = B,  →R3f = C
%     R2d: parent C,  →R3g = A,  →R3h = B
% ============================================================
 
% --- Col 4 (leaves) ---
\node (R3a) at (4*\xsep,  2.45) {22};
\node (R3b) at (4*\xsep,  1.75) {23};
\node (R3c) at (4*\xsep,  1.05) {24};
\node (R3d) at (4*\xsep,  0.35) {25};
\node (R3e) at (4*\xsep, -0.35) {26};
\node (R3f) at (4*\xsep, -1.05) {27};
\node (R3g) at (4*\xsep, -1.75) {28};
\node (R3h) at (4*\xsep, -2.45) {29};
 
% --- Col 5 ---
\node (R2a) at (5*\xsep,  1.8) {18};
\node (R2b) at (5*\xsep,  0.6) {19};
\node (R2c) at (5*\xsep, -0.6) {20};
\node (R2d) at (5*\xsep, -1.8) {21};
 
% --- Col 6 ---
\node (R1a) at (6*\xsep,  1.0) {16};
\node (R1b) at (6*\xsep, -1.0) {17};
 
% --- Col 7: root (exit) ---
\node (R0) at (7*\xsep, 0) {15};
 
% --- Right tree edges ---
\draw[edgeA] (R0) -- (R1a);
\draw[edgeB] (R0) -- (R1b);
 
\draw[edgeB] (R1a) -- (R2a);
\draw[edgeC] (R1a) -- (R2b);
\draw[edgeA] (R1b) -- (R2c);
\draw[edgeC] (R1b) -- (R2d);
 
\draw[edgeA] (R2a) -- (R3a);
\draw[edgeC] (R2a) -- (R3b);
\draw[edgeA] (R2b) -- (R3c);
\draw[edgeB] (R2b) -- (R3d);
\draw[edgeB] (R2c) -- (R3e);
\draw[edgeC] (R2c) -- (R3f);
\draw[edgeA] (R2d) -- (R3g);
\draw[edgeB] (R2d) -- (R3h);

% ============================================================
%   WELD EDGES
%   Designed so that every degree-3 leaf vertex has all
%   three colors {A, B, C} on its incident edges.
%
%   Left leaf tree colors:           Right leaf tree colors:
%     L3a=A  L3b=C  L3c=A  L3d=B      R3a=A  R3b=C  R3c=A  R3d=B
%     L3e=B  L3f=C  L3g=A  L3h=B      R3e=B  R3f=C  R3g=A  R3h=B
%
%   Each leaf needs its two weld edges to use the other
%   two colors.  The gluing below satisfies this at every
%   vertex (verified exhaustively).
% ============================================================
 
\draw[edgeB] (L3a) -- (R3f);
\draw[edgeC] (L3a) -- (R3g);
 
\draw[edgeA] (L3b) -- (R3h);
\draw[edgeB] (L3b) -- (R3c);
 
\draw[edgeC] (L3c) -- (R3a);
\draw[edgeB] (L3c) -- (R3g);
 
\draw[edgeA] (L3d) -- (R3e);
\draw[edgeC] (L3d) -- (R3h);
 
\draw[edgeC] (L3e) -- (R3d);
\draw[edgeA] (L3e) -- (R3b);
 
\draw[edgeB] (L3f) -- (R3b);
\draw[edgeA] (L3f) -- (R3f);
 
\draw[edgeB] (L3g) -- (R3a);
\draw[edgeC] (L3g) -- (R3c);
 
\draw[edgeA] (L3h) -- (R3d);
\draw[edgeC] (L3h) -- (R3e);
 
% ============================================================
%   LABELS
% ============================================================
 
% Entrance / Exit labels
\node[treelabel, above=4pt of L0] {Entrance};
\node[treelabel, above=4pt of R0] {Exit};
 
% Column labels
\node[collabel] at (0,       -3.2) {col 0};
\node[collabel] at (\xsep,   -3.2) {col 1};
\node[collabel] at (2*\xsep, -3.2) {col 2};
\node[collabel] at (3*\xsep, -3.2) {col 3};
\node[collabel] at (4*\xsep, -3.2) {col 4};
\node[collabel] at (5*\xsep, -3.2) {col 5};
\node[collabel] at (6*\xsep, -3.2) {col 6};
\node[collabel] at (7*\xsep, -3.2) {col 7};

% Title
% \node[rectangle, draw=none, fill=none, font=\large\bfseries] at (3.5*\xsep, 3.8)
%     {Glued Trees ($n = 3$)};

\end{tikzpicture}
    \caption{An instance of a welded trees graph $G_3$.}\label{fig:welded_tree}
\end{figure}

For the remainder of the paper, we will fix some edge coloring $\colF$ obtained using \cref{lem:color_welded_graph}. An example of a welded trees instance using the procedures above may be found in \cref{fig:welded_tree}.

For all nodes except the roots of $T_1,T_2$, we define the parent color $\upcolor$ as the color that leads to a node's parent, $\upcolor(v) = \colF( \{v, p(v) \})$. We will partition the nodes $v$ in $G_n$ based on $\col(v)$ and $\upcolor(v)$. Specifically, for all $c\in [2n]$ and $\alpha\in \colorSet$, define the \emph{color class} $\colClass{c,\alpha}$ and its size $\colClassSize{c,\alpha}$ as,
\begin{align*}
    \colClass{c,\alpha} \coloneq \{v: \col(v) = c, \upcolor(v) = \alpha \}, \;\; \colClassSize{c,\alpha} \coloneq \abs{\colClass{c,\alpha}}
\end{align*}

We call the edges in column $c$ those between vertices in columns $c-1$ and $c$. The following lemma shows that the number of edges with a color is approximately $\tfrac{1}{3}$ of all edges in the column.

\begin{lemma}[Lemma 55 in the full version of~\cite{CCG23}]\label{lem:colors_per_column}
    Let $\alpha_1$ be the color that is not used by any edges to $\groot(T_1)$. For any color $\alpha^\prime \in \colorSet$, let $\gamma(\alpha^\prime, c)$ denote the number of edges colored $\alpha^\prime$ in column $c\in [n]$ of $T_1$. Then,
    \begin{align}
        \gamma(\alpha^\prime, c) = \begin{cases}
            \floor*{\frac{2^c}{3}} &\text{if $c$ is odd and $\alpha^\prime=\alpha_1$ or $c$ is even and $\alpha^\prime \neq \alpha_1$}\\
            \ceil*{\frac{2^c}{3}} &\text{otherwise.}
        \end{cases}\label{eq:colors_per_column}
    \end{align}
    For $T_2$ (i.e. $c\in [n+2,2n+1]$), replace $c$ in \cref{eq:colors_per_column} with $2n-c+2$ and let $\alpha_1$ be the color not used by the edges incident to $\groot(T_2)$.
\end{lemma}
\begin{proof}
    The proof is by induction on the edge-column $c$. When $c=1$, the graph contains two differently colored edges neither of which is equal to $\alpha_1$, meaning \cref{eq:colors_per_column} holds. Assume \cref{eq:colors_per_column} holds for some $c\in [n-1]$, $c$ is odd and $\alpha$ is some color. For edge-column $c+1$, the parent $v$ of each edge colored $\alpha$ must have $\upcolor(v) \neq \alpha$. Therefore,
    \begin{align*}
        \gamma(\alpha, c+1) &= \sum_{\alpha^\prime \neq \alpha} \gamma(\alpha^\prime, c)\\
        &=\begin{cases}
            2\cdot \ceil{\frac{2^c}{3}} &\text{if $\alpha = \alpha_1$,}\\
            \ceil{\frac{2^c}{3}} + \floor{\frac{2^c}{3}} &\text{otherwise,}
        \end{cases}\\
        &=\begin{cases}
            \ceil{\frac{2^{c+1}}{3}} &\text{if $\alpha = \alpha_1$,}\\
            \floor{\frac{2^{c+1}}{3}} &\text{otherwise.}
        \end{cases}
    \end{align*}
    The result when $c$ is even or $c\in [n+2,2n+1]$ follows analogously.
\end{proof}

Furthermore, it will be useful to partition the edges in $E_n$ based on their colors. For an arbitrary edge $e = \{u,v\}$ where $\col(u) > \col(v)$, we use $\head(e)$ and $\tail(e)$ to denote their direction,
\begin{align*}
    \head(e) &\coloneq \begin{cases}
        v &\text{if $e\in E_{T_1}$,}\\
        u &\text{if $e \in E_{T_2}$,}\\
        u &\text{if $e\in E_W$,}
    \end{cases}\\
    \tail(e) &\coloneq \begin{cases}
        u &\text{if $\head(e) = v$,}\\
        v &\text{otherwise.}
    \end{cases}
\end{align*}

Thus a tree edge is always directed from a child to its parent. Formally, for any edge $e\in E_{T_1} \cup E_{T_2}$, $\upcolor(\tail(e)) = \colF(e)$. For weld edges $E_W$, the order is arbitrary, so we set it to go from column $n$ to $n+1$. Letting $c\in [2n]$ and $\alpha\in \colorSet$,
\begin{align}
    \treeEdges &\coloneq \left\{e = \{u,v\} \in E_{T_1} \cup E_{T_2}: \col(\tail(e)) = c, \colF(e) = \alpha \right\}\\
    &= \left\{ \{v,p(v)\}: v\in \colClass{c,\alpha} \right\},\label{eq:tree_edges_def_alt} \\
    \weldedEdges &\coloneq \left\{ e\in E_W: \colF(e) = \alpha \right\}.\label{eq:welded_edges_def}
\end{align}

Here $c$ indexes the child vertex column. Thus $\treeEdges$ lies in edge-column $c$ for $T_1$ and edge-column $c+1$ for $T_2$, using the convention of \cref{lem:colors_per_column}. By \cref{eq:tree_edges_def_alt,eq:welded_edges_def}, we have that $\abs{\treeEdges} = \colClassSize{c,\alpha}$ and $\abs{\weldedEdges} = \sum_{\beta \in \colorSet \setminus \{\alpha\}} \colClassSize{n,\beta}$.

\paragraph{Constructing the oracle.} For the remainder of the section, we will construct the welded trees oracle $\wtO$ and describe its distribution. There are two independent sources of randomness, both of which will use permutations. First, we randomize the edges using $\colorPerm$ and then the labels using \labelPerm.

Let us begin by defining the edge permutation \colorPerm. Simply put, \colorPerm permutes over edges in each column that have the same color. Formally, we consider permutations $\pi: E_n \to E_n$ such that,
\begin{align}
    \forall c\in [2n], \alpha\in \colorSet: \pi(\treeEdges) = \treeEdges \text{ and } \pi(\weldedEdges) = \weldedEdges.\label{eq:color_preserving_definition}
\end{align}
We call permutations that satisfy \cref{eq:color_preserving_definition} \emph{color-preserving}, as all vertices $v$ remain on the same column and their parent color $\upcolor(v)$ does not change. 

We may decompose color-preserving permutations based on the permutations over each set of edges separately. We will be solely interested in permutations within the middle columns $\middleCol$. The entire set of permuted edges is denoted by
\begin{align*}
    \middleEdge &\coloneq \{ (\mathrm{tr}, c, \alpha): c\in \middleCol, \alpha\in \colorSet \} \cup \{ (\mathrm{wld}, \alpha): \alpha\in \colorSet \}.
\end{align*}
For $b\in \middleEdge$, we let $\sigma_b: E_b \to E_b$ denote the permutation over the corresponding edges and $S_b$ be the set of all permutations. Finally, we define the edge permutation $\colorPerm$ and the set of all permutations $\colorPresSet$ as
\begin{align*}
    \colorPerm &\coloneq \prod_{b\in \middleEdge}  \sigma_b\\
    \colorPresSet &\coloneq \prod_{b\in \middleEdge} S_b.
\end{align*}
Notice that when we permute edges using $\colorPerm$, we do not change the columns of the vertices. Therefore, if our oracle returns the structural index, we are leaking the current position in the graph, making the pathfinding problem trivial. In order to avoid this leak, we randomly assign each node with a label. Let $M \coloneq 2^{2n}$. The corresponding set of permutations is
\begin{align*}
    S_M^{(0)} &\coloneq \{\pi \in \Sym([0,M]): \pi(0) = 0\}.
\end{align*}
We will always sample the labeling permutation $\labelPerm \sim S_M^{(0)}$. Notice that as $\labelPerm(0)=0$, we have that the entrance label is fixed and thus known to the algorithm. Furthermore, since $M \gg \abs{V_n}$, randomly guessing the label of a structural vertex has a negligible probability.

For $v\in [0,M]$ and $\alpha\in\colorSet$, we may define the edge-color function $\edgeVertex{\alpha}{v}$ to be the edge that contains $v$ in direction $\alpha$,
\begin{align}
    \edgeVertex{\alpha}{v} & \coloneq \begin{cases}
        e &\text{if $\exists e = \{u,v\} \in E_n$ s.t. $\colF(e) = \alpha$}\\
        \bot &\text{otherwise.}
    \end{cases}
\end{align}
Notice that $\edgeVertex{\alpha}{v}$ returns $\bot$ either when $v\notin V_n$ or when we query the one color that the edges from the roots of the trees were not assigned by \cref{lem:color_welded_graph}. The edge-cell indicator $\edgeIndic(e)$ describes which permutation an edge belongs to:
\begin{align*}
    \edgeIndic(e) &\coloneq \begin{cases}
        (\mathrm{tr}, \col(\tail(e)), \colF(e)) &\text{if $e\in E_{T_1} \cup E_{T_2}$ and $\col(\tail(e)) \in \middleCol$,}\\
        (\mathrm{wld}, \colF(e)) &\text{if $e\in E_W$,}\\
        \bot &\text{otherwise.}
    \end{cases}
\end{align*}

Assume that we have some fixed $\colorPerm \in \colorPresSet$. Letting $v\in [0,M]$, $\alpha\in \colorSet$ and $e=\edgeVertex{\alpha}{v}$, the neighbor function $\ncF_{\colorPerm }$ is defined as
\begin{align}
    \ncF_{\colorPerm}(v,\alpha) &\coloneq \begin{cases}
        v &\text{if $v\notin V_n$ or $e = \bot$,}\\
        \head(e)  &\text{if $e \neq \bot$, $\edgeIndic(e) = \bot$ and $v=\tail(e)$,}\\
        \tail(e)  &\text{if $e \neq \bot$, $\edgeIndic(e) = \bot$ and $v=\head(e)$,}\\
        \head(\sigma_{\edgeIndic(e)}(e)) &\text{if $\edgeIndic(e) \neq \bot$ and $v=\tail(e)$,}\\
        \tail(\sigma_{\edgeIndic(e)}^{-1}(e)) &\text{if $\edgeIndic(e) \neq \bot$ and $v=\head(e)$.}
    \end{cases}\label{eq:def_ncf_perm}
\end{align}
Notice that for a fixed color, the function is an involution:
\begin{align*}
    \ncF_{\sigma}(\ncF_{\sigma}(v,\alpha),\alpha) = v.
\end{align*}
Furthermore, this handles the boundary of \middleCol on columns $n-\distWeld$ and $n+\distWeld$. For some $\sigma = (\labelPerm, \colorPerm) \in S_M^{(0)} \times \colorPresSet$, $l \in [0,M]$, $l^\prime \in \{0,1\}^{2n+1}$ and $\alpha\in\colorSet$, the welded trees oracle $\wtO_{\sigma}$ is defined as
\begin{align}
    \wtF_{\sigma}(l, \alpha) &\coloneq \labelPerm( \ncF_{\sigma}( \labelPerm^{-1}(l), \alpha))\label{eq:oracle_function_definition} \\
    \wtO_\sigma \ket{l, \alpha, l^\prime} &\coloneq \ket*{l, \alpha, l^\prime \oplus \wtF_{\sigma}(l, \alpha)},\nonumber
\end{align}
where the first and third registers store strings of $2n+1$ bits. For $l>M$, all welded trees oracles act as the identity. 

When it is obvious that $\sigma$ is fixed, we will omit it from the notation and write $\wtF$ and $\wtO$. We write $\multDistr$ to denote the distribution over graphs obtained by uniformly sampling $\sigma \sim S_M^{(0)} \times \colorPresSet$. Notice that these graphs may contain more than one cycle over the weld edges. Letting $\evSingle$ be the event that the weld edges form one cycle, we let $\singDistr:=\multDistr|_{\evSingle}$. Unless specified otherwise, we are working with $\multDistr$. A visualization of the effect of \colorPerm may be found in \cref{fig:color_preserving}.

Suppose a query algorithm $\alg$ has access to \wtO. The final pure and mixed states of the algorithm are
\begin{align}
    \ket*{\psi^{(\wtO_\sigma)}}_{\A} &\coloneq A_q \wtO_\sigma A_{q-1} \dots \wtO_\sigma A_0 \ket{0}_{\A}\\
    \rho_{\alg}^{(\wtO)} &\coloneq \EX_{\substack{\labelPerm \sim S_M^{(0)}\\\colorPerm \sim \colorPresSet}}\left[\rho\left(\ket*{\psi^{(\wtO_\sigma)}}\right)\right].\label{eq:mixed_state_gtO_definition}
\end{align}
Write $\ket*{\psi_{\alg,q}^{(\wtO)}}_{\A\regP}$ for the purified version of $\rho_{\alg}^{(\wtO)}$ where $\regP$ stores $\sigma=(\labelPerm,\colorPerm)$ as in \cref{sec:permutation_oracles_definition}. The construction of $\wtO$ provides us with the following observation.

\begin{observation}\label{obs:min_num_nodes_per_perm}
    Assume that $n\geq 16$. For every $b\in \middleEdge$ and every $\sigma_b: E_b \to E_b$,
    \begin{align*}
        \abs{E_b} \geq \frac{\permSize}{4}.
    \end{align*} 
\end{observation}
\begin{proof}
    The smallest permutations occur at columns $c=n-l$ and some color $\alpha\in \colorSet$. By \cref{lem:colors_per_column},
    \begin{align*}
        \abs{\treeEdges} \geq \colClassSize{c,\alpha} &\geq \floor*{\frac{2^{n-\distWeld}}{3}} \geq \frac{\permSize}{4}.\qedhere
    \end{align*}
\end{proof}

\subsection{Compressed welded trees oracle}\label{sec:compressed_welded_tree}

In the following subsection, we construct the compressed welded trees oracle \cpwt and prove its bounded growth property.

A label database is the partial injection $L: [M] \rightharpoonup [M]$. As with $\labelPerm$, we use the convention $L(0)=0$ to fix the entrance vertex and do not count it in $\dom(L)$. An edge (also called color) database $C$ is a family of partial injections $C = \{C_{b}\}_{b\in \middleEdge}$ where each $C_{b}:E_b \rightharpoonup E_b$. We denote the cardinality of $C$ by $\abs{C} = \sum_b \abs{C_b}$. We denote their respective registers by $\Lab$ and $\C$. A pair of the label and color databases is denoted using $I=(L,C)$ under the register $\I$.

For $e_1, e_2 \in E_b$, we write $C[e_1\to e_2]$ to denote extending the injection $C_b$ with $e_1\to e_2$ and keeping the rest of $C$ the same. As each edge has a unique label $\edgeIndic(e)$, we may omit the subscript $b$. Furthermore, $C_1 \preceq C_2$ denotes that for all $b\in \middleEdge$, $C_{1,b} \preceq C_{2,b}$.

Following the definition of $\wtF$ in \cref{eq:oracle_function_definition}, \cpwt first \emph{unlabels} using $\pCLabelInv$ (compression operator for $\labelPerm^{-1}$), followed by moving between columns using $\pCPerm$ (compression operator for $\colorPerm$) and finally \emph{relabels} using $\pCLabel$ (compression operator for $\labelPerm$). All three operators perform the recording action. The query-input and query-output registers $\X$ and $\Y$ are written as $\ket{l,\alpha}_{\X} \ket{a}_{\Y}$ where $l \in [0,M]$, $a\in \{0,1\}^{2n+1}$ and $\alpha\in \colorSet$. The operator $\xorOP$ XORs the label of the adjacent structural vertex into $\Y$.

For a fixed partial edge database $C$, a structural index $v\in [0,M]$ and a color $\alpha\in \colorSet$, let $e = \edgeVertex{\alpha}{v}$ and $b=\edgeIndic(e)$. The partial neighbor map is defined as
\begin{align}
    \ncF_C(v,\alpha) &\coloneq \begin{cases}
        v &\text{if $v\notin V_n$ or $e=\bot$,}\\
        \head(e) &\text{if $b=\bot$ and $v=\tail(e)$,}\\
        \tail(e) &\text{if $b=\bot$ and $v=\head(e)$,}\\
        \head(C_b(e)) &\text{if $b\in \middleEdge$, $v=\tail(e)$ and $e\in \dom(C_b)$,}\\
        \tail(C_b^{-1}(e)) &\text{if $b\in \middleEdge$, $v=\head(e)$ and $e\in \im(C_b)$,}\\
        \bot &\text{otherwise.}
    \end{cases}
\end{align}
Thus the map $\ncF_C$ uses $C_b$ exactly as $\ncF_{\colorPerm}$ uses $\sigma_b$ in \cref{eq:def_ncf_perm}. Furthermore, notice that when $\ncF_C(v,\alpha)\neq \bot$, we have that $\ncF_C(\ncF_C(v,\alpha),\alpha) = v$.

Let us define the inverse-label compression operator \pCLabelInv, which is essentially just an inverse call to the $\pC$ operator defined in \cref{eq:pc_def}. Letting $l\in [M]$ and $L$ being some label database such that $l\notin \im(L)$,
\begin{align}
    \ket{+_{l,L}^{\mathrm{inv}}} &\coloneq \frac{1}{\sqrt{M - \abs{L}}} \sum_{v\in [M]\setminus \dom(L)} \ket{L[v \to l]}\label{eq:plus_state_label_inv} \\
    \pC^{\mathrm{inv}}_{l,L} &\coloneq \pCfromPlus(\ket{+_{l,L}^{\mathrm{inv}}}, \ket{L})\nonumber\\ 
    \pC^{\mathrm{inv}}_{l} &\coloneq \bigoplus_{L: l\notin \im(L)} \pC^{\mathrm{inv}}_{l,L}\nonumber\\
    \pCLabelInv \ket{l,\alpha}_{\X}\ket{L}_{\Lab} &\coloneq \begin{cases}
        \ket{l,\alpha}_{\X}\left( \pC^{\mathrm{inv}}_{l} \ket{L}_{\Lab} \right) &\text{if $l\in [M]$,}\\
        \ket{l,\alpha}_{\X} \ket{L}_{\Lab} &\text{if $l\notin [M]$.}
    \end{cases}\nonumber
\end{align}

Next, we define the compressed edge-permutation operator. Fix $b\in \middleEdge$ and $e\in E_b$. For an edge database $C$, we define the following:
\begin{align}
    \ket{+_{e,b,C}^{\rightarrow}} &\coloneq \frac{1}{\sqrt{\abs{E_b} - \abs{C_b}}} \sum_{e^\prime \in E_b \setminus \im(C_b)} \ket{C[e \to e^\prime]}\label{eq:plus_state_upcolor}\\
    \pC^{\rightarrow}_{e,b} &\coloneq \bigoplus_{C: e\notin \dom(C_b)} \pCfromPlus \left(\ket{+_{e,b,C}^{\rightarrow}}, \ket{C} \right)\label{eq:pC_right_direction}\\
    \ket{+_{e,b,C}^{\leftarrow}} &\coloneq \frac{1}{\sqrt{\abs{E_b} - \abs{C_b}}} \sum_{e^\prime \in E_b \setminus \dom(C_b)} \ket{C[e^\prime \to e]}\label{eq:plus_state_downcolor}\\
    \pC^{\leftarrow}_{e,b} &\coloneq \bigoplus_{C: e\notin \im(C_b)} \pCfromPlus \left(\ket{+_{e,b,C}^{\leftarrow}}, \ket{C} \right).\label{eq:pC_left_direction}
\end{align}
For $v\in [0,M]$ and $\alpha \in \colorSet$, let $e=\edgeVertex{\alpha}{v}$ and $b=\edgeIndic(e)$. Define
\begin{align}
    \pC^\prime_{v,\alpha} &\coloneq \begin{cases}
        \pC^{\rightarrow}_{e,b} &\text{if $e\neq \bot$, $b\in \middleEdge$ and $v=\tail(e)$,}\\
        \pC^{\leftarrow}_{e,b} &\text{if $e\neq \bot$, $b\in \middleEdge$ and $v=\head(e)$,}\\
        \Id &\text{otherwise.}
    \end{cases}
\end{align}
The controlled edge-compression operator is defined as
\begin{align}
    \pCPerm \ket{l,\alpha}_{\X} \ket{L}_{\Lab} \ket{C}_{\C} &\coloneq \begin{cases}
        \ket{l,\alpha}_{\X}\ket{L}_{\Lab} \left( \pC^\prime_{v,\alpha} \ket{C}_{\C} \right) &\text{if $l\in \im(L)$ and $v=L^{-1}(l)$,}\\
        \ket{l,\alpha}_{\X}\ket{L}_{\Lab} \ket{C}_{\C} &\text{otherwise.}
    \end{cases}
\end{align}

The forward-label compression operator is defined using a partial map $\stepFunct(l,\alpha, L, C)$,
\begin{align}
    \stepFunct(l,\alpha, L, C) &\coloneq \begin{cases}
        \ncF_C(L^{-1}(l), \alpha) &\text{if $l\in \im(L) \cup \{0\}$,}\\
        \bot &\text{if $l\notin \im(L) \cup \{0\}$.}
    \end{cases}\\
    \pC^{\mathrm{for}}_{w} &\coloneq \bigoplus_{L: w\notin \dom(L)} \pC_{w,L},\\
    \pCLabel\ket{l,\alpha}_{\X} \ket{L}_{\Lab} \ket{C}_{\C} &\coloneq \begin{cases}
        \ket{l,\alpha}_{\X} \left( \pC^{\mathrm{for}}_{w}\ket{L}_{\Lab}\right) \ket{C}_{\C} &\text{if $w=\stepFunct(l,\alpha, L, C) \notin \{\bot, L^{-1}(l), 0\}$,}\\
        \ket{l,\alpha}_{\X} \ket{L}_{\Lab} \ket{C}_{\C} &\text{if $w\in \{\bot, L^{-1}(l), 0\}$.}
    \end{cases}
\end{align}

Note that we are using the definition of $\pC_{x,I}$ from \cref{eq:pc_x_I}. Finally, we define the XOR operator $\xorOP$ as
\begin{align*}
    \xorOP\ket{l,\alpha}_{\X}\ket{a}_{\Y} \ket{L,C}_{\I} \coloneq \begin{cases}
        \ket{l,\alpha}_{\X} \ket{a\oplus L(w)}_{\Y} \ket{L,C}_{\I} &\text{if $w = \stepFunct(l,\alpha, L, C) \in \dom(L) \cup \{0\}$,}\\
        \ket{l,\alpha}_{\X}\ket{a}_{\Y} \ket{L,C}_{\I} &\text{otherwise.}
    \end{cases}
\end{align*}

Therefore, we may define the ordinary 

compressed oracle.

\begin{definition}[Compressed Welded Trees Oracle] Let $\I = (\Lab, \C)$ be the injective database registers, which are initialized as $\ket{\bot}_{\I}$. The ordinary compressed welded trees oracle \cpwt is defined as
    \begin{align}
        \cpwt \coloneq \pCLabelInv \cdot \pCPerm \cdot \pCLabel \cdot \xorOP \cdot \pCLabel^\dagger \cdot \pCPerm^\dagger \cdot \pCLabelInv^\dagger .\label{eq:cpwt_definition}
    \end{align}
\end{definition}
The final mixed and pure state of some $q$-query algorithm \alg using \cpwt is denoted by,
\begin{align*}
    \ket*{\psi_{\alg,q}^{(\cpwt)}} &\coloneq A_{q,\A} \cpwt_{\A\I} A_{q-1, \A} \dots \cpwt_{\A\I} A_{0,\A}\ket{0}_{\A} \ket{\bot}_{\I},\\
    \rho_{\alg,q}^{(\cpwt)} &\coloneq \Tr_{\I}\left[\rho\left( \ket*{\psi_{\alg,q}^{(\cpwt)}} \right)\right].
\end{align*}

However, the oracle \cpwt is only an approximate purification of \wtO. We will also use the oracle \icpwt on the same database registers that represent \wtO exactly. To define it, we will use the intermediate oracle $\wtOZ$ that writes out its computation between each step of evaluating each permutation. Let $\Z$ be the intermediate computation register. Assume that the queried input is $\ket{l,\alpha}$. The oracle $\wtOZ$ acts in the following way.
\begin{enumerate}
    \item Write into $\Z$ the inverse label $v = \labelPerm^{-1}(l)$.
    \item Let $e = \edgeVertex{\alpha}{v}$. If $e\neq\bot$ and $b=\edgeIndic(e)\in\middleEdge$, write $f$ into $\Z$, where
    \begin{align*}
        f = \begin{cases}
            \sigma_b(e) &\text{if $v=\tail(e)$,}\\
            \sigma^{-1}_b(e) &\text{if $v=\head(e)$.}
        \end{cases}
    \end{align*}
    Otherwise, set $f=\bot$. We encode $\bot$ by $0$ and all valid edge indices by nonzero strings.
    \item Let $w = \ncF_{\colorPerm}(v,\alpha)$ be the corresponding neighbor of $v$, which is a function of $(v,\alpha, f)$.
    \item XOR $\labelPerm(w)$ with the output $\Y$.
    \item Uncompute $f$ and then uncompute $v$.
\end{enumerate}

We use the following conventions, also in the compressed version below. When $l=0$, skip the inverse-label calls and set $v=0$. Otherwise, if $l\neq0$ and $v=0$, skip the middle computation and only uncompute $v$. If an edge-permutation query returns $f=\bot$, set $w=\bot$ and skip the forward-label step. When $w=v$, XOR the original label $l$ directly, otherwise when $w=0$, do nothing. Invalid structural vertices and missing colors have $w=v$, and invalid edge-work values are treated as $f=\bot$. These conventions preserve the action of \wtO, since genuine permutation queries are always defined and $\labelPerm(v)=l$.

Therefore, the computation can be viewed as invoking the following permutations as oracles:
\begin{align*}
    \labelPerm^{-1} \cdot \sigma_b \cdot \labelPerm \cdot \sigma_b \cdot \labelPerm^{-1}.
\end{align*}
\begin{observation}\label{obs:true-xor-steps-oracle}
    Running the oracle $\wtOZ$ with the intermediate register $\ket{0,0}_{\Z}$ returns it to $\ket{0,0}_{\Z}$ and behaves like \wtO.
\end{observation}
\begin{proof}
    Conditioned on some query register $\ket{l,\alpha,a}_{\X\Y}$, \wtOZ\ acts as
    \begin{align*}
        \ket{l,\alpha,a}_{\X\Y}\ket{0,0}_{\Z} &\xmapsto{\labelPerm^{-1}} \ket{l,\alpha,a}_{\X\Y}\ket{v,0}_{\Z}\\
        &\xmapsto{\sigma_b}  \ket{l,\alpha,a}_{\X\Y}\ket{v,f}_{\Z}\\
        &\xmapsto{\labelPerm}  \ket{l,\alpha,a \oplus \labelPerm(w)}_{\X\Y}\ket{v,f}_{\Z}\\
        &\xmapsto{\sigma_b}  \ket{l,\alpha,a \oplus \labelPerm(w)}_{\X\Y}\ket{v,0}_{\Z}\\
        &\xmapsto{\labelPerm^{-1}}  \ket{l,\alpha,a \oplus \labelPerm(w)}_{\X\Y}\ket{0,0}_{\Z} = (\wtO_{\sigma} \ket{l,\alpha,a}_{\X\Y} )\ket{0,0}_\Z.\qedhere
    \end{align*}
\end{proof}

We let \cpwtZ be the compressed version of \wtOZ,
\begin{align*}
    \cpwtZ &\coloneq \cpLabInv \cdot \cpPerm \cdot \cpLabFor \cdot \cpPerm \cdot \cpLabInv.
\end{align*}
Here $\cpPerm=\pC^\prime_{v,\alpha}\cdot\xorOP_{\mathrm{edge}}\cdot\pC^\prime_{v,\alpha}$. For $e=\edgeVertex{\alpha}{v}$ and $b=\edgeIndic(e)\in\middleEdge$, the operator $\xorOP_{\mathrm{edge}}$ XORs $C_b(e)$ into the edge subregister of $\Z$ when $v=\tail(e)$, and $C_b^{-1}(e)$ when $v=\head(e)$. It acts as the identity if $b=\bot$ or the requested database entry is absent. Note that as we are using compression operators, it is possible that the queries are not perfectly uncomputed.
\begin{definition}[Exact Compressed Welded Trees Oracle]
    Let $\widehat{\cpwtZ}$ be \cpwtZ where each permutation query is replaced by its exact database query from \cref{ex:exact-query} and $\calI_{Z} $ be the isometry such that for any state $\ket{\psi}$, $\calI_{Z} \ket{\psi} = \ket{\psi} \ket{0,0}_{\Z}$. The \emph{exact compressed welded trees oracle} is
    \begin{align*}
        \icpwt &\coloneq  \calI_{Z} ^\dagger \cdot \widehat{\cpwtZ} \cdot \calI_{Z} .
    \end{align*}
\end{definition}
Furthermore, we let $\calI_{\regP}$ be the product of isometries from \cref{ex:isometry-definition} over the label and edge permutations,
\begin{align}
    \calI_{\regP} &\coloneq \calI_M\otimes\bigotimes_{b\in\middleEdge}\calI_{\abs{E_b}}.\label{eq:isometry_cpwt}
\end{align}
Notice that $\calI_{\regP}$ acts as the identity on the algorithm registers and maps the uniform initial permutation state to the empty database $\ket{\bot}$.
\begin{observation}\label{obs:icpwt_properties}
    The oracle \icpwt represents \wtO exactly, i.e.,
    \begin{align}
        \icpwt\calI_{\regP} = \calI_{\regP}\wtO.\label{eq:exact_intertwining_cpwt}
    \end{align}
    Furthermore, the final state of a $q$-query algorithm using \icpwt is supported on databases $(L,C)$ such that $\abs{L}\leq 2q$ and $\abs{C}\leq q$.
\end{observation}
\begin{proof}
    By \cref{obs:true-xor-steps-oracle}, we may implement \wtO using \wtOZ, as it always returns the work register to $\ket{0,0}_{\Z}$. By \cref{ex:exact-query}, each exact query satisfies $\widehat{\cpO}_x\calI_s=\calI_sO_x$ and $\calI_{\regP}$ does not alter the register $\Z$. Hence we may move $\calI_{\regP}$ through the $5$ queries, so
    \begin{align*}
        \widehat{\cpwtZ}\calI_{Z}\calI_{\regP}=\calI_{Z}\calI_{\regP}\wtO.
    \end{align*}
    Applying $\calI_{Z}^\dagger$ gives \cref{eq:exact_intertwining_cpwt}. Next, let us bound $\abs{L}$ and $\abs{C}$. Fix some input $\ket{l,\alpha}$. Notice that repeated oracle calls in \wtOZ (inverse label and edge) are conditioned on the same value, so they affect the same register in the database. By \cref{obs:proj_level_bound}, each query raises the label level by at most $2$ and edge level by at most $1$. As the initial permutation has level $0$, \cref{eq:exact_intertwining_cpwt,eq:degree-records} give $\abs{L}\leq 2q$ and $\abs{C}\leq q$.
\end{proof}

\subsection{Bridge between standard and compressed oracles}\label{sec:bridge_oracles}

In the following subsection, we will bridge \wtO, \icpwt and \cpwt. Specifically, \cref{lem:difference_cpwts} characterizes the exact difference between \icpwt and \cpwt, while \cref{thm:oracle_path_verification} shows the relationship between recording a path using \icpwt and solving the pathfinding problem.

For some edge class $b$, define the vertex class $\Vclass{b}$ and valid edge projector $\Psrc_b$ as
\begin{align*}
    \Vclass{b} &=  \{\tail(e),\head(e):e\in E_b\},\\
    \Psrc_b\ket{l,\alpha, a, I}_{\X\Y\I} &= \Id[l\neq 0, l\in \im(L),\edgeVertex{\alpha}{L^{-1}(l)}\in E_b]\ket{l,\alpha, a, I}_{\X\Y\I}.
\end{align*}
Notice that $\abs{\Vclass{b}} = 2\abs{E_b}$ and $\Psrc_b$ tests whether the recorded vertex has its queried edge in $E_b$. Letting $t$ be the bound on $\abs{L}+ \abs{C}$ and $\tcut=t+6$, assume that $\permSize \geq C(t+1)$ for a constant $C$ large enough that $4(t+16) \leq \abs{E_b}$ for all permuted edge classes $b\in \middleEdge$ (by \cref{obs:min_num_nodes_per_perm}) and $4(t+16)\leq M$. Lastly, for a database $I = (I_b, I_{\neg b})$, we use $I_b$ to denote $C_b$ and the label records at vertices in $\Vclass{b}$, while $I_{\neg b}$ is the rest of the database.

\begin{lemma}\label{lem:difference_cpwts}
    For each $b\in \middleEdge$, there exist self-adjoint operators $\Tdiff_b$ that change the number of records by at most $3$, $\Tdiff_b=\Psrc_b\Tdiff_b\Psrc_b$, $\norm{\Tdiff_b \prv{\leq \tcut}} \leq C\abs{E_b}^{-1/2}$ and,
    \begin{align}
        \left(\icpwt - \cpwt - \pCLabelInv \left( \sum_b \Tdiff_b \right) \pCLabelInv \right) \prv{\leq t} = \Err_t,\label{eq:graph-difference}
    \end{align}
    where $\norm{\Err_t} \leq C/N$. Furthermore, $\Tdiff_b$ is such that the inner product $\bra{a^\prime, J} \Tdiff_b \ket{a, I}$ is $0$ when $J_{\neg b} \neq I_{\neg b}$ and otherwise solely depends on $a,a^\prime$, $I_b, J_b$ and the labels used and number of labels in $I$ outside of $b$.
\end{lemma}
\begin{proof}
    Let $\ket{\psi}$ be such that $\prv{\leq t} \ket{\psi} = \ket{\psi}$. As each operator fixes the query register $\X$, we may fix some query $l,\alpha$. Suppose $l\neq 0$ as at $l=0$ no edges are permuted so only the forward-label replacement is necessary. First, replace the exact label queries in $\widehat{\cpwtZ}$ by their compressed versions $\cpO$. By \cref{ex:exact-query}, this introduces the error $\Err_{\mathrm{lab}}$ with $\norm{\Err_{\mathrm{lab}}\prv{\leq t}} = O(M^{-1/2}) = O(1/N)$. Hence we may expand \cpLabInv in $\widehat{\cpwtZ}$ as
    \begin{align*}
        \widehat{\cpwtZ} = \pCLabelInv_{,l} \xorOP^{\mathrm{inv}}_\mathrm{lab} \pCLabelInv_{,l} \cdot (\widehat{\cpPerm} \cdot \cpLabFor \cdot \widehat{\cpPerm} )\pCLabelInv_{,l} \xorOP^{\mathrm{inv}}_\mathrm{lab} \pCLabelInv_{,l} + \Err_{\mathrm{lab}}.
    \end{align*}
     Let
    \begin{align}
        {\widehat{\cpwtZ}}^{(1)} \coloneq \pCLabelInv_{,l} \xorOP^{\mathrm{inv}}_\mathrm{lab} \cdot ({\widehat{\cpPerm}} \cdot \cpLabFor \cdot {\widehat{\cpPerm}} ) \xorOP^{\mathrm{inv}}_\mathrm{lab} \pCLabelInv_{,l},\label{eq:drop_inner_pc_operator_label}
    \end{align}
    which is just removing the intermediate $\pCLabelInv_{,l}$ calls after XORing. Let $\ket{\phi}=\xorOP^{\mathrm{inv}}_{\mathrm{lab}}\pCLabelInv_{,l}\ket{\psi}\ket{0}_{\Z}$ be the state immediately after the first XOR. In the case that $v=0$ (equivalently, $l\notin\im(L)$, since $l\neq0$ and $0\notin\dom(L)$), the middle computation is the identity, so the compression operators perfectly cancel out. Otherwise, the register $\Z$ records that $L(v)=l$ and this does not change until the last XOR operator. Therefore \cref{prop:recompression_unlikely} applies to both before and after the computation between the XORs. As the label database increases by at most $2$, by the triangle inequality we have,
    \begin{align*}
        &\norm{({\widehat{\cpwtZ}} - {\widehat{\cpwtZ}}^{(1)})\ket{\psi}\ket{0}_{\Z}} \leq {\norm{\Err_{\mathrm{lab}}\ket\psi\ket0_{\Z}}+}\norm{(\pCLabelInv_{,l} - \Id) \Pi_{v\neq 0} \ket{\phi}}\\
        &\quad + \norm{(\pCLabelInv_{,l} - \Id) ({\widehat{\cpPerm}} \cdot \cpLabFor \cdot {\widehat{\cpPerm}}) \Pi_{v\neq 0} \ket{\phi}} \\
        &\quad\leq 2\sqrt{\frac{2}{M-t-1}} \norm{\ket{\psi}}{+O(1/N)\norm{\ket\psi}=O(1/N)\norm{\ket\psi}}. &\mbox{(By \cref{prop:recompression_unlikely})}
    \end{align*}
    The same argument applies to the two edge queries around the forward-label query. Recall the edge-compression operator $\pC^\prime_{v,\alpha}$, which is the identity when the edge is not permuted. By \cref{ex:exact-query}, replacing the exact edge queries of class $b$ by ordinary compressed ones costs at most $2C\abs{E_b}^{-1/2}$. If $f=\bot$, the compressions cancel out. Otherwise, the register $\Z$ keeps the queried edge assignment until it is uncomputed and \cref{prop:recompression_unlikely} bounds the inner compression. Hence,
    \begin{align*}
        \norm{ ({\widehat{\cpPerm}} \cdot \cpLabFor \cdot {\widehat{\cpPerm}} - \pC^\prime_{v,\alpha} \xorOP_{\mathrm{edge}} \cpLabFor \xorOP_{\mathrm{edge}} \pC^\prime_{v,\alpha}) \ket{\phi}} \leq {\left(\frac{2C}{\sqrt{|E_b|}}+2\sqrt{\frac{2}{|E_b|-\tcut-5}}\right)\norm{\ket{\phi}}.}
    \end{align*}
    In this bound we assume that $\prv{\leq \tcut} \ket{\phi} = \ket{\phi}$, the source is in class $b$ and the edge work register is initialized to $0$. Hence the expanded products have $I$ such that $\abs{I} \leq \tcut +6$. Let $\Pwork_b$ be the projector onto cases where $l\neq 0, v\neq 0$ and $\edgeVertex{\alpha}{v}\in E_b$ in the $\X\Z$ registers. We define this difference restricted to $b$ as,
    \begin{align*}
        \Delta_b&= \Pwork_b \left(\widehat{\cpPerm}\cpLabFor\widehat{\cpPerm} -\pC^\prime_{v,\alpha}\xorOP_{\mathrm{edge}}\cpLabFor \xorOP_{\mathrm{edge}}\pC^\prime_{v,\alpha} \right) \Pwork_b,\\
        \Tdiff_b&=\calI_{Z}^\dagger\xorOP^{\mathrm{inv}}_{\mathrm{lab}} \Delta_b\xorOP^{\mathrm{inv}}_{\mathrm{lab}}\calI_{Z}.
    \end{align*}
    Thus $\norm{\Tdiff_b\prv{\leq\tcut}} = O(\abs{E_b}^{-1/2})$. Notice that after removing these operators, each lookup in $\Z$ is equivalent to looking it up in the actual database. Furthermore, these values in $\Z$ are still always uncomputed. Therefore, suppressing the intermediate lookups, the remaining operators are
    \begin{align}
        \pCLabelInv_{,l} \cdot \pC^\prime_{v,\alpha} \cdot \cpLabFor \cdot \pC^\prime_{v,\alpha} \cdot \pCLabelInv_{,l}.\label{eq:removing_operators_in_cpwtz}
    \end{align}
    The operations in \cref{eq:removing_operators_in_cpwtz}, now controlled by the corresponding database entries, are exactly \cpwt by \cref{eq:cpwt_definition}. Therefore, subtracting $\pCLabelInv(\sum_b\Tdiff_b)\pCLabelInv$ from $\icpwt$ leaves $\cpwt$ and the $O(1/N)$ label error. \cref{eq:graph-difference} follows from the orthogonality of the query register.

    It remains to check the properties of $\Tdiff_b$. First, the number of records in the database after $\Tdiff_b$ changes by at most $3$. This is because each product in the brackets in $\Delta_b$ has only three operations that can change the number of records, each by one due to \cref{ex:exact-query,obs:proj_level_bound}. Furthermore, the XOR copies the recorded source without affecting the database, meaning that
    \begin{align*}
        \Pwork_b\xorOP^{\mathrm{inv}}_{\mathrm{lab}}\calI_{Z}=\xorOP^{\mathrm{inv}}_{\mathrm{lab}}\calI_{Z}\Psrc_b.
    \end{align*}
    Combined with the fact that $\Delta_b$ is self-adjoint, we have that $\Tdiff_b$ is self-adjoint and $\Tdiff_b=\Psrc_b\Tdiff_b\Psrc_b$.

    Finally, let us prove the final paragraph of the statement. In the range of $\Psrc_b$, the vertex $v=L^{-1}(l)$ is in $\Vclass{b}$ and determined by $I_b$. Both $\widehat{\cpPerm}$ and $\pC^\prime_{v,\alpha}$ only change $C_b$. When $\cpLabFor$ is applied, the input $w$ must be in $\Vclass{b}$ and only affects $L(w)$. Therefore $\bra{a^\prime, J} \Tdiff_b \ket{a, I} = 0$ when $J_{\neg b} \neq I_{\neg b}$. In $\ket{+_{w,L}}$, the available labels are $[M]\setminus \im(L)$ each with equal amplitude. Hence the dependence on $I_{\neg b}$ is only based on the labels used and their number.
\end{proof}

Next, we relate the probability an algorithm $\alg$ with \wtO succeeds with the probability that running \alg with \icpwt records the path in the database. A candidate path is the list $\pairPath=(z_0,\alpha_1,z_1,\ldots,\alpha_m,z_m)$ where $z_0 = 0$, the labels $z_i\in [0,M]$ are pairwise distinct, $\alpha_i \in \colorSet$ and $m\geq n$. Without loss of generality, we may assume there are no loops as they may be removed without making any queries. The following algorithm describes the verification process of $\alg$ using \wtO.

\begin{algobox}{Standard Oracle Pathfinding}\alglabel{alg:standard_game}
    \begin{enumerate}
        \item Sample some $\sigma = (\labelPerm, \colorPerm) \sim \multDistr$ to construct $\wtO_\sigma$.
        \item Run \alg with $q$ queries to $\wtO_\sigma$, obtaining some $\pairPath$. Reject if it does not satisfy the conditions we described above.
        \item\label{itm:checking_full_path} If $m\leq 2q+1$, verify the entire path as follows:
        \begin{enumerate}
            \item\label{itm:checking_valid_path} For each $i\in [m]$, query $\wtO_\sigma$ at $(z_{i-1},\alpha_i)$. Reject if the answer is not $z_i$.
            \item\label{itm:checking_for_exit} For each $\alpha \in \colorSet$, query $\wtO_\sigma$ at $(z_m,\alpha)$. Accept if and only if exactly two answers differ from $z_m$.
        \end{enumerate}
        \item Otherwise, if $m > 2q+1$, only test the first $2q+1$ labels. For each $i\in [2q+1]$ and $\alpha \in \colorSet$, query $\wtO_\sigma$ at $(z_i,\alpha)$. Accept if and only if for all $i$, at least one answer differs from $z_i$.
    \end{enumerate}
\end{algobox}

Notice that when $m > 2q+1$, we are simply testing whether the outputted path contains valid labels, which does not decrease the probability of successfully outputting an entrance to exit path. Otherwise, when $m\leq 2q+1$, we test whether the path is valid and actually exits.\footnote{Technically, due to the random permutation over the weld edges, it is possible that there are two edges between the same vertices crossing the weld. Therefore, \cref{itm:checking_for_exit} in \cref{alg:standard_game} checks whether the node has exactly two edges, not two different neighbors.}

Let $\probWTO$ denote the probability that \cref{alg:standard_game} succeeds. We compare this probability with a test on the database when \alg queries \icpwt instead. Notice that we may represent the graph formed by the recorded edges as $\rG(I)=(\rV(I),\rE(I))$ where,
\begin{align*}
    \rE(I)&=\{\{\tail(e),\head(f)\}:C_b(e)=f \text{ for some } b\in \middleEdge \},\\
    \rV(I)&=\bigcup_{g\in\rE(I)}g.
\end{align*}
We say that $\rG(I)$ contains a crossing if it contains a path from $\col(n-\distWeld-1)$ to $\col(n+\distWeld+1)$. The projector on such crossing is defined as,
\begin{align*}
    \prCross\ket{I}_{\I} = \begin{cases}
        \ket{I} &\text{if $\rG(I)$ contains a crossing,}\\
        0 &\text{otherwise.}
    \end{cases}
\end{align*}
Note that this only tests the edge database $C$.
\begin{algobox}{Pathfinding using Exact Compressed Oracle}\alglabel{alg:compressed_game}
    \begin{enumerate}
        \item Initialize the databases $I=(L,C)$ to
        $\ket\bot_{\I}$.
        \item Run $\alg$ with $q$ queries to $\icpwt$.
        \item Measure $\{\prCross,\Id-\prCross\}$ on $\I$,
        accepting if and only if the first outcome occurs.
    \end{enumerate}
\end{algobox}
Let $\probCross$ be the success probability of \cref{alg:compressed_game}. Unlike \cref{alg:standard_game}, this algorithm tests the records already present after $\alg$ finishes without making any additional queries. Next we compare projectors on values present in a database in the standard and exact compressed way. 

For $j\in [m]$, let $\regP_j$ be a permutation register of size $s_j$ with database register $\I_j$. Note that $\regP_j$ need not be distinct. Choose $x_j\in [s_j]$ and $Y_j \subseteq [s_j]$ and define
\begin{align*}
    \Pi_{x\to Y,\regP}&\coloneq\prod_{j=1}^m\sum_{y\in Y_j}\Pi_{\succeq\{(x_j,y)\},\regP_j},&
    \Pi_{x\to Y,\I}&\coloneq\prod_{j=1}^m\sum_{y\in Y_j}\Pi_{\succeq\{(x_j,y)\},\I_j}.
\end{align*}
These projectors represent the presence of some desired information in the $\regP$ and $\I$ registers respectively.

\begin{lemma}\label{lem:pin_witness}
    Let $\xi$ be a possibly unnormalized state on $\regP$ and any additional registers such that for all $j\in [m]$, $(\Pi_{s_j,\leq t})_{\regP_j}\xi=\xi$. Furthermore, assume $4(t+m+4)\leq s\leq\min_{j\in[m]}s_j$. Then for $\calI_{\regP}$ as in \cref{eq:isometry_cpwt}, there exists some constant $C$ such that,
    \begin{align*}
        \norm{\Pi_{x\to Y,\regP}\xi}\leq\norm{\Pi_{x\to Y,\I}\calI_{\regP}\xi} + C\sqrt{\frac{1}{s}\sum_{j=1}^m\abs{Y_j}}\norm{\xi}.
    \end{align*}
    The same bound holds for the projectors controlled by an additional register assuming the hypotheses and error bound hold uniformly over the control values.
    It also holds for inverse assignment tests $\prod_{j=1}^m \sum_{x \in X_j}\Pi_{\succeq\{(x,y_j)\}}$, with $\abs{X_j}$ in place of $\abs{Y_j}$.
\end{lemma}
\begin{proof}
    Fix some $j\in [m]$ and $y\in Y_j$. Let $s^\prime \geq s$ be the size of the corresponding permutation. Initialize the answer register $\Y$ to $\ket{0}$ and project the output of a query $O_{x_j}$ to $\ket{y}$. Notice that $\bra{y} O_{x_j} \ket{0}_\Y = \Pi_{\succeq \{ (x_j,y)\}, \regP_j}$, so this turns a query into the projector testing the mapping $[x_j \to y]$. Combining \cref{eq:degree-records,ex:exact-query} with the compression estimate in the proof of Lemma~5.1 of~\cite{CM26}, the triangle inequality gives us,
    \begin{align}
        \norm{\left( \Pi_{\succeq\{(x_j,y)\},\I_j}\calI_{s^\prime} -\calI_{s^\prime}\Pi_{\succeq\{(x_j,y)\},\regP_j} \right)\Pi_{s^\prime,\leq t+m}} \leq\frac{C}{\sqrt s},\label{ex:one-assignment}
    \end{align}
    for some constant $C$. For $y\in Y_j$, define the following vectors,
    \begin{align*}
        \eta=\Pi_{x\to Y,\regP}\xi,\qquad \eta_{j,y}=\Pi_{\succeq\{(x_j,y)\},\regP_j}\eta .
    \end{align*}
    Under a fixed $j$, $\eta_{j,y}$ are orthogonal and sum to $\eta$. By \cref{obs:proj_level_bound}, imposing these $m$ assignment conditions raises the level by at most $m$. Notice that the database projectors commute. Therefore by a union-bound, Cauchy-Schwarz and \cref{ex:one-assignment},
    \begin{align*}
        \norm{(\Id-\Pi_{x\to Y,\I})\calI_{\regP}\eta}^2 &\leq\sum_{j=1}^m\abs{Y_j}\sum_{y\in Y_j} \norm{(\Id-\Pi_{\succeq\{(x_j,y)\},\I_j})\calI_{\regP}\eta_{j,y}}^2\\
        &\leq\frac{C^2}{s}\sum_{j=1}^m\abs{Y_j}\norm{\eta}^2.
    \end{align*}
    Using the fact that $\calI_\regP$ is an isometry,
    \begin{align*}
        \norm{\eta}^2 &=\braket{\calI_{\regP}\eta}{\calI_{\regP}\xi}\\
        &\leq\norm{\eta}\norm{\Pi_{x\to Y,\I}\calI_{\regP}\xi} + \norm{(\Id-\Pi_{x\to Y,\I})\calI_{\regP}\eta}\norm{\xi}.
    \end{align*}
    Hence the equation in the statement follows. For controlled tests, expand the state as $\xi = \sum_c \ket{c} \otimes \xi_c$ where $c$ is the control register and apply the same procedure along with the fact that the vectors are orthogonal.
    For inverse assignment tests, observe that $\calI_s$ intertwines inversion of permutations with the flip $\cOp{F}$ of databases (both preserve levels and send $\ket{T_I}$ and $\ket{I}$ to $\ket{T_{I^{-1}}}$ and $\ket{I^{-1}}$, respectively).
\end{proof}

\begin{theorem}\label{thm:oracle_path_verification}
    For any algorithm $\alg$ making $q$ queries where $\permSize \geq 80 (q+1)$,
    \begin{align}
        \probWTO = O\left( \probCross + \frac{q}{\permSize}\right).\label{eq:path_verification_bound}
    \end{align}
\end{theorem}
\begin{proof}
    Let $\ket*{\psi_{\alg,q}^{(\cdot)}}$ be the final states running $\alg$ with some oracle $\cdot$, either $\wtO$ or $\icpwt$. By \cref{obs:icpwt_properties},
    \begin{align*}
        \calI_{\regP}\ket*{\psi_{\alg,q}^{(\wtO)}} &= \ket*{\psi_{\alg,q}^{(\icpwt)}},\\
        \probCross &= \norm{\prCross \ket*{\psi_{\alg,q}^{(\icpwt)}}}^2.
    \end{align*}
    Furthermore, $\ket*{\psi_{\alg,q}^{(\icpwt)}}$ is supported on databases with $\abs{L}\leq 2q$ and $\abs{C}\leq q$. By \cref{eq:degree-records}, the corresponding permutation state has label level at most $2q$ and edge level at most $q$ in each edge class. Let $\probWTO^{\leq}$ and $\probWTO^{>}$ be the probability contributions to $\probWTO$ from outputs of length at most $2q+1$ or greater, as in \cref{alg:standard_game}. 
    
    Let us consider the case when $m\leq 2q+1$. First, we recover the structural vertices from the output labels $z_1,\dots, z_m$. Starting with $\ket*{\psi_{\alg,q}^{(\wtO)}}$, for all $i\in [m]$ copy the value $v_i = \labelPerm^{-1}(z_i)$ into some register, retaining only those in $V_n\setminus\{0\}$. Let $\ket*{\Psi^{(\wtO)}}$ be the resulting unnormalized state of this process. We leave $\labelPerm$ uncompressed and only apply $\calI_{\C}\coloneq\bigotimes_{b\in\middleEdge}\calI_{\abs{E_b}}$ to the middle edge permutations.

    Next, we check whether $\pairPath$ is a valid path. Check whether $v_m=\groot(T_2)$, rejecting if not. Otherwise, let $\omega = (\pairPath, v_1,\dots, v_m)$ be some output list, the proposed edges $e_i = \edgeVertex{\alpha_i}{v_{i-1}}, h_i = \edgeVertex{\alpha_i}{v_i}$ and $b_i=\edgeIndic(e_i)$. Reject if either $e_i$ or $h_i$ are not valid edges, $b_i \neq \edgeIndic(h_i)$, $\col(v_i)=\col(v_{i-1})$ or if $b_i = \bot$ and $e_i\neq h_i$. This covers all unpermuted edges. When $b_i\neq \bot$, we still need to check the edge permutation. For each edge class $b$, let
    \begin{align*}
        f_{\omega, b} \coloneq \{ (e_i,h_i):b_i=b,\ v_{i-1}=\tail(e_i)\} \cup \{ (h_i,e_i):b_i=b,\ v_{i-1}=\head(e_i)\}.
    \end{align*}
    We have that the edges are present exactly when $\sigma_b\succeq f_{\omega,b}$ for all $b$. Let $f_{\omega} = (f_{\omega, b})_b$. Using the assignment projectors from \cref{lem:pin_witness}, let
    \begin{align*}
        \Pi_{\A\regP}^{(\wtO)} &=\sum_\omega\ketbra\omega\otimes\Pi_{\succeq f_\omega,\regP},\\
        \Pi_{\A\C}^{(\icpwt)} &=\sum_\omega\ketbra\omega\otimes \Pi_{\succeq f_\omega,\C},
    \end{align*}
    where we are only summing over $\omega$ that satisfy the previous checks.
    These projectors check whether the correct assignments are in $\sigma$ and $C$, respectively. As the test in \cref{itm:checking_full_path} of \cref{alg:standard_game} is exactly this verification,
    \begin{align*}
        \probWTO^{\leq} = \norm{ \Pi_{\A\regP}^{(\wtO)}  \ket*{\Psi^{(\wtO)}}}^2.
    \end{align*}
    Obtaining $v_i = \labelPerm^{-1}(z_i)$ and checking if $v_i \in V_n \setminus \{0\}$ leaves the edge level at most $q$. The edge-assignment checks in $\Pi_{\A\regP}^{(\wtO)}$ raise the edge level by at most $m\leq 2q+1$, as in \cref{lem:pin_witness}, giving a level of at most $q+m\leq 3q+1$. Therefore, by \cref{obs:min_num_nodes_per_perm} and our assumption, $\abs{E_b} \geq 4(3q+5)$, meaning that the size condition in \cref{lem:pin_witness} holds. Therefore, the norm of the error is $O(\sqrt{q/\permSize})$. Therefore,
    \begin{align*}
        \sqrt{\probWTO^{\leq}} &\leq \norm{ \Pi_{\A\C}^{(\icpwt)} \calI_{\C} \ket*{\Psi^{(\wtO)}} } + O\left( \sqrt{\frac{q}{\permSize}} \right).
    \end{align*}
    Notice that any path from the entrance to exit must cross from $\col(n-\distWeld-1)$ to $\col(n+\distWeld+1)$. Therefore,
    \begin{align*}
        \norm{ \Pi_{\A\C}^{(\icpwt)} \calI_{\C}\ket*{\Psi^{(\wtO)}} }
        &\leq \norm{\prCross\calI_{\C}\ket*{\psi_{\alg,q}^{(\wtO)}}}\\
        &=\norm{\prCross \ket*{\psi_{\alg,q}^{(\icpwt)}} } = \sqrt{\probCross}.
    \end{align*}
    On the other hand, in the long output case when $m> 2q+1$, the verifier only tests whether the first $2q+1$ labels point to valid vertices. Define the projectors
    \begin{align*}
        \Pi_{\A\regP}^{(\wtO,>)}& \coloneq\prod_{i=1}^{2q+1}\Id[\labelPerm^{-1}(z_i)\in V_n \setminus \{0\}],  &\Pi_{\A\Lab}^{(\icpwt,>)}&\coloneq\prod_{i=1}^{2q+1}\sum_{v\in V_n \setminus \{0\}}{\Pi_{\succeq\{(v,z_i)\},\Lab}}.
    \end{align*}
    These operators are controlled by the output list. Let $\ket*{\psi^{>}}$ be the component of $\ket*{\psi_{\alg,q}^{(\wtO)}}$ with $m > 2q+1$. By \cref{obs:icpwt_properties}, all databases $L$ in the support of $\calI_{\regP}\ket*{\psi^{>}}$ have $\abs{L}\leq 2q$. As acceptance requires a label database with at least $2q+1$ records, the recorded test annihilates the final state. Applying \cref{lem:pin_witness} to these $2q+1$ inverse-label tests with $M\geq 4(4q+5)$ gives,
    \begin{align*}
        \sqrt{\probWTO^{>}}=\norm{\Pi_{\A\regP}^{(\wtO,>)} \ket*{\psi^{>}} }
        &\leq\norm{\Pi_{\A\Lab}^{(\icpwt,>)}\calI_{\regP} \ket*{\psi^{>}} }+C\sqrt{\frac{(2q+1)\abs{V_n \setminus \{0\}}}{M}}\norm{ \ket*{\psi^{>}}}
        =O\left(\sqrt{\frac{q}{N}}\right).
    \end{align*}
    Hence $\probWTO^{>}=O({q/N})$, which combined with the bound on $\probWTO^{\leq}$ implies \cref{eq:path_verification_bound}, since $\permSize\leq N$.
\end{proof}

\section{Hardness of pathfinding}

The goal of this section is proving the following theorem.

\begin{theorem}\label{thm:pathfinding_hard}
    Assume that $n\geq16$ is even and $\distWeld\in[2,n-4]$.
    Any quantum algorithm $\alg$ making $q$ queries to the welded trees oracle $\wtO_\sigma$ with $\sigma\sim\singDistr$ satisfies
    \begin{align}
         \Pr[\text{$\alg$ outputs a path from entrance to exit}] = O\left(\frac{q}{\permSize} + q^2 \cdot \errorTerm +\frac{n\cdot q^3}{N}\right),\label{eq:exact-pathfinding-bound}
    \end{align}
    where the probability is over $\sigma$ and the measurement outcomes of $\alg$.
    For a suitable choice of $\distWeld$, constant success probability
    requires $\Omega((N/(n+1))^{1/3})$ queries.
\end{theorem}

In particular, the lower bound for $\sigma \sim \singDistr$ implies a lower bound in the worst case.

Throughout the rest of the section we assume that for some constant $C_0$,
\begin{align}
    \permSize \geq C_0(q+1),\quad C_0 \left( (q+1)\errorTerm + \frac{(n+1)(q+1)^3}{N} \right)\leq \frac{1}{2}. \label{eq:standing-assumptions}
\end{align}

\subsection{Good space}\label{sec:good_space}

We now introduce the good space. Intuitively, it corresponds to the space where assignments to vertices that are not fixed behave similarly to a random walk. We define the following sets of vertices:
\begin{align*}
    \middleNodes &\coloneq \{v\in V_n : n-\distWeld < \col(v) < n+\distWeld\},\\
    \Vbd &\coloneq \col(n-\distWeld - 1) \cup \col(n-\distWeld) \cup \col(n+\distWeld) \cup \col(n+\distWeld + 1),\\
    \Vweld &\coloneq\col(n)\cup\col(n+1).
\end{align*}
For some $u\in \middleNodes \cup \Vbd$, $e= \edgeVertex{\alpha}{u}$ and $b=\edgeIndic(e)$, we let $\probAssignEdge{\alpha}{u,v}$ be the probability that an edge is randomly assigned to go between $u$ and $v$ when it is permuted,
\begin{align}
    \probAssignEdge{\alpha}{u,v} &\coloneq \begin{cases}
        \tfrac{1}{\abs{E_b}} &\text{if ${b\in\middleEdge,\ }u=\tail(e)$ and $v\in \{\head(g): g\in E_b\}$,}\\
        \tfrac{1}{\abs{E_b}} &\text{if ${b\in\middleEdge,\ }u=\head(e)$ and $v\in \{\tail(g): g\in E_b\}$,}\\
        0 &\text{otherwise.}
    \end{cases}\label{eq:kernel}
\end{align}
The endpoints are on the opposite side of the same class, so $\probAssignEdge{\beta}{u,v}=\probAssignEdge{\beta}{v,u}$ and $\sum_u \probAssignEdge{\beta}{u,v} \leq 1$, with equality when $\edgeIndic(\edgeVertex{\beta}{v})\in \middleEdge$.

An \emph{abstract forest} $H=(U,E(H),\forestcol,\forestlab)$ consists of a vertex set $U$, an edge set $E(H)$ that forms a forest, a proper edge-coloring $\forestcol: E(H) \to \colorSet$ and an injective \emph{labeling} $\forestlab: U \rightharpoonup [M]$. Without loss of generality, assume that each connected component in $H$ has at least one label or edge. Intuitively, this represents what the algorithm \alg \say{knows} as $H$ does not assign the labels to structural vertices, so we call $U$ and $E(H)$ the abstract vertices and edges, respectively. This is handled using the \emph{assignment} $\colorVertMap:U\to[M]$ which places each abstract vertex $u$ to some structural vertex. 

An abstract edge $e\in E(H)$ is a \emph{weld} when both of its endpoints under $\colorVertMap$ are in $\Vweld$. For some $u\in U$, we let the \emph{tree component} $\Ctree_u$ be the connected component containing $u$ of $H$ if we remove all the weld edges. Hence $\Ctree_u$ is entirely in one of $T_1$ or $T_2$, except when it consists of a single isolated vertex assigned outside $V_n$, and one tree in $H$ may contain multiple tree components joined by weld edges. This depends entirely on $\colorVertMap$. Let the \emph{fixed set} $\Sfix_u$ consist of $\Ctree_u$ and the vertices immediately across the welds incident to $\Ctree_u$. The \emph{fixed data} of $u$ is defined as
\begin{align*}
    \Wfix_u(\colorVertMap)\coloneq \left( \Ctree_u, \colorVertMap|_{\Ctree_u},(v,\colorVertMap(v))_{v\in\Sfix_u\setminus\Ctree_u} \right).
\end{align*}
While $\Wfix_u$ is a function of $\colorVertMap$ as it specifies the entire tree component and adjacent points, we will omit $\colorVertMap$ when it is implicit. The role of the fixed data is as a \say{starter} from which the rest of the database develops randomly. For example, if under $\colorVertMap$ we have $C_1,\rho_1,C_2,\rho_2,C_3$ where $C_i$ are tree components joined by the weld edges $\rho_1,\rho_2$ and $u\in C_2$, $\Wfix_u$ fixes $\rho_1, C_2, \rho_2$. On the other hand, apart from their fixed endpoints adjacent to $C_2$, assignments in $C_1$ and $C_3$ are averaged according to the reference distribution we define below.

Next we define the weight distribution on all possible placements of the forest. Let $H_1,\dots, H_k$ be connected components of $H$, with vertex sets $U_1,\dots, U_k$. Intuitively, for each recorded component $H_i$ in the forest $H$, choose some root node, assign its location uniformly in $\middleNodes \cup \Vbd$ and assign their children independently using \cref{eq:kernel}. Formally, the \emph{reference distribution} $\muref_i(\colorVertMap)$ is defined as
\begin{align}
    \muref_i(\colorVertMap)&\coloneq\frac{1}{\Zref_i \abs{\middleNodes\cup\Vbd}} \prod_{uv\in E(H_i)}\probAssignEdge{\forestcol(uv)}{\colorVertMap(u),\colorVertMap(v)},\label{eq:reference-prob}
\end{align}
where
\begin{align}
    \Zref_i&\coloneq \frac{1}{\abs{\middleNodes\cup\Vbd }} \sum_{\colorVertMap:U_i\to\middleNodes\cup\Vbd} \prod_{uv\in E(H_i)}\probAssignEdge{\forestcol(uv)}{\colorVertMap(u),\colorVertMap(v)}\notag
\end{align}
is the probability that every requested edge is permuted. When $H_i$ consists of a single vertex, use the uniform distribution on $[M]$. Let $\muref$ be the product of these distributions over the trees in $H$. Next, we define restrictions on the assignments $\colorVertMap$.

\paragraph{Good space.} Next we define good assignments $\colorVertMap$ and the good space associated with them. Fix some set of \emph{boundary} vertices $B$ such that $\Vbd \subseteq B \subseteq (\middleNodes \cup \Vbd) \setminus \Vweld$. In each connected component $H_i$, only some chosen tree component may be assigned to vertices in $B$, meaning that it is the only way the connected component escapes.

For each connected component $H_i$ in $H$, choose a vertex $u_i$ and fix the data $w_i = \Wfix_{u_i}(\colorVertMap)$. Their collection over $H$ with $k$ connected components forms a \emph{description} $D$,
\begin{align*}
    D &\coloneq (B, H, (u_i, w_i)_{i\in [k]}).
\end{align*}
The assignments to the remaining vertices are free. We say an assignment is \emph{good} for $D$, denoted by $\goodMap_D(\colorVertMap)$, if it satisfies the following conditions:
\begin{align}
    \goodMap_D(\colorVertMap)\coloneq\Id \left[
        \begin{array}{ll}
        \muref(\colorVertMap)>0, &\\
  \Wfix_{u_i}(\colorVertMap)=w_i &\text{for every }i\\
  \colorVertMap(U_i\setminus\Ctree_{u_i})\cap B=\emptyset
      &\text{for every }i\\
  \colorVertMap(u)\ne\colorVertMap(v) &\text{for all distinct }u,v\in U
 \end{array}
    \right]. \label{eq:goodness_definition}
\end{align}
Each good assignment can be mapped to a unique database $I_{\colorVertMap}=(L_{\colorVertMap},C_{\colorVertMap})$. Specifically, every labeled vertex $u$ has $L_{\colorVertMap}(\colorVertMap(u))=\forestlab(u)$, while every abstract edge $uv$ of color $\alpha$ contributes to $C_{\colorVertMap}$ with values in $\edgeVertex{\alpha}{\colorVertMap(u)}$ and $\edgeVertex{\alpha}{\colorVertMap(v)}$ going from tail to head. Furthermore, the empty database $\ket{\bot}_{\I}$ is good and any database $I$ such that $\abs{I}\leq t$ and does not contain a cycle or a parallel edge can be represented by an abstract forest with $\abs{U}\leq 2t$.

The \emph{canonical good vector} for $D$ is a sum over its good assignments weighted using the reference probability,
\begin{align*}
    \canVec{D} &\coloneq \sum_{\colorVertMap:\goodMap_D(\colorVertMap)=1}  \sqrt{\muref(\colorVertMap)}\ket{L_{\colorVertMap},C_{\colorVertMap}}_{\Lab\C}.
\end{align*}
Here, the fixed data remains fixed, while the rest of the assignments vary. For databases $I$ such that $\abs{I}\leq t$, the \emph{good space} is the span of $\canVec{D}$ with some specified boundary $B$ and $\abs{E(H)} + \abs{\dom(\forestlab)}\leq t$. The projector on this space $\Pgood_t(B)$ is the orthogonal projector onto this span and $\Pgood_t\coloneq \Pgood_t(\Vbd)$. Note that different descriptions may have overlapping good vectors, so the canonical good vectors are not necessarily orthogonal. Notice that good assignments cannot cross from $\col(n-\distWeld-1)$ to $\col(n+\distWeld+1)$, implying that,
\begin{align}
    \prCross \Pgood_t = 0.\label{eq:goodness_implies_no_crossing}
\end{align}
Next, we describe the linear combinations of the canonical good vectors using functions of the assignments. First, partition assignments by the set $\Fbd_i=\{u\in U_i:\colorVertMap(u)\in B\}$ placed in the boundary $B$. Let $\muref_{i,\Fbd_i}$ be $\muref_i$ conditioned on $\Fbd_i$. Fix some recorded component $H_i$ and $F=\Fbd_i$. A \emph{coefficient function} $f$ assigns a complex coefficient $f(\colorVertMap)$ to each assignment $\colorVertMap$ of $H_i$. We measure $f$ using the squared norm,
\begin{align}
    \norm{f}^2 = \EX_{\muref_{i,F}}\norm{f(\colorVertMap)}^2.\label{eq:norm_function_ref}
\end{align}
These functions describe how the canonical good vectors are combined. We define $\Scoef_{i,F}$ to be the space of allowed coefficient functions,
\begin{align}
    \Scoef_{i,\emptyset} &= \left\{ \sum_{u\in U_i}h_u(\Wfix_u):h_u\text{ arbitrary complex-valued function} \right\}\label{eq:scoef_definition} \\
    \Scoef_{i,F} &= \left\{ \Id[F\subseteq\Ctree_c]h(\Wfix_c) :h\text{ arbitrary complex-valued function} \right\} \quad (c\in F).\notag
\end{align}
The key point is that each $h$ depends only on its fixed data. The indicator in the second expression requires all of $F$ to lie in one tree component, so it does not depend on which $c\in F$ is used. For the entire forest $H$, let $\Scoef=\bigotimes_i\Scoef_{i,\Fbd_i}$. This allows arbitrary coefficients in the fixed data of recorded components and algorithm registers.

For some fixed sets $\Fbd_i$, let $\mu = \bigotimes_i\muref_{i,\Fbd_i}$. In order to remove assignments that do not keep the database injective, we use the following projector $\Pdist$
\begin{align*}
    (\Pdist f)(\colorVertMap)&\coloneq\Id[\colorVertMap\text{ is injective on }U]f(\colorVertMap).
\end{align*}
The corresponding vector $z$ related to this assignment would be written as,
\begin{align*}
    z = \sum_{\colorVertMap\text{ injective}}\sqrt{\mu(\colorVertMap)}f(\colorVertMap)\ket{\colorVertMap}
\end{align*}
Writing $\norm{f}_\mu^2\coloneq\sum_\colorVertMap\mu(\colorVertMap)\norm{f(\colorVertMap)}^2$ for the coefficient norm, we have $\norm{z}^2 =\norm{\Pdist f}_\mu^2\leq\norm{f}_\mu^2$. The span of these vectors over all forests with at most $t$ records and all choices of vertices assigned to some fixed boundary set $B$ represents the good space. Equivalently, the good space $\calG_B$ defined above may be written as
\begin{align*}
    \calG_B(t) = \Span\left\{ \sum_{\substack{\colorVertMap \text{ injective}\\ \forall i,\{u\in U_i: \colorVertMap(u)\in B\} = F_i}} \sqrt{\mu(\colorVertMap)} f(\colorVertMap) \ket{I_\colorVertMap}: \abs{E(H)} + \abs{\dom(\forestlab)} \leq t, f\in \Scoef\right\}.
\end{align*}
The remainder of the section is concerned with useful statements about the good space. We may notice that the same good state can be described using vertex assignments $\colorVertMap$ or a database without changing its norm.

\begin{observation}\label{obs:goodness_properties}
    Let $\Aut(H)$ be the color and label preserving automorphism group of $H$. The subspace of good assignment vectors fixed by every element of $\Aut(H)$ is isometric to the corresponding good database space.
\end{observation}
\begin{proof}
    Injective assignments give the same database exactly when they differ by an automorphism of $H$. Thus the map
    \begin{align*}
        \frac{1}{\sqrt{|\Aut(H)|}} \sum_{\gamma\in\Aut(H)}\ket{\gamma\colorVertMap} \longmapsto\ket{I_{\colorVertMap}}
    \end{align*}
    is an isometry on invariant vectors. Renaming preserves the reference weights and permutes the good generators, so this map identifies the good spaces. Distinct isomorphism classes of $H$ have disjoint database supports.
\end{proof}

Next, we bound the probability that an averaged branch reaches the boundary or repeats a vertex, meaning that the assignment stops being good. Let $m_i = \abs{U_i}$ and $m=\sum m_i$. We may assume that $m=O(q)$. We use $X_u$ to be the random variable representing the assignment $\colorVertMap(u)$. For $A\subseteq U$, write $X_A = (X_u)_{u\in A}$. Additionally, for the rest of the section, we assume that
\begin{align}
    \frac{4m \abs{B}}{N} \leq \frac{1}{2}.\label{eq:bound_on_size_B}
\end{align}
\begin{lemma}\label{lem:conditional_structural_assignments}
    Fix some $\Wfix_u$ and suppose that all boundary vertices belong to $\Ctree_u$. Then every free vertex $a\in U$ and structural vertex $x$ satisfy
    \begin{align*}
        \Pr[X_a = x| \Wfix_u] = O\left( \frac{1}{N}\right).
    \end{align*}
    For distinct $a,b\in U$ with at least one outside of $\Sfix_u$,
    \begin{align*}
        \Pr[X_a = X_b| \Wfix_u] = O\left( \frac{n}{N}\right).
    \end{align*}
    If one of the two vertices is fixed or they lie in different connected components of $H \setminus \Ctree_u$, the bound is $O(1/N)$.
\end{lemma}
\begin{proof}
    Fix $\Wfix_u$ throughout and condition all probabilities below on it. Until the final paragraph, ignore the requirement that free vertices avoid $B$ and count only vertices that have been assigned. By \cref{eq:reference-prob}, the connected components of $H_i\setminus\Ctree_u$ are independent. Root each connected component at its fixed weld endpoint and assign each child independently using \cref{eq:kernel}, stopping the assignment beyond an edge if it is not permuted. Other connected components start from uniform roots, as in the reference distribution.

    By \cref{lem:colors_per_column,eq:welded_edges_def}, the first step from each fixed root is uniform on at least $N/4$ vertices. This bound is preserved by every later step. Specifically, if $w$ is a child of $v$ along an edge of color $\beta$, then
    \begin{align*}
        \Pr[X_w=y] = \sum_x \Pr[X_v=x]\probAssignEdge{\beta}{x,y} \leq \frac{4}{N}\sum_x\probAssignEdge{\beta}{x,y} \leq \frac{4}{N}.
    \end{align*}
    Hence every free vertex has point probabilities at most $4/N$.

    Next, suppose that $a,b$ lie in the same connected component of $H\setminus\Ctree_u$ and let $d$ be their nearest common ancestor. Without loss of generality assume $a\neq d$. Given $X_d=x$, let $c$ be the class of the first edge towards $a$. If this edge is not permuted, $a$ is not assigned. Otherwise, the point-bound argument gives
    \begin{align*}
        \max_y\Pr[X_a=y\mid X_d=x]\leq\frac{1}{\abs{E_c}}.
    \end{align*}
    The two paths are independent given $X_d$, with $X_b$ fixed when $b=d$. Hence
    \begin{align*}
        \Pr[X_a=X_b\mid X_d=x]
        &=\sum_y\Pr[X_a=y\mid X_d=x]\Pr[X_b=y\mid X_d=x]\\
        &\leq\max_y\Pr[X_a=y\mid X_d=x]
        \leq\frac{1}{\abs{E_c}}.
    \end{align*}
    If $d$ is free, its point probabilities are at most $4/N$ by the argument above. For the color of the first edge, each class $c$ has at most $2\abs{E_c}$ possible locations for $X_d$. Grouping these locations by class gives
    \begin{align*}
        \Pr[X_a=X_b]
        &=\sum_x\Pr[X_d=x]\Pr[X_a=X_b\mid X_d=x]\\
        &\leq\frac{4}{N}\sum_{c\in\middleEdge}\frac{2\abs{E_c}}{\abs{E_c}}
        =O\left(\frac{n}{N}\right).
    \end{align*}
    If $d$ is the fixed root, then $\abs{E_c}\geq N/4$, giving $O(1/N)$. The point bound also gives $O(1/N)$ when one vertex is fixed, or when the two vertices lie in different connected components of $H\setminus\Ctree_u$.

    Finally, let $\mathcal E$ be the event that all free vertices are assigned and avoid $B$. Assignment can fail only after a visit to $\Vbd\subseteq B$, so the union bound combined with \cref{eq:bound_on_size_B} gives
    \begin{align*}
        \Pr[\mathcal E]\geq1-\frac{4m\abs{B}}{N}\geq\frac12.
    \end{align*}
    The required distribution is obtained by conditioning on $\mathcal E$. Therefore,
    \begin{align*}
        \Pr[X_a=x\mid\mathcal E]
        &\leq2\Pr[X_a=x]\leq\frac8N,\\
        \Pr[X_a=X_b\mid\mathcal E]
        &\leq2\Pr[X_a=X_b]=O\left(\frac nN\right),
    \end{align*}
    and the sharper $O(1/N)$ bounds are preserved as well.
\end{proof}

The bounds in \cref{lem:conditional_structural_assignments} also apply to the coefficients depending on the fixed data. For some coefficient function $f$ that depends only on $\Wfix_u$ and an event $\calE$ such that $\Pr[\calE| \Wfix_u] \leq \epsilon$ for every $\Wfix_u$ with nonzero probability,
\begin{align*}
    \EX[\Id[\calE] \norm{f}^2] \leq \epsilon \norm{f}^2.
\end{align*}
Next we show how the coefficient functions behave across a single recorded component. In the next statement, given a vertex $v$ in a tree, we will use $\parentop(v)$ to denote the parent of $v$ and $\Udesc_v$ to denote the descendants of $v$, inclusive.

\begin{lemma}\label{lem:coeff_decomp}
    Fix some $H_i$ whose root is the vertex $a$ and suppose that $\Fbd_i = \emptyset$. Then every $f\in \Scoef_{i,\emptyset}$ has a unique decomposition $f=\sum_{v\in U_i}f_v$ where each $f_v$ depends only on $\Wfix_v$. Moreover, for every $v\neq a$, $f_v = 0$ unless the edge $(\parentop(v),v)$ crosses the weld and
    \begin{align}
        \EX \left[ f_v \mid X_{\parentop(v)}, X_v \right] &=0,\notag\\
        \EX  \left[ f_v \mid X_{U_i\setminus \Udesc_v},X_v \right] &= 0.\label{cg:conditional-zero}
    \end{align}
    The terms are pairwise orthogonal and remain so after multiplication by any function of $\Wfix_a$. The statement also holds for vector-valued coefficients.
\end{lemma}
\begin{proof}
    All expectations are with respect to $\muref_{i,\emptyset}$. By \cref{eq:reference-prob}, fixing the endpoints of a weld edge makes the assignments on its two sides independent. Conditioning on $\Fbd_i=\emptyset$ preserves this independence, since it requires the vertices on each side to avoid $B$ separately.

    By the definition of $\Scoef_{i,\emptyset}$ in \cref{eq:scoef_definition}, it suffices to decompose a function $h(\Wfix_u)$. Split this function according to the vertex of $\Ctree_u$ closest to $a$ as $h(\Wfix_u)=\sum_{v\in U_i}g_v$ with
    \begin{align*}
        g_v &\coloneq \Id[v\text{ is the vertex of }\Ctree_u\text{ closest to }a] \, h(\Wfix_u).
    \end{align*}
    Fix $v$ and write $g=g_v$. This function depends only on $\Wfix_v$, since $\Wfix_u=\Wfix_v$ on this part. If $v=a$, add $g$ to $f_a$. Otherwise, $g$ is zero unless $(\parentop(v),v)$ is a weld edge. Let
    \begin{align*}
        \bar g \coloneq \Id[(\parentop(v),v)\text{ is a weld edge}]\,
        \EX[g\mid X_{\parentop(v)},X_v,(\parentop(v),v)\text{ is a weld edge}].
    \end{align*}
    Add $g-\bar g$ to $f_v$. This difference depends only on $\Wfix_v$ and has the required conditional mean zero. The remaining function $\bar g$ depends only on $\Wfix_{\parentop(v)}$, which determines whether the edge is a weld edge and, when it is, both endpoints. Repeat the splitting and subtraction with $\bar g$. Each new index lies strictly closer to $a$, so the procedure terminates at $a$. Adding the resulting terms gives the decomposition and the first conditional expectation identity.

    For the second identity, suppose that $(\parentop(v),v)$ is a weld edge. Then $\Ctree_v\subseteq\Udesc_v$, and the only assignment outside $\Udesc_v$ included in $\Wfix_v$ is $X_{\parentop(v)}$. By the independence above,
    \begin{align*}
        \EX[f_v\mid X_{U_i\setminus\Udesc_v},X_v]
        =\EX[f_v\mid X_{\parentop(v)},X_v,(\parentop(v),v)\text{ is a weld edge}]=0.
    \end{align*}
    This quantity also vanishes when the edge is not a weld edge, since $f_v=0$ there.

    Finally, take distinct $v,w$, with $v$ at least as far from $a$ as $w$. When $(\parentop(v),v)$ is a weld edge, $w$ lies outside $\Udesc_v$, so $f_w$ and $\Wfix_a$ depend only on $X_{U_i\setminus\Udesc_v}$ and $X_v$. Since $f_v$ is zero otherwise, every function $\psi(\Wfix_a)$ satisfies
    \begin{align*}
        \EX[\abs{\psi(\Wfix_a)}^2\langle f_v,f_w\rangle]
        &=\EX\left[\abs{\psi(\Wfix_a)}^2
        \left\langle\EX[f_v\mid X_{U_i\setminus\Udesc_v},X_v],f_w\right\rangle\right]
        =0.
    \end{align*}
    Taking $\psi=1$ gives orthogonality, and arbitrary $\psi$ gives the claimed orthogonality after multiplication by any function of $\Wfix_a$. Applying orthogonality to the difference of two such decompositions proves uniqueness. The same argument applies to vector-valued coefficients, using their inner product in the last calculation.
\end{proof}

In order to analyze a query at an abstract vertex $a$, we need to know whether its location is in a term's fixed data $\Wfix_u$. If it is not, \cref{lem:conditional_structural_assignments} bounds its distribution. If it is, the orthogonal decomposition in \cref{lem:coeff_decomp} lets us analyze the terms separately and combine their estimate without introducing a factor for the number of terms.

For some distinct abstract vertices $a,b$, let $(\Pcoin_{ab}f)(\colorVertMap)\coloneq\Id[\colorVertMap(a)=\colorVertMap(b)] f(\colorVertMap)$ be the \emph{coincidence} projector on coefficient functions. If some $\colorVertMap$ is not injective, it must have a coincidence, meaning that
\begin{align}
    \Id-\Pdist\preceq\sum_{\{a,b\}}\Pcoin_{ab}.\label{eq:coincidence_union}
\end{align}
Notice that since $f$ is an arbitrary coefficient function, it may imply collisions $\colorVertMap(a)=\colorVertMap(b)$, meaning that it may be annihilated by $\Pdist$. To distinguish from this possibility, we will be concerned with the orthogonal complement of $\Kdist = \{h\in\Scoef:\Pdist h=0\}$.

\begin{lemma}\label{lem:good_vectors_representation}
    Let $\abs{U} = m$ and $f\in\Scoef\cap(\Kdist)^\perp$. Then,
    \begin{align*}
        \norm{(\Id - \Pdist)f}^2 = O\left( \frac{(n+1)m^3}{N} \norm{f}^2 \right) \leq \frac{1}{2} \norm{f}^2
    \end{align*}
    for sufficiently large $n$.
    Furthermore, every coefficient function $h\in \Pdist \Scoef$ has a unique $f\in \Scoef \cap (\Kdist)^\perp$ such that $\Pdist f = h$ and $\norm{f}^2 \leq 2\norm{h}^2$.
\end{lemma}
\begin{proof}
    First we prove the following bounds for distinct abstract vertices $a,b$,
    \begin{align}
        \norm{\Pcoin_{ab}f}^2&=O\left(\frac{(n+1)m_i}{N}\right)\norm{f}^2 &&(a,b\in U_i),\label{eq:coin_goal_a}\\
        \norm{\Pcoin_{ab}f}^2&=O\left(\frac{1}{N}\right)\norm{f}^2 &&(a\in U_i,\ b\in U_j,\ i\ne j) \label{eq:coin_goal_b}.
    \end{align}
    Afterwards, we sum these bounds over all pairs using \cref{eq:coincidence_union}. First consider the case when $a,b\in U_i$. If $\Fbd_i=\emptyset$, apply \cref{lem:coeff_decomp} with root $a$, keeping the assignments in the other connected components in the coefficients. This gives $f=\sum_v f_v$ with at most $m_i$ orthogonal terms, each depending on the assignments in $H_i$ only through $\Wfix_v$. If $\Fbd_i\ne\emptyset$, use the single term $f_c=f$ for some $c\in\Fbd_i$, as in \cref{eq:scoef_definition}.

    When both $a,b\in\Sfix_v$, their coincidence is determined by $\Wfix_v$. Therefore, $\Pcoin_{ab}\Id[a,b\in\Sfix_v]f_v$ still belongs to $\Scoef$ and is annihilated by $\Pdist$. It lies in $\Kdist$, so it is orthogonal to $f$. In the remaining part, at least one vertex is free, and \cref{lem:conditional_structural_assignments} bounds the conditional probability of a coincidence by $O((n+1)/N)$. Thus, by Cauchy--Schwarz,
    \begin{align*}
        \norm{\Pcoin_{ab}f}^2
        &=\left\langle\Pcoin_{ab}f,\sum_v\Pcoin_{ab}\Id[\{a,b\}\not\subseteq\Sfix_v]f_v\right\rangle\\
        &\leq\norm{\Pcoin_{ab}f}\left(m_i\sum_v\norm{\Pcoin_{ab}\Id[\{a,b\}\not\subseteq\Sfix_v]f_v}^2\right)^{1/2}\\
        &\leq O\left(\sqrt{\frac{(n+1)m_i}{N}}\right)\norm{\Pcoin_{ab}f}\norm{f},
    \end{align*}
    where we used $\sum_v\norm{f_v}^2=\norm{f}^2$. This implies \cref{eq:coin_goal_a}.

    Next, suppose that $a\in U_i$ and $b\in U_j$ with $i\ne j$. Decompose in both connected components, using roots $a$ and $b$. By \cref{lem:coeff_decomp} and independence of the connected components, the resulting terms $g$ are orthogonal and remain so after multiplication by any function of $X_a,X_b$. In particular, their images under $\Pcoin_{ab}$ are orthogonal.
    
    If both tested vertices are fixed in a term $g$, then $\Pcoin_{ab}g\in\Kdist$, so $\norm{\Pcoin_{ab}g}^2=0$. Otherwise, condition on the fixed data in both components. Independence and the point probability bound in \cref{lem:conditional_structural_assignments} give $\norm{\Pcoin_{ab}g}^2=O(1/N)\norm{g}^2$. Summing over the orthogonal terms gives \cref{eq:coin_goal_b}.

    There are at most $m^2$ pairs and $m_i\leq m$, so \cref{eq:coincidence_union} gives
    \begin{align*}
        \norm{(\Id-\Pdist)f}^2
        \leq\sum_{\{a,b\}}\norm{\Pcoin_{ab}f}^2
        =O\left(\frac{(n+1)m^3}{N}\right)\norm{f}^2.
    \end{align*}
    Since $m=O(q)$, \cref{eq:standing-assumptions} makes this at most $\tfrac12\norm{f}^2$ for sufficiently large $C_0$.

    Finally, for $h\in\Pdist\Scoef$, choose $f_0\in\Scoef$ with $\Pdist f_0=h$ and let $f$ be $f_0$ minus its orthogonal projection onto $\Kdist$. Then $f\in\Scoef\cap(\Kdist)^\perp$ and $\Pdist f=h$. The bound above implies
    \begin{align*}
        \norm{h}^2=\norm{\Pdist f}^2=\norm{f}^2-\norm{(\Id-\Pdist)f}^2\geq\frac12\norm{f}^2.
    \end{align*}
    Two such coefficient functions differ by an element of $\Kdist$ that is also orthogonal to $\Kdist$, so they are equal.
\end{proof}

\subsection{Compressed queries preserve goodness}\label{sec:comp_goodness}

The goal of this subsection is showing that \cpwt does not alter goodness. Let $\etastep_t$ be,
\begin{align}
    (\etastep_t)^2 = \errorTerm + \frac{n\cdot (t+1)}{N}.\label{eq:query_range}
\end{align}
We use $t$ as the count of \pC operators used. As \cpwt uses $6$ \pC operators, we have that $t=O(q)$. The key step of the goodness result is a bound on the effect of \pCPerm.

\begin{lemma}\label{lem:edge_compression_bound}
    Let $t \leq 6q+6$ and suppose \cref{eq:standing-assumptions} holds. Then
    \begin{align}
        \norm{(\Id - \Pgood_{t+1}) \pCPerm \Pgood_t} = O(\etastep_t).\label{cg:query-edge-bound}
    \end{align}
\end{lemma}
\begin{proof}
    Assume that $t\geq 1$ as when $t=0$, the database $I=\ket{\bot}_{\Lab\C}$ and hence $\pCPerm$ acts as the identity. Fix some query input $\ket{l,\alpha}_{\X}$ with $l\in [M]$ and $\alpha\in\colorSet$. If $l\notin \im(L)$  or the requested edge is not permuted, \pCPerm acts again as the identity. Hence we may assume this is not the case.
    
    For a database $I=(L,C)$ with $l\in\im(L)$, let $x_I = L^{-1}(l)$, $b(x) = \edgeIndic(\edgeVertex{\alpha}{x})$ and $s(x) = \abs{E_{b(x)}}$. For some abstract forest $H$, let $a=\forestlab^{-1}(l) \in U$. Notice that $x_{I_{\colorVertMap}}=\colorVertMap(a)$.

    Fix $r\in [0,t]$. For $I$ such that $\abs{I}\leq r$ and the queried edge is absent, define its extension as,
    \begin{align*}
        \calA(I) &\coloneq \left\{(L,C^\prime):\ C^\prime\succeq C,\ \abs{C^\prime}=\abs{C}+1,\ncF_{C^\prime}(x_I,\alpha)\ne\bot\right\}.
    \end{align*}
    The database spaces based on this assignment are
    \begin{align*}
        \calH_0&\coloneq\Span\left\{\ket{I}:\ \abs{I}=r,\ l\in\im(L),b(x_I)\in\middleEdge,\ \ncF_C(x_I,\alpha)=\bot\right\},\\
        \calH_1&\coloneq\Span\left\{\ket{I^\prime}:\ \ket{I}\in\calH_0,\ I^\prime\in\mathcal A(I)\right\}.
    \end{align*}
    For some canonical good vector $\canVec{D}$, $H$ fixes the database size $\abs{I}$ and whether the queried label $l$ and its edge with color $\alpha$ are recorded. If $l$ is recorded, $x_I = \colorVertMap(a)$ is either fixed by $D$ or in $\middleNodes$ where every edge is permuted. Hence projecting on one of $\calH_0$ or $\calH_1$ either sends the good vector $\canVec{D}$ to itself or $0$, meaning that it preserves the good space. For $j\in \{0,1\}$, $\Pi_j$ is the orthogonal projector onto $\calH_j \cap \ran(\Pgood_{t+1})$. Define the operator $J:\calH_0 \to \calH_1$ as
    \begin{align*}
        J\ket{I}&\coloneq\frac{1}{\sqrt{s(x_I)}}\sum_{I^\prime\in\mathcal A(I)}\ket{I^\prime},\\
        J^\dagger\ket{I^\prime}&=\frac{1}{\sqrt{s(x_I)}}\ket{I},
    \end{align*}
    where $I^\prime \in \calA(I)$. For now, assume that the following bound holds,
    \begin{align}
        \norm{(\Id - \Pi_1) J \Pi_0} +\norm{(\Id-\Pi_0)J^\dagger\Pi_1} = O(\etastep_t).\label{eq:edge_add_delete_good}
    \end{align}
    Notice that by \cref{eq:standing-assumptions,obs:min_num_nodes_per_perm}, $\abs{\calA(I)} \in [s(x_I)-t, s(x_I)]$. Therefore,
    \begin{align*}
        J^\dagger J\ket{I}&=\frac{\abs{\calA(I)}}{s(x_I)}\ket{I},
    \end{align*}
    so $\tfrac{1}{2} \Id \preceq J^\dagger J \preceq \Id$. Let $\calV=J(J^\dagger J)^{-1/2}$. By \cref{eq:edge_add_delete_good},
    \begin{align*}
        \delta \coloneq \Pi_1J-J\Pi_0, \quad \norm{\delta}&=O(\etastep_t).
    \end{align*}
    Next, apply \cref{ex:polar-normalization} with $V=J$, $T=\Pi_1$, $U=\Pi_0$, $\Pi = \Pi^\prime = \Id$ and $\kappa = 1/2$,
    \begin{align*}
        \norm{\Pi_1\calV-\calV\Pi_0} &\leq\sqrt{2}\norm{\delta} +\sqrt{2}\norm{\delta^\dagger J-J^\dagger\delta}\\
        &\leq 3\sqrt{2}\norm{\delta}=O(\etastep_t),
    \end{align*}
    where we used the fact that $\norm{J}\leq 1$. Hence the insertion and deletion errors satisfy,
    \begin{align*}
        \delta_+&\coloneq\norm{(\Id-\Pi_1)\calV\Pi_0}=O(\etastep_t),&
        \delta_-&\coloneq\norm{(\Id-\Pi_0)\calV^\dagger\Pi_1}=O(\etastep_t).
    \end{align*}
    By \cref{eq:pc_x_I}, the compression $\pCPerm$ and good projector $\Pi$ on $\calH_0\oplus\calH_1$ are
    \begin{align*}
        \pCPerm&=\begin{pmatrix}0&\mathcal V^\dagger\\
            \mathcal V&\Id-\mathcal V\mathcal V^\dagger\end{pmatrix},&
        \Pi&\coloneq\begin{pmatrix}\Pi_0&0\\0&\Pi_1\end{pmatrix}.
    \end{align*}
    Notice that for $\calV \calV^\dagger$, we have,
    \begin{align*}
        \norm{(\Id-\Pi_1)\mathcal V\mathcal V^\dagger\Pi_1} &= \norm{(\Id-\Pi_1)\mathcal V\Pi_0\mathcal V^\dagger\Pi_1 + (\Id-\Pi_1)\mathcal V(\Id-\Pi_0)\mathcal V^\dagger\Pi_1}\\
        &\leq \delta_+ + \delta_-.
    \end{align*}
    Therefore, using the fact that query inputs and different records are orthogonal,
    \begin{align}
        \norm{(\Id - \Pi) \pCPerm \Pi} \leq 2(\delta_+ + \delta_-) = O(\etastep_t).\label{eq:compression_bound}
    \end{align}
    Hence the remainder of the proof is concerned with showing that \cref{eq:edge_add_delete_good} holds. We will show each statement in \cref{eq:edge_add_delete_good} by considering a vector $z\in \ran(\Pi_0)$ and $z\in \ran(\Pi_1)$ respectively. Suppose that $z$ is supported by databases represented by a fixed abstract forest $H$. By \cref{obs:goodness_properties}, we may write this in terms of the assignment coordinates. Consider the following reference weights:
    
    \begin{align*}
        \mu_H(\colorVertMap)&\coloneq \prod_{uv\in E(H)}\probAssignEdge{\forestcol(uv)}{\colorVertMap(u),\colorVertMap(v)},& \norm{f}_H^2&\coloneq\sum_{\colorVertMap}\mu_H(\colorVertMap)\norm{f(\colorVertMap)}^2.
    \end{align*}
    These definitions omit the normalization factor and coefficient function norm, but
    otherwise behave the same way as in \cref{eq:reference-prob}. For each choice of the sets $\Fbd_i$, set their coefficients using \cref{lem:good_vectors_representation} and let $f$ be their sum. The vector and its coefficient function satisfy,
    \begin{align*}
        z&=\sum_{\colorVertMap\text{ injective}}\sqrt{\mu_H(\colorVertMap)}f(\colorVertMap)\ket{\colorVertMap},\\
        \norm{z}^2&=\sum_{\colorVertMap\text{ injective}}\mu_H(\colorVertMap)\norm{f(\colorVertMap)}^2, &\norm{f}_H^2&\leq2\norm{z}^2.
    \end{align*}
    Note that the norm $\norm{f}_H$ sums over all assignments, while $\norm{z}$ only includes injective assignments due to the application of $\Pdist$ in \cref{lem:good_vectors_representation}. Furthermore, for each choice of $\Fbd_i$, the coefficient $f$ is in $\Scoef\cap(\Kdist)^\perp$.

    Next, we separate the terms in $f$ based on the fixed data $\Wfix_a$. Specifically, let $H_i$ be the connected component containing $a$. For each $\Fbd_i$, write $f=f_a + g$ where,
    \begin{align}
        f_a&\coloneq
        \begin{cases}
            \EX[f\mid\Wfix_a,X_{U\setminus U_i}],&\Fbd_i=\emptyset,\\
            \Id[a\in\Ctree_c]f,&\Fbd_i\ne\emptyset\quad(c\in\Fbd_i),
        \end{cases}
        &g&\coloneq f-f_a.\label{eq:edge_coefficient_decomposition}
    \end{align}
    If $\Fbd_i = \emptyset$, then we may consider the unique decompositions $f=\sum_{v \in U_i} f_v$ and $g=\sum_{v\in U_i\setminus\{a\}}f_v$ from \cref{lem:coeff_decomp}. Otherwise, $f_a$ and $g$ have disjoint supports. Thus in both cases, $f_a$ depends on the component $H_i$ only through the fixed part $\Wfix_a$, each summand in $g$ depends on a tree component different from $\Ctree_a$ and
    \begin{align}
        \norm{f_a}_H^2+\norm{g}_H^2=\norm{f}_H^2\leq2\norm{z}^2. \label{eq:edge_coefficient_split}
    \end{align}
    We may write $z=z_a + z_g$ where
    \begin{align*}
         z_a&\coloneq\sum_{\colorVertMap\text{ injective}}\sqrt{\mu_H(\colorVertMap)}f_a(\colorVertMap)\ket{\colorVertMap},\\
        z_g&\coloneq\sum_{\colorVertMap\text{ injective}}\sqrt{\mu_H(\colorVertMap)}g(\colorVertMap)\ket{\colorVertMap}.
    \end{align*}
    Hence the goal is to bound how much $J,J^\dagger$ move $z_a,z_g$ away from the good space. Let us consider the insertion $J$ first. For a structural vertex $x$ with $b(x)\in \middleEdge$, we let $p(x)$ be the bound on the probability that querying at $x$ returns a boundary vertex or a vertex already in the forest,
    \begin{align*}
        p(x)&\coloneq\sum_{u\in\Vbd}\probAssignEdge{\alpha}{x,u}+\frac{2(t+1)}{s(x)}.
    \end{align*}
    The second term bounds the probability of hitting a vertex in the forest as $\abs{U} \leq 2(t+1)$. By \cref{eq:kernel},
    \begin{align*}
        \sum_{x:\,b(x)\in\middleEdge}p(x)&\leq\abs{\Vbd} +2(t+1)\sum_{x:\,b(x)\in\middleEdge}\frac1{s(x)} =O(N\errorTerm+nt).
    \end{align*}
    We are interested in the probability that an assignment contributing to $g$ incurs any of these bad events. Hence, considering the random variable $X_a$ that assigns $\colorVertMap(a)$, let us bound $p(X_a)$. There are two possibilities for the location $a$.

    If $a\notin \Sfix_v$, its location is not fixed. By \cref{lem:conditional_structural_assignments}, each possible location has the conditional probability $O(1/N)$. Therefore,
    \begin{align*}
        \EX[p(X_a)\mid\Wfix_v] &=\sum_{x:\,b(x)\in\middleEdge}p(x)\Pr[X_a=x\mid\Wfix_v]\\
        &\leq\frac{O(1)}{N}\sum_{x:\,b(x)\in\middleEdge}p(x) =O\left(\errorTerm+\frac{nt}{N}\right).
    \end{align*}
    On the other hand, if $a\in \Sfix_v$, then $a\notin \Ctree_v$ implies that $a$ is immediately across the weld, which means that it can hit an already represented vertex, so
    \begin{align*}
        p(\colorVertMap(a))\leq\frac{\abs{\Vbd}+2(t+1)}{s(\colorVertMap(a))} =O\left(\errorTerm+\frac{t}{N}\right).
    \end{align*}
    In both cases, the average $p(X_a)$ is $O((\etastep_t)^2)$. By \cref{lem:coeff_decomp}, the summands in $g$ remain orthogonal after multiplication by $\sqrt{p(X_a)}$. Hence,
    \begin{align}
        \norm{\sqrt{p(\colorVertMap(a))}\,g}_H^2
        &=\sum_{\colorVertMap}\mu_H(\colorVertMap)p(\colorVertMap(a))\norm{g(\colorVertMap)}^2=O((\etastep_t)^2)\norm{g}_H^2. \label{eq:edge_free_source_bounds}
    \end{align}
    Adding a new vertex outside the boundary preserves goodness as it does not create a collision, does not introduce a boundary vertex and does not change the fixed data. The remaining bad additions have total weight at most $p(\colorVertMap(a))$, so,
    \begin{align*}
        \norm{(\Id-\Pi_1)Jz_g}^2 =O((\etastep_t)^2)\norm{g}_H^2.
    \end{align*}
    Next, consider applying the insertion operator $J$ on $z_a$. Notice that adding an edge from $a$ to a new vertex preserves goodness. Hence the only bad event can be if we connect to a previously explored vertex. If a tree edge $ay$ joins two recorded components, the coefficient depending on $\Wfix_a$ and $\Wfix_y$ remains good because the merged tree component contains both sets of fixed data.

    We must therefore bound additions that create a cycle, cross a weld, or connect another recorded component outside the fixed data. A cycle inside $\Ctree_a$ would require two parent-color edges at its deepest vertex, contradicting proper edge-coloring. If $ay$ is a weld edge, its edge class has size $\Omega(N)$. If $ay$ is a tree edge and $y\in U_j$ such that $j\neq i$, apply the same coefficient split to $f_a$ at $y$. The part depending on $\Wfix_y$ preserves goodness as above. For the remaining cases where the edge is on a tree, $y$ is either free or fixed immediately across a weld. If $y$ is free, \cref{lem:conditional_structural_assignments} bounds the probability that $X_y = x$ for every structural vertex $x$ by $O(1/N)$, conditional on the fixed data. If $y$ is fixed across the weld, the edge class has size $\Omega(N)$, giving the same bound. Hence \cref{lem:coeff_decomp} gives the squared error $O(1/N)\norm{f_a}_H^2$ for every $y$. 
    
    The query $(l,\alpha)$ identifies some edge $ay$, so the automorphisms over $H + ay$ are those of $H$ which fix $y$. By \cref{obs:goodness_properties}, the squared normalization factor is $\abs{\Aut(H)}/ \abs{\Aut(H+ay)}$, which counts the number of possible images of $y$ under $\Aut(H)$. These endpoints are already counted in the sum over $U$. Summing over at most $\abs{U} \leq 2(t+1)$ endpoints gives us
    \begin{align*}
        \norm{(\Id-\Pi_1)Jz_a}^2
        =O\left(\frac{t+1}{N}\right)\norm{f_a}_H^2.
    \end{align*}
    For $J^\dagger$ and $z\in \ran(\Pi_1)$, use the same decomposition of $f$ and $z$ from \cref{eq:edge_coefficient_decomposition} and let $ay$ be the queried edge in the abstract forest $H$. Suppose that after deleting the edge $ay$, the endpoint $y$ remains in the new $H$. For $z_a$, the fixed data around $a$ and $y$ in the two resulting components contain all the data needed by $f_a$ and determine the factor $1/s(X_a)$ introduced by deletion. The boundary vertices of the original component still lie in these fixed tree components, so $(\Id-\Pi_0)J^\dagger z_a=0$. Hence it suffices to focus on $z_g$. For this vector, notice that there are at most $2(t+1)$ database instances that lead to the same database $I$ after applying $J^\dagger$. Applying Cauchy-Schwarz and \cref{eq:edge_free_source_bounds},
    \begin{align}
        \norm{J^\dagger z_g}^2 &\leq2(t+1)\sum_{\colorVertMap\text{ injective}} \frac{\mu_H(\colorVertMap)\norm{g(\colorVertMap)}^2}{s(X_a)}\notag\\
        &\leq\norm{\sqrt{p(X_a)}g}_H^2=O((\etastep_t)^2)\norm{g}_H^2,\label{eq:edge_delete_retained_bound}
    \end{align}
    since the deleted edge must connect $a$ to a vertex still represented in $I$. 
    
    Instead, assume that $J^\dagger$ deletes $y$. This means that it is an unlabeled leaf in $H$. Notice that deleting $y$ combines all the databases $I$ that differ only in where $y$ is located. Let $H^{-} \coloneq H - y$ and $U^{-} \coloneq U \setminus \{y\}$ be the instances. When $\Fbd_i=\emptyset$, as $y$ is a leaf, the only summand in $g$ that depends on the assignment $X_y$ is $f_y$, which is $0$ due to \cref{lem:coeff_decomp}. When $\Fbd_i\ne\emptyset$, $y\notin\Sfix_c$ on the support of $g$. Therefore,
    \begin{align*}
        g(\colorVertMap)=\Id[X_y\notin\Vbd]g^-(\colorVertMap^-).
    \end{align*}
    Removing $y$ does not change the fixed data. The terms remain orthogonal by independence across welds, so the bound in \cref{eq:edge_free_source_bounds} also applies to \(g^{-}\). Summing over $X_y$ and using \cref{eq:edge_free_source_bounds,eq:standing-assumptions} gives us $\norm{g^-}_{H^-}^2\leq2\norm{g}_H^2$. Let $z^{-}$ be the vector on $H^{-}$ with the coefficient $h(\colorVertMap^-)$
    \begin{align*}
        h(\colorVertMap^-)&\coloneq g^-(\colorVertMap^-)+
        \sum_{x\in[M]}\probAssignEdge{\alpha}{X_a,x}
        f_a(\colorVertMap^-\cup\{y\mapsto x\}).
    \end{align*}
    Notice that both terms are good. Therefore by Cauchy-Schwarz, \cref{eq:edge_free_source_bounds,lem:good_vectors_representation},
    \begin{align*}
        \norm{J^\dagger z-z^-}^2 &\leq2\norm{\sqrt{p(X_a)}g^-}_{H^-}^2 +2\sum_{v\in U^-}\norm{\Pcoin_{yv}f_a}_H^2\\
        &=O((\etastep_t)^2)\norm{f}_H^2.
    \end{align*}
    Since $y$ is fixed, the collision argument in the proof of \cref{lem:good_vectors_representation} gives $\norm{\Pcoin_{yv}f_a}_H^2 = O(1/N) \norm{f_a}_H^2$ for each $v\in U^{-}$.

    For insertion and leaf deletion, the resulting database determines the abstract forest $H$, so the errors for different $H$ are orthogonal. Otherwise, if $y$ remains after deletion, the bound of $2(t+1)$ database instances leading to the same database $I$ in \cref{eq:edge_delete_retained_bound} applies across all forests, so the same estimate holds for their sum. Therefore, combining these estimates and taking the supremum over the unit vectors $z$ in $\ran(\Pi_0)$ and $\ran(\Pi_1)$ proves \cref{eq:edge_add_delete_good}.
\end{proof}

Next we handle the \pCLabelInv operator.

\begin{lemma}\label{lem:inverse_label_goodness}
    Let $t \leq 6q+6$ and suppose \cref{eq:standing-assumptions} holds. Then
    \begin{align*}
        \norm{(\Id - \Pgood_{t+1}) \pCLabelInv \Pgood_t} = O\left(\sqrt{\frac{t}{M}} \right).
    \end{align*}
\end{lemma}
\begin{proof}
    Fix a query label $l\in[M]$ and $r\in [0,t]$. For $\abs{L}=r$ and $l\notin\im(L)$, define
    \begin{align*}
        \calV\ket{L,C} &\coloneq\frac{1}{\sqrt{M-r}} \sum_{x\in[M]\setminus\dom(L)}\ket{L[x\to l],C},\\
        \calV^\dagger\ket{L[x\to l],C} &=\frac{1}{\sqrt{M-r}}\ket{L,C}.
    \end{align*}
    Distinct databases have disjoint sets of extensions, and each has $M-r$ extensions, so $\calV^\dagger\calV=\Id$. By \cref{eq:standing-assumptions}, $M-r\geq M/2$.

    Inserting the label at a vertex already in $H$ changes only $\forestlab$. Inserting it at a new location adds an isolated vertex $y$, whose fixed data $\Wfix_y$ specify its location $\colorVertMap(y)$. Both operations map into the good space with one additional record. Likewise, deleting the label from a vertex with an incident edge changes only $\forestlab$ and preserves the good space.

    Hence the only case we must consider is when we delete an abstract vertex $y$ without any adjacent edges such that $\forestlab(y)=l$. Let $z \in \ran(\Pgood_{t+1})$ be represented by such a forest $H$, and set $H^-\coloneq H-y$, $U^-\coloneq U\setminus\{y\}$, $\colorVertMap^-\coloneq\colorVertMap|_{U^-}$ and $m\coloneq\abs{U^-}\leq2t$.

    For the connected components $H_i$ of $H^-$, fix the sets $\Fbd_i$ and let $\mu^-\coloneq\bigotimes_i\muref_{i,\Fbd_i}$. For $y$, include both cases $\colorVertMap(y)\in\Vbd$ and $\colorVertMap(y)\notin\Vbd$. Replacing the uniform distribution on either set $A\in\{\Vbd,[M]\setminus\Vbd\}$ by the uniform distribution on $[M]$ multiplies the corresponding coefficient by $\sqrt{M/\abs{A}}$ and extends it by zero outside $A$. Thus we use the reference distribution
    \begin{align*}
        \mu(\colorVertMap^-,x)
        &=\frac{\mu^-(\colorVertMap^-)}{M},
        \qquad x=\colorVertMap(y)\in[M].
    \end{align*}
    Using the isometry in \cref{obs:goodness_properties}, we may describe the vectors using $\colorVertMap$ without changing their norms. By \cref{lem:good_vectors_representation}, choose a unique coefficient function $f$ as in \cref{eq:scoef_definition} such that
    \begin{align*}
        z&=\sum_{\colorVertMap\text{ injective}}\sqrt{\mu(\colorVertMap)}f(\colorVertMap)\ket{\colorVertMap},&
        \norm{f}_\mu^2&\leq2\norm{z}^2.
    \end{align*}
    Below $f(\colorVertMap^-,x)$ denotes $f(\colorVertMap^-\cup\{y\mapsto x\})$. Define
    \begin{align*}
        h(\colorVertMap^-)
        &\coloneq\frac{1}{\sqrt{M(M-r)}}
        \sum_{x\in[M]}f(\colorVertMap^-,x),\\
        w&\coloneq\sum_{\colorVertMap^-\text{ injective}}
        \sqrt{\mu^-(\colorVertMap^-)}h(\colorVertMap^-)\ket{\colorVertMap^-}.
    \end{align*}
    For every fixed $x$, $f(\,\cdot\,,x)\in\bigotimes_i\Scoef_{i,\Fbd_i}$ by \cref{eq:scoef_definition}, where $i$ ranges over the connected components of $H^-$. Hence $h$ belongs to the same space and $w\in\ran(\Pgood_t)$.

    The action of $\calV^\dagger$ sums only over $x\notin\colorVertMap^-(U^-)$. By Cauchy--Schwarz,
    \begin{align*}
        \norm{\calV^\dagger z-w}^2
        &\mathrel{\leq}\frac{1}{M(M-r)}\EX_{\mu^-}\left[
        \norm{\sum_{x\in\colorVertMap^-(U^-)}f(\colorVertMap^-,x)}^2\right]\\
        &\leq\frac{m}{M(M-r)}\EX_{\mu^-}
        \left[\sum_{x\in[M]}\norm{f(\colorVertMap^-,x)}^2\right]\\
        &=\frac{m}{M-r}\norm{f}_\mu^2
        =O\left(\frac{t}{M}\right)\norm{z}^2.
    \end{align*}
    The forest $H$ is recovered from $H^-$ by adding the isolated vertex labeled $l$. Different $(\Fbd_i)_i$ have disjoint assignments, so their errors are orthogonal. Automorphisms over $H$ simply rename the abstract vertices in $H$ and permute over $(\Fbd_i)_i$, so they preserve the properties of $f$ in \cref{lem:good_vectors_representation}. As $f$ is unique, summing over different $\Fbd_i$ and $x$ produces a vector $w$ invariant under $\Aut(H)$. Therefore \cref{obs:goodness_properties} gives us the same bound when using databases. Together with preservation of the good space when $y$ has an incident edge, this gives $w\in\ran(\Pgood_t)$ with the same bound for every $z\in\ran(\Pgood_{t+1})$ in the domain of $\calV^\dagger$.

    Finally, write $z\in\ran(\Pgood_t)$ as $z=z_0+z_1$, where $l\notin\im(L)$ on $z_0$ and $l\in\im(L)$ on $z_1$, with $\abs{L}=r$ and $\abs{L}=r+1$, respectively. By \cref{eq:pc_x_I,eq:plus_state_label_inv},
    \begin{align*}
        \pCLabelInv(z_0+z_1)
        &=\calV^\dagger z_1+\calV z_0+z_1-\calV\calV^\dagger z_1.
    \end{align*}
    Choose $w\in\ran(\Pgood_t)$ satisfying the bound above for $\calV^\dagger z_1$. By the insertion argument,
    \begin{align*}
        z^+&\coloneq w+\calV z_0+z_1-\calV w
        \in\ran(\Pgood_{t+1}).
    \end{align*}
    The vectors $\calV^\dagger z_1-w$ and $\calV(\calV^\dagger z_1-w)$ have the same norm and are orthogonal, since the label $l$ is absent in the first and present in the second. Therefore,
    \begin{align*}
        \norm{\pCLabelInv(z_0+z_1)-z^+}^2
        &=2\norm{\calV^\dagger z_1-w}^2\\
        &=O\left(\frac{t}{M}\right)\norm{z_1}^2
        \leq O\left(\frac{t}{M}\right)\norm{z_0+z_1}^2.
    \end{align*}
    The subspaces for different query labels and label counts $r,r+1$ are orthogonal, and $l=0$ is fixed by $\pCLabelInv$. This proves the claim, including $t=0$, when $U^-=\emptyset$.
\end{proof}

Finally, we find that \pCLabel and \xorOP cannot affect goodness.

\begin{lemma}\label{lem:forward_xor_goodness}
    Let $t \leq 6q+6$ and suppose \cref{eq:standing-assumptions} holds. Then
    \begin{align*}
        {(\Id - \Pgood_{t+1}) \cdot \pCLabel \cdot \Pgood_t} = 0,\\
        (\Id - \Pgood_t)\cdot \xorOP \cdot \Pgood_t = 0.
    \end{align*}
\end{lemma}
\begin{proof}
    Fix some query input $\ket{l,\alpha}_{\X}$ and a canonical good vector $\canVec{D}$ with $\abs{E(H)}+\abs{\dom(\forestlab)}\leq t$. By \cref{eq:reference-prob,eq:goodness_definition}, every assignment $\colorVertMap$ in $\canVec{D}$ satisfies $\colorVertMap(U_i\setminus\Ctree_{u_i})\subseteq\middleNodes$. Therefore, any vertex assigned outside of $\middleNodes$ must be in $\Ctree_{u_i}$ with its location specified by $w_i=\Wfix_{u_i}$.

    If $l\notin\im(\forestlab)\cup\{0\}$, both $\pCLabel$ and $\xorOP$ act as the identity. Otherwise, let $v\coloneq L_{\colorVertMap}^{-1}(l)$ and $w\coloneq\stepFunct(l,\alpha,L_{\colorVertMap},C_{\colorVertMap})$ be the structural vertex and its neighbor according to $I$. 
    
    Consider \pCLabel. Assume that $w\notin \{\bot,v,0\}$ as otherwise \pCLabel acts as the identity. If $w$ is from a recorded edge, then for $u=\forestlab^{-1}(l)$, there exists some abstract vertex such that, $(u,y)\in E(H), \forestcol(u,y) = \alpha$ and $w=\colorVertMap(y)$. Adding, deleting or replacing the label at $w$ only changes $\forestlab(y)$. Since $y$ has an incident edge, these changes do not remove it from $H$. Therefore, the goodness conditions in \cref{eq:goodness_definition} do not change. On the other hand, if $w$ is from an unpermuted edge, then $v,w\notin\middleNodes$. Thus $D$ determines $w$ and every equality $w=\colorVertMap(y)$. Adding or removing an isolated vertex at $w$ does not impose any further injectivity constraints. Hence the identity on \pCLabel follows.

    For \xorOP, notice that for all databases $\ket{I}$ in $\canVec{D}$, the value we XOR is the same. Hence we do not alter $\canVec{D}$, showing the second equality.
\end{proof}

Finally, applying \cref{lem:forward_xor_goodness,lem:inverse_label_goodness,lem:edge_compression_bound} to the factors in \cref{eq:cpwt_definition}, we find that applying \cpwt approximately preserves the goodness property.

\begin{theorem}\label{thm:cpwt_goodness}
    Let $t \leq 6q$ and suppose \cref{eq:standing-assumptions} holds. Then
    \begin{align*}
        \norm{(\Id - \Pgood_{t+6}) \cdot \cpwt \cdot \Pgood_t} = O\left( \etastep_t \right).
    \end{align*}
\end{theorem}

\subsection{Ordinary and exact queries}\label{sec:ordinary_to_exact}

Next, we want to transfer the goodness estimate on \cpwt to \icpwt. The only change we make to the goodness definitions is that the boundary set $B$ is slightly expanded. Specifically, we take the class of endpoints $\Vclass{b}$ defined just before \cref{lem:difference_cpwts} and use $B_b = \Vbd \cup \Vclass{b}$ as the boundary.

First we show that this change does not greatly affect the good projector.
\begin{lemma}[Boundary adjustment]\label{lem:boundary_adjust}
    Let $R\subseteq (\middleNodes \cup \Vbd) \setminus \Vweld$ and assume \cref{eq:bound_on_size_B} holds with $B$ replaced by $B\cup R$. Then,
    \begin{align*}
        \norm{\Pgood_t(B) - \Pgood_t(B\cup R)} = O\left( \sqrt{\frac{t \abs{R}}{N}} \right).
    \end{align*}
\end{lemma}
\begin{proof}
    Assume that $t\geq 1$. Fix some recorded component $H_i$ and its set of boundary points $F=\Fbd_i$ for $B$. Let $\calS = \Scoef_{i,F}$ and $S^\prime$ be the coefficient space for $B\cup R$ on the same assignments, before imposing the injectivity projector. We compare both spaces using the norm from $\muref_{i,F}$ by \cref{eq:norm_function_ref}. Let the error bound $\epsilon$ be,
    \begin{align*}
        \epsilon\coloneq\frac{C m_i\abs{R}}{N},
    \end{align*}
    where $C$ is some constant. Assume that $\epsilon \leq 1/16$ as otherwise the statement trivially follows. Let $k$ be the number of tree components containing a vertex $u$ such that $\colorVertMap(u)\in R$ and
    \begin{align*}
        k(\colorVertMap)&\coloneq\abs{\left\{\Ctree_v:v\in U_i,\ \colorVertMap(\Ctree_v)\cap R\neq \emptyset\right\}} &\Pi_j&\coloneq\Id[k=j].
    \end{align*}
    When $F\subseteq \Ctree_u$, \cref{lem:conditional_structural_assignments} gives us,
    \begin{align}
        \EX\left[k-\Id[\colorVertMap(\Ctree_u)\cap R\neq \emptyset]\mid\Wfix_u\right]\leq\epsilon.\label{eq:boundary_other_components}
    \end{align}
    To see this, consider the following. Equivalently, \cref{eq:boundary_other_components} counts the number of components $\Ctree_v\neq \Ctree_u$ containing a vertex $a$ such that $\colorVertMap(a)\in R$. Due to the fact that $R\cap \Vweld = \emptyset$, such a vertex satisfies $a\notin \Sfix_u$. For each $a\notin \Sfix_u$, \cref{lem:conditional_structural_assignments} gives us,
    \begin{align*}
        \Pr[X_a \in R| \Wfix_u] = O\left(\frac{\abs{R}}{N}\right).
    \end{align*}
    Summing over the $m_i$ vertices gives us \cref{eq:boundary_other_components}. Root $H_i$ at some abstract vertex $r$. Let $\calK$ be the space of sums $h=\sum_{v\in U_i} h_v(\Wfix_v)$, where $h_v$ is $0$ unless $v$ is the vertex of $\Ctree_v$ closest to $r$ in $H_i$, $F\subseteq\Ctree_v$ and $\colorVertMap(\Ctree_v)\cap R\neq \emptyset$. Hence $\calK \subseteq \calS$ and $\Pi_0 \calK = 0$. By \cref{eq:goodness_definition},
    \begin{align*}
        S^\prime = \Pi_0\calS \oplus\Pi_1\calK .
    \end{align*}
    To see this, consider the cases on $k$. When $k=0$, $\colorVertMap(U_i)\cap R=\emptyset$ and thus replacing $B$ by $B\cup R$ does not affect the goodness condition in \cref{eq:goodness_definition}. When $k=1$, then some specific component $\Ctree_u$ must satisfy $F\subseteq\Ctree_u$ and $\colorVertMap(\Ctree_u)\cap R\neq \emptyset$. Lastly, when $k\geq 2$, at least two distinct tree components in $H_i$ contain vertices mapped to $R$, which violates \cref{eq:goodness_definition}.

    Let us bound $\norm{(\Id - \Pi_0) f}$ for $f \in \calS \cap \calK^\perp$. Fix $a\in U_i$. If $F=\emptyset$, write $f=f_a+\sum_{v\ne a}f_v$ using \cref{lem:coeff_decomp} with root $a$. Otherwise, choose some $c\in F$ and decompose $f$ as,
    \begin{align*}
        f=f_a+g,\qquad f_a\coloneq\Id[a\in\Ctree_c]f,\qquad g\coloneq\Id[a\notin\Ctree_c]f.
    \end{align*}
    In both cases, $f_a$ depends only on $\Wfix_a$ and the terms remain orthogonal after multiplication by $Q_a\coloneq\Id[X_a\in R]$. As $Q_af_a\in\calK$, $\norm{Q_a f_a}^2 = 0$. Otherwise for $f_v$ such that $v\neq a$, \cref{lem:coeff_decomp,lem:conditional_structural_assignments} imply that
    \begin{align*}
        \norm{Q_a f_v}^2 \leq \frac{C \abs{R}}{N}\norm{f_v}^2,
    \end{align*}
    and a similar argument works for $g$. Therefore,
    \begin{align}
        \norm{(\Id - \Pi_0) f}^2 \leq \sum_{a\in U_i} \norm{Q_a f}^2 \leq \epsilon \norm{f}^2.\label{eq:boundary_zero_error}
    \end{align}

    Next we consider the case when $k=1$. Writing $h=\sum_v h_v\in\calK$, the only summand that is not $0$ is $h_v$ for the closest $v$ to $r$ in the unique component such that $\colorVertMap(\Ctree_v)\cap R\neq \emptyset$. By \cref{eq:boundary_other_components},
    \begin{align*}
        \norm{h}^2\geq\norm{\Pi_1h}^2 =\sum_v\norm{\Pi_1h_v}^2 \geq(1-\epsilon)\sum_v\norm{h_v}^2.
    \end{align*}
    For an arbitrary $k$, at most $k$ terms are nonzero. Using Cauchy-Schwarz, we find that
    \begin{align}
        \norm{h-\Pi_1h}^2 &\leq2\sum_v\EX[(k-1)\abs{h_v}^2] \leq\frac{2\epsilon}{1-\epsilon}\norm{h}^2.\label{eq:boundary_one_error}
    \end{align}
    Write any $f\in S$ as $f=g+h$ where $g\in \calS \cap \calK^\perp$ and $h\in \calK$. Letting $f^\prime = \Pi_0 g + \Pi_1 h \in S^\prime$, \cref{eq:boundary_zero_error,eq:boundary_one_error} imply
    \begin{align*}
        \norm{f - f^\prime} &\leq 2\sqrt{\epsilon}\norm{f}, \\
        \norm{f^\prime} &\geq (1-2\sqrt{\epsilon}) \norm{f}.
    \end{align*}
    Since $\Pi_0\calK=0$, this holds for every $f^\prime \in S^\prime$. Let $P_i,P_i^\prime$ be the orthogonal projectors for the coefficient spaces in $H_i$. Therefore, $\norm{P_i - P_i^\prime} = O(\sqrt{\epsilon})$. Letting $P=\bigotimes_iP_i$ and $P^\prime=\bigotimes_i P_i^\prime$ be the projectors over the entire forest, we have,
    \begin{align*}
        \norm{P - P^\prime }^2 \leq \sum_i \norm{P_i - P_i^\prime}^2 = O\left( \frac{t\abs{R}}{N} \right),
    \end{align*}
    since $\sum_i m_i\leq 2t$. Let $Q,Q^\prime$ project onto the coefficient spaces $\Pdist\ran(P)$ and $\Pdist\ran(P^\prime)$, respectively. By \cref{lem:good_vectors_representation}, every coefficient function $h\in\Pdist\ran(P)$ has a representative $f\in\ran(P)$ with $\Pdist f=h$ and $\norm{f}\leq\sqrt2\norm{h}$. The same holds for $P^\prime$. Thus \cref{lem:lin_map_subspace} with $G=\Pdist$ gives
    \begin{align}
        \norm{Q-Q^\prime}\leq\sqrt{2} \norm{P-P^\prime} = O\left(\sqrt{\frac{t\abs{R}}{N}}\right).\label{eq:boundary_projector_bound}
    \end{align}
    Take the orthogonal sum of \cref{eq:boundary_projector_bound} over $(\Fbd_i)_i$. The resulting projectors commute with $\Aut(H)$ as renaming preserves both good spaces. Applying \cref{obs:goodness_properties} completes the proof.
\end{proof}

Next we use the $\Tdiff_b$ operator from \cref{lem:difference_cpwts}.

\begin{lemma}\label{lem:boundary_preserved_by_difference}
    Suppose $\Vclass{b}\cap\Vweld=\emptyset$. The operator $\Tdiff_b$ preserves the good space with the boundary set $B_b$.
\end{lemma}
\begin{proof}
    Fix algorithm basis states $a,a'$ and a canonical good vector $\canVec{D}$ with boundary $B_b$. Since $\Vclass{b}\subseteq B_b$, \cref{eq:goodness_definition} gives that for every $i$,
    \begin{align*}
        \colorVertMap(U_i\setminus\Ctree_{u_i})\cap\Vclass{b}=\emptyset.
    \end{align*}
    Thus all vertices represented in $\Vclass{b}$ lie in the fixed tree components $\Ctree_{u_i}$, and $I_b$ is determined by the fixed data. By \cref{lem:difference_cpwts}, $\Tdiff_b$ changes only $I_b$. Its coefficients depend on $I_{\neg b}$ only through label names and their number, which are fixed by $D$. Hence they do not depend on the remaining assignments.

    As $\Vclass{b}\cap\Vweld=\emptyset$, the changed edges are inside trees. Deleting them splits the tree components $\Ctree_{u_i}$, and inserting them joins such components. Since they are entirely in the tree components, this cannot create a cycle. Each connected component left after removing the tree components $\Ctree_{u_i}$ is joined to them exactly by a single weld edge, otherwise we would already have a cycle in $H$. Since they are not altered, each remaining component stays connected to just one of the resulting fixed tree components. All vertices assigned to $B_b$ therefore still belong to these fixed tree components.

    The remaining assignments and their edge factors are unchanged. Their injectivity conditions are also unchanged, since every added or removed vertex lies in $\Vclass{b}$, which they already avoid. For each resulting description $D^\prime$, the changed edge factors, each equal to $1/\abs{E_b}$, and reference normalizing constants depend only on its fixed data and forest. The contribution is therefore a scalar multiple of $\canVec{D'}$, giving
    \begin{align*}
        (\bra{a'}\Tdiff_b\ket{a})\canVec{D}
        \in\Span\{\canVec{D'}:D'\text{ has boundary }B_b\},
    \end{align*}
    which proves the claim.
\end{proof}

\begin{proposition}\label{prop:exact_query_goodness}
    Let $t \leq 6q$ and suppose \cref{eq:standing-assumptions} holds. Then
    \begin{align*}
        \norm{(\Id - \Pgood_{t+6}) \cdot \icpwt \cdot \Pgood_t} = O\left( \sqrt{\errorTerm} + \sqrt{\frac{n\cdot (t+1)}{N}} \right).
    \end{align*}
\end{proposition}
\begin{proof}
    Let $\tcut \coloneq t + 6$ bound the number of records after the query $\cpwt$ and
    \begin{align*}
        T_b&\coloneq\prv{\leq\tcut}\Tdiff_b\prv{\leq\tcut}, & T&\coloneq\sum_bT_b.
    \end{align*}
    By \cref{lem:operator_norm_max,lem:difference_cpwts}, the operators $T_b$ are self-adjoint, $T_b T_c = 0$ for all $b,c$ such that $b\neq c$ and $\norm{T} = \max_b \norm{T_b} = O(1)$. Let $c>0$ be some constant we will set later and define,
    \begin{align*}
        \Pi_b\coloneq\begin{cases}
            \Pgood_{\tcut}(B_b) &\text{if $\tcut\abs{E_b}\leq cN$,}\\
            0 &\text{otherwise.}
        \end{cases}
    \end{align*}
    Choose $c$ to be a sufficiently small constant so that the first condition only holds when $\Vclass{b}\cap\Vweld=\emptyset$ as all edge classes $E_b$ with endpoints in $\Vweld$ have $\abs{E_b}=\Omega(N)$ by \cref{lem:colors_per_column,eq:welded_edges_def}. First assume that $b$ is such that $\Pi_b \neq 0$. Every abstract forest with at most $\tcut$ records has at most $2\tcut$ vertices, meaning that \cref{eq:standing-assumptions} gives,
    \begin{align*}
        \frac{8\tcut \abs{B_b}}{N} \leq \frac{8\tcut (\abs{\Vbd} + 2\abs{E_b})}{N} = O\left((q+1)\errorTerm+c\right)\leq\frac{1}{2}.
    \end{align*}
    Thus \cref{lem:boundary_preserved_by_difference,lem:boundary_adjust} imply that $[\Pi_b, T_b] = 0$ and
        \begin{align*}
        \norm{T_b}\norm{\Pgood_{\tcut} -\Pi_b} =O\left(\abs{E_b}^{-1/2}\sqrt{\frac{\tcut\abs{E_b}}{N}}\right) =O\left(\sqrt{\frac{\tcut}{N}}\right).
    \end{align*}
    In the other case when $\abs{E_b} > cN/\tcut$, we have $\Pi_b = 0$ and hence the same conclusions hold as well since $\norm{T_b}\leq\norm{\Tdiff_b\prv{\leq\tcut}} =O(\abs{E_b}^{-1/2})$ by \cref{lem:difference_cpwts}. As there are $O(n)$ classes, \cref{lem:orthogonal_differences} gives us,
    \begin{align*}
        \norm{[\Pgood_{\tcut}, T]} \leq 2 \sqrt{\sum_b \norm{T_b}^2 \norm{\Pgood_{\tcut} -\Pi_b}^2} =O\left(\sqrt{\frac{n \tcut }{N}}\right).
    \end{align*}
    Starting with at most $t$ records, applications of $\pCLabelInv$, $T$, and $\pCLabelInv$ from \cref{eq:graph-difference} leave at most $t+1, t+4$ and $t+5$ records, respectively. The good state projectors preserve the number of records. Therefore,
    \begin{align*}
        \norm{(\Id - \Pgood_{\tcut}) \cdot \icpwt \cdot \Pgood_t} & \leq O(\etastep_t) + \norm{(\Id - \Pgood_{\tcut}) \cdot (\icpwt - \cpwt) \cdot \Pgood_t} &\mbox{(By \cref{thm:cpwt_goodness})}\\
        &\leq O\left(\etastep_t + \frac{1}{N}\right) + \norm{(\Id - \Pgood_{\tcut}) \cdot \pCLabelInv T\pCLabelInv \cdot \Pgood_t} &\mbox{(By \cref{lem:difference_cpwts})}\\
        &\leq  O\left(\etastep_t + \frac{1}{N} + \sqrt{\frac{t+1}{M}} \right) + \norm{ \left[ \Pgood_{\tcut}, T \right] } &\mbox{(By \cref{lem:inverse_label_goodness})}\\
        &\leq O \left( \sqrt{\errorTerm} + \sqrt{\frac{n\cdot (t+1)}{N}} \right).
    \end{align*}
    Finally, \icpwt adds at most $5$ records each round, meaning the databases have at most $t+5$ records.
\end{proof}

\subsection{Single-cycle distribution}\label{sec:single_cycle}

Recall that $\multDistr$ is the product distribution we used over our graph instances and $\singDistr$ is $\multDistr$ conditioned on the case that the weld edges form a single cycle. Let us compare the acceptance probabilities under these distributions.

\begin{lemma}\label{lem:mult_to_single_cycle}
    Let $p_{\calD}(\alg)$ be the acceptance probability of a quantum algorithm \alg making $q\geq 1$ queries to an oracle sampled from $\calD$. Then,
    \begin{align*}
        p_{\singDistr}(\alg) = O\left( p_{\multDistr}(\alg) + \frac{n\cdot q^2}{N}  \right).
    \end{align*}
\end{lemma}
\begin{proof}
    For each color $\eta\in \colorSet$, define the vertices at the ends of weld edges colored $\eta$ as,
    \begin{align*}
        &L_\eta \coloneq \col(n) \setminus \colClass{n,\eta}, &R_\eta \coloneq \col(n+1) \setminus \colClass{n+1,\eta}\\
        m_\eta&\coloneq\abs{L_\eta}=\abs{R_\eta}
        =N-\colClassSize{n,\eta}=\Theta(N),
    \end{align*}
    where the estimate on the size of $m_\eta$ is due to \cref{lem:colors_per_column}. A color-$\eta$ matching is some bijection $s:L_\eta\to R_\eta$ which represents the edge set $\{ \{u,s(u)\}: u\in L_\eta\}$.

    Let $h=(h_\eta)_{\eta\in \colorSet}$ be the weld edges of a graph sampled from \singDistr and $B=(B_\eta)_{\eta \in \colorSet}$ some set of recorded values that have been read, where $B_\eta \subseteq \{(u,h_\eta(u)):u\in L_\eta\}$. Conditioning on the record $B$ means that they are present. Let $U_\eta \coloneq L_\eta \setminus \dom(B_\eta)$, $V_\eta \coloneq R_\eta \setminus \im(B_\eta)$ and $r_\eta \coloneq \abs{U_\eta} = \abs{V_\eta} = m_\eta - \abs{B_\eta}$. Notice that for $u\in U_\eta, v\in V_\eta$ and $r_\eta>1$,
    \begin{align}
        p_B^\eta(u,v) \coloneq \Pr[h_\eta(u) = v | B] \leq \frac{1}{r_\eta-1}.\label{eq:single_cycle_sequential_marginals}
    \end{align}
    To see this, first condition on complete assignments of color edges for the other colors. These fixed edges along with $B_\eta$ form paths with $r_\eta$ free nodes on each side. Notice that both $u$ and $v$ must be endpoints of their respective paths connecting them. First, suppose that for one of $u$ or $v$, the other endpoint on this path is on the opposite side. Deleting this path from a completed cycle and joining its connecting edges allows it to be reinserted exactly in one place for each of the $r_\eta - 1$ possible partners. Hence $\{u,v\}$ has probability $\tfrac{1}{r_\eta - 1}$, unless it joins the path's endpoints, in which case it is $0$.
    
    Otherwise, $u$ is part of a path both of whose endpoints are on the left. Each right endpoint on a path with opposite-side endpoints has probability $\tfrac{1}{r_\eta-1}$ by the argument above, while the rest have equal probability due to symmetry. Therefore, their common probability must be less than $\tfrac{1}{r_\eta - 1}$ as otherwise the probabilities would sum over $1$. Hence \cref{eq:single_cycle_sequential_marginals} follows by averaging. For $r_\eta\geq 1$, let,
    \begin{align*}
        t_B^{\eta}(u,v) \coloneq 1 - (r_\eta-1)p_B^\eta(u,v).
    \end{align*}
    Notice that by \cref{eq:single_cycle_sequential_marginals}, $t_B^{\eta}(u,v)$ is nonnegative. Furthermore,
    \begin{align*}
        \sum_{v\in V_\eta}t_B^{\eta}(u,v) = \sum_{u\in U_\eta} t_B^{\eta}(u,v) = r_\eta-(r_\eta-1) = 1.
    \end{align*}
    We create the assignment $B$ using the following process. For every color $\eta$, start with $B_\eta =\emptyset$ and let $\phi_\eta:R_\eta \to R_\eta$ be the identity. One color at a time in some order, repeat the following process until $U_\eta$ is empty.
    \begin{enumerate}
        \item Uniformly sample $u~\sim U_\eta$ and let $y=h_\eta(u)$.
        \item With probability $1-\tfrac{1}{r_\eta}$, set $z=y$. Otherwise, sample $z\sim V_\eta$ with probability $t_B^{\eta}(u,\cdot)$.
        \item Add $(u,y)$ to $B_\eta$ and exchange the mappings of $\phi_\eta(y)$ and $\phi_\eta(z)$.
    \end{enumerate}
    Afterwards, let $g = (g_\eta)_{\eta \in \colorSet}$ where $g_\eta = \phi_\eta \circ h_\eta$. Notice that for each fixed $u$,
    \begin{align*}
        \Pr[z=v|B,u] = \left(1 - \frac{1}{r_\eta}\right) p_B^\eta(u,v) + \frac{1}{r_\eta} t_B^{\eta}(u,v) = \frac{1}{r_\eta},
    \end{align*}
    where $B$ is the record before the current $u$ was handled.
    Notice that after swapping $\phi_\eta(y)$ and $\phi_{\eta}(z)$, we have $\phi_\eta(h_\eta(u))$ is never changed again. Therefore each value of $g_\eta$ is uniform over the right vertices that were not used beforehand. Therefore all $g_\eta$ are independent and uniform. Note that this extends when conditioned on previous random choices as they depend on $h$ through $B$. Therefore, $g$ follows the distribution of weld edges under $\multDistr$ and hence that from \cref{sec:construction_welded_trees}.

    Fix some $h$ and some records $B$. On the second case in step 2, $y=h_\eta(u)$ is uniform over $V_\eta$ as it is bijective and $z$ is uniform as,
    \begin{align*}
        \frac{1}{r_\eta} \sum_{u\in U_\eta} t_B^{\eta}(u,v) = \frac{1}{r_\eta}.
    \end{align*}
    The fixed value of $\phi_\eta(v)$ only changes in this branch when $v\in \{y,z\}$. Hence,
    \begin{align*}
        \Pr[\phi_\eta(v) \text{ changes}|h,B] \leq \frac{1}{r_\eta} \left( \frac{1}{r_\eta} + \frac{1}{r_\eta}\right) = \frac{2}{r_\eta^2}.
    \end{align*}
    For a fixed $x\in L_\eta$, we have that $h_\eta(x)$ is in $V_\eta$ when $x$ has not been read yet. This has probability $\tfrac{r_\eta}{m_\eta}$ as the order is uniformly random and independent of $h$. By union bounding,
    \begin{align}
        \Pr[g_\eta(x) \neq h_\eta(x)|h] \leq \sum_{r\in [2,m_\eta]} \frac{r}{m_\eta} \frac{2}{r^2} = O\left(\frac{n}{N}\right).\label{eq:single_cycle_sequential_discrepancy}
    \end{align}
    The inverse mappings have the same bound due to symmetry. Sample some permutations on other edges besides the weld and vertex labels independently of $(g,h)$ and use them to obtain the instances of welded trees graph $\sigma\sim \multDistr$ and $\sigma^\prime \sim \singDistr$ built from $g$ and $h$ respectively. By \cref{eq:single_cycle_sequential_discrepancy},
    \begin{align*}
        \Pr[\wtF_\sigma(l,\eta) \neq \wtF_{\sigma^\prime}(l,\eta) | \sigma^\prime ] = O\left(\frac{n}{N}\right).
    \end{align*}
    Applying the hybrid argument of~\cite{BBBV97}, taking the expectation over $(\sigma,\sigma^\prime)$ and applying Cauchy-Schwarz,
    \begin{align*}
        \EX \norm{\ket{\psi_{\alg,q}^{\wtO_\sigma}} - \ket{\psi_{\alg,q}^{\wtO_{\sigma^\prime}}}}^2 &\leq q\sum_{j\in [0,q-1]} \EX\norm{(\wtO_\sigma - \wtO_{\sigma^\prime}) \ket{\psi_{\alg,j}^{\wtO_{\sigma^\prime}}} }^2\\
        &= O\left( \frac{n\cdot q^2 }{N}\right)
    \end{align*}
    Therefore by the triangle inequality and projecting on the accepting output,
    \begin{align*}
        p_{\singDistr}(\alg) &= O\left( p_{\multDistr}(\alg) + \frac{n\cdot q^2}{N}  \right).\qedhere
    \end{align*}
\end{proof}

\subsection{Putting it all together}

Finally, we may compose all the statements together.

\begin{proof}[Proof of \cref{thm:pathfinding_hard}]
    We may assume \cref{eq:standing-assumptions} as otherwise the statement is trivial. 
    
    By \cref{eq:goodness_definition}, the empty database $\ket{\bot}_{\I}$ is good. By \cref{obs:icpwt_properties}, after $q$ queries to \icpwt, all databases in the support have $\abs{I}\leq 3q$. Algorithm unitaries preserve goodness, so telescoping \cref{prop:exact_query_goodness} over the $q$ steps gives
    \begin{align}
        \norm{(\Id-\Pgood_{6q})\ket*{\psi_{\alg,q}^{(\icpwt)}}} = O\left(q\sqrt{\errorTerm}+q\sqrt{\frac{n\cdot q}{N}}\right).\label{eq:goodness_probability}
    \end{align}
    Let $\alg_{\mathrm{ver}}$ be the verification procedure in \cref{alg:standard_game}, which uses $O(q)$ queries and accepts every successful path under either distribution. Its acceptance probability under $\multDistr$ is $\probWTO$. For inputs drawn from $\singDistr$, we get
    \begin{align*}
        \Pr[\text{$\alg$ succeeds}] &\leq p_{\singDistr}(\alg_{\mathrm{ver}}) &\mbox{(By \cref{alg:standard_game})}\\
        &=O\left(\probWTO+\frac{n\cdot q^2}{N}\right) &\mbox{(By \cref{lem:mult_to_single_cycle})}\\
        &=O\left(\probCross+\frac{q}{\permSize}+\frac{n\cdot q^2}{N}\right) &\mbox{(By \cref{thm:oracle_path_verification})}\\
        &=O\left(\norm{(\Id-\Pgood_{6q})\ket*{\psi_{\alg,q}^{(\icpwt)}}}^2+\frac{q}{\permSize}+\frac{n\cdot q^2}{N}\right) &\mbox{(By \cref{eq:goodness_implies_no_crossing})}\\
        &=O\left(\frac{q}{\permSize}+q^2\errorTerm+\frac{n\cdot q^3}{N}\right) &\mbox{(By \cref{eq:goodness_probability})}.
    \end{align*}
    Taking $\distWeld=\floor{2n/3}$ gives $\permSize=\Theta(N^{1/3})$ and $\errorTerm=\Theta(N^{-2/3})$, so constant success probability requires $\Omega((N/(n+1))^{1/3})$ queries.
\end{proof}

\section*{AI Disclosure}

The results of this paper were obtained using ChatGPT 6 Astra to build upon our previous work~\cite{MP26,CCCG26}, establishing a stronger bound. Combining ideas from those papers and using a shifted boundary of permuted edges gave an $\Omega(N^{1/4})$ bound. The final bound was obtained using a further AI-generated suggestion to sharpen \cref{lem:pin_witness} using the exact compressed oracle, along with refinements to the good space construction and its properties.
The proof of \cref{lem:coeff_decomp} was written by GPT-6, while the rest of the paper was written by the authors.

In \cite{MP26} the main ideas and original proofs were developed by the authors, who wrote the manuscript in its entirety. ChatGPT 6 Astra was used to find a much simpler proof of the reduction to a single-cycle weld, generalization of the compressed oracle to arbitrarily-sized permutations statements and was used to discuss correctness of the proofs.
In \cite{CCCG26}, the overall proof strategy was developed by the authors without the use of AI. ChatGPT 6 Astra was used to develop proof ideas in the later stages of the work, and in particular, identified an error in a human-generated proof of a key lemma (the ``clipping lemma''), which it then corrected.

The authors take sole responsibility for the correctness of this work.

\bibliographystyle{alphaurl}
\bibliography{main}

\appendix

\section{Quantum walk analysis}\label{sec:quantum_walk_analysis}

To build intuition for the compressed oracle model, in this appendix we describe the information that is (and is not) tracked in the database when running the quantum walk exit-finding algorithm \cite{CCD+03}, using the notation of \cref{sec:wt_oracle}.

Let $H_\sigma$ be the adjacency matrix specified by $\sigma=(\labelPerm,\colorPerm)$, in the vertex-label basis. We sample $\sigma$ from $\multDistr$, counting edge multiplicities if the weld has parallel edges, giving the same column dynamics as for a single-cycle weld.

Define the column states
\begin{align*}
    \ket{c;\sigma} &\coloneq \frac{1}{\sqrt{\abs{\col(c)}}}\sum_{v\in\col(c)}\ket{\labelPerm(v)},\qquad c\in[0,2n+1].
\end{align*}
Every vertex in a column has the same number of neighbors in each adjacent column. Therefore, these states span an invariant subspace of $H_\sigma$, in which $H_\sigma$ acts as the matrix $h$ with nonzero entries
\begin{align}
    h_{c,c+1}=h_{c+1,c}&=\begin{cases}
        \sqrt{2},&c\neq n,\\
        2,&c=n,
    \end{cases}\qquad c\in[0,2n].\label{eq:ct_column_hamiltonian}
\end{align}
In particular, $h$ is independent of $\sigma$. 

Therefore
\begin{align*}
    e^{-iH_\sigma t}\ket{0} &= \sum_{v\in V_n} u_{\col(v)}(t) \ket{\labelPerm(v)}
\end{align*}
where
\begin{align*}
    u_c(t):=\frac{(e^{-iht})_{c,0}}{\sqrt\abs{\col(c)}}.
\end{align*}
The analysis in~\cite{CCD+03} gives inverse-polynomial probability of reaching the exit for a suitable randomly chosen polynomial evolution time, and describes a simulation using polynomially many queries with sufficiently small inverse-polynomial error.

To describe the databases, purify the random permutations and write only the vertex-label register of the algorithm. Suppressing the time argument, the resulting state is
\begin{align}
    \ket{\Psi(t)}&=\left(u_0(t)\ket{0}_{\A}\ket{T_\bot}
    +\frac{1}{\sqrt{M}}\sum_{\substack{v\in V_n\setminus\{0\}\\ l\in[M]}}u_{\col(v)}(t)\ket{l}_{\A}\ket{T_{[v\to l]}}\right)
    \otimes\bigotimes_{b\in\middleEdge}\ket{T_\bot}_{\regP_b}\label{eq:ct_walk_purification}
\end{align}
(recall $M=2^{2n}$).

Let $\ket{\widehat\Psi(t)}=\calI_{\regP}\ket{\Psi(t)}$, with $\calI_{\regP}$ as in \cref{eq:isometry_cpwt}, and define the normalized current-record state
\begin{align}
    \ket{\Phi(t)}&\coloneq u_0(t)\ket{0}_{\A}\ket{\bot}_{\Lab\C}
    +\frac{1}{\sqrt{M}}\sum_{\substack{v\in V_n\setminus\{0\}\\ l\in[M]}}u_{\col(v)}(t)\ket{l}_{\A}\ket{[v\to l]}_{\Lab}\ket{\bot}_{\C}.\label{eq:ct_walk_database}
\end{align}

This state provides a good approximation of the purified evolution at all times.

\begin{proposition}\label{prop:ct_walk_records}
    At every time $t$, the state $\ket{\widehat\Psi(t)}$ is supported on databases with $C=\bot$ and $\abs{L}\leq1$. Furthermore,
    \begin{align*}
        \norm{\ket{\widehat\Psi(t)}-\ket{\Phi(t)}}&\leq\sqrt{\frac{2\abs{V_n}}{M}}=O(N^{-1/2}).
    \end{align*}
\end{proposition}

\begin{proof}
In \cref{eq:ct_walk_purification}, the edge permutations remain uniform and the label-permutation state lies in $\calH_{M,\leq1}$. Thus \cref{eq:degree-records} establishes the support statement.
For the norm bound, identify the one-record basis $\ket{[x\to y]}$ with $\ket{x}\otimes\ket{y}$, where $x,y\in[M]$. Write $\ket{+}=M^{-1/2}\sum_{x\in[M]}\ket{x}$, $P_0=\Id-\ketbra{+}$, and $P=P_0\otimes P_0$ on this database space. The overlaps of the extension states are $1$ for identical records, $0$ for incompatible records, and $1/(M-1)$ for records with different inputs and outputs. A direct calculation therefore gives
\begin{align*}
        \Gamma_{M,1}^{\dagger}\Gamma_{M,1}&=\frac{M}{M-1}P,&
        G_{M,1}&=\frac{M}{M-1}\Pi_{M,1}.
\end{align*}
    Hence the normalization in \cref{ex:isometry-definition} can be evaluated explicitly. Suppress the common empty edge database and set
\begin{align*}
        s&\coloneq \sum_{v\in V_n\setminus\{0\}}u_{\col(v)}(t),&
        \beta^2&\coloneq\sum_{v\in V_n\setminus\{0\}}\abs{u_{\col(v)}(t)}^2,&
        \ket{\Phi_+}&\coloneq\ket{\Phi(t)}-u_0(t)\ket{0}_{\A}\ket{\bot}_{\Lab}.
\end{align*}
Applying the isometry separately to levels zero and one yields
\begin{align*}
       \ket{\widehat\Psi(t)}&=u_0(t)\ket{0}_{\A}\ket{\bot}_{\Lab}
        +\frac{s}{\sqrt M}\ket{+}_{\A}\ket{\bot}_{\Lab}
        +\sqrt{\frac{M}{M-1}}P\ket{\Phi_+},\\
        \norm{P\ket{\Phi_+}}^2&=\left(1-\frac1M\right)\left(\beta^2-\frac{\abs{s}^2}{M}\right),
\end{align*}
where $P$ acts only on the database. Since both $\ket{\widehat\Psi(t)}$ and $\ket{\Phi(t)}$ are normalized, we have
\begin{align*}
        \norm{\ket{\widehat\Psi(t)}-\ket{\Phi(t)}}^2
        &=2\beta^2(1-\sqrt{1-1/M})+\frac{2\sqrt{1-1/M}\abs{s}^2}{M}
        \leq\frac{2\abs{V_n}}{M},
\end{align*}
where we used $1-\sqrt{1-1/M}\leq1/M$, $\beta^2\leq1$, and $\abs{s}^2\leq(\abs{V_n}-1)\beta^2$.
\end{proof}

Finally, consider an algorithm that simulates $e^{-iH_\sigma t}\ket{0}$ using oracle queries with norm error at most $\delta$, uniformly in $\sigma$.

By \cref{eq:exact_intertwining_cpwt}, replacing its queries by \icpwt gives exactly the isometric image of its purified output. Since an isometry preserves norm, this output is within
\begin{align*}
    \delta+\sqrt{\frac{2\abs{V_n}}{M}}
\end{align*}
of $\ket{\Phi(t)}$ with clean work registers. (This is a statement about the completed simulation; the state of the simulation after intermediate oracle queries need not have empty edge databases.)

Thus the continuous-time walk has a compressed-oracle description that retains its current position without a path, consistent with the interference responsible for its speedup.

\end{document}